\documentclass[transmag]{IEEEtran}
\usepackage{latexsym}
\usepackage{graphicx}
\usepackage{subfigure}
\usepackage{multirow}
\usepackage{amsfonts,amssymb,amsmath}
\usepackage{hyperref}
\usepackage{bm}
\usepackage{booktabs}
\def\BibTeX{{\rm B\kern-.05em{\sc i\kern-.025em b}\kern-.08em T\kern-.1667em\lower.7ex\hbox{E}\kern-.125emX}}

\usepackage{algorithm} 		
\usepackage{algorithmic} 	
\usepackage{bm}
\usepackage{graphicx}
\usepackage{amsmath} %
\usepackage{flushend}
\usepackage{cite}
\usepackage{amssymb}
\usepackage{multirow}

\usepackage{epstopdf}
\usepackage{exscale}
\usepackage{relsize}
\usepackage{color}
\usepackage{diagbox}
\usepackage{slashbox}
\usepackage{booktabs}
\usepackage{array}
\usepackage{makecell}
\usepackage{multicol}

\newcolumntype{I}{!{\vrule width 1.5pt}}

\newlength\savewidth

\usepackage{subeqnarray}
\usepackage{enumerate}
\usepackage{flushend}
\usepackage{subfigure}

\begin{document}

\title{A Survey on Artificial Noise for Physical Layer Security: Opportunities, Technologies, Guidelines, Advances, and Trends}

\author{Hong Niu, Yue Xiao, Xia Lei, Jiangong Chen, Zhihan Xiao, Mao Li, and Chau Yuen,~\IEEEmembership{Fellow, IEEE}
\thanks{
This work has been accepted for publication in IEEE Communications Surveys \& Tutorials, 2025.

H. Niu, Y. Xiao, X. Lei, J. Chen, Z. Xiao, and M. Li are with the National Key Laboratory of Science and Technology on Communications, University of Electronic Science and Technology of China, 611731, Sichuan, China (Email: niuhong@std.uestc.edu.cn;
xiaoyue@uestc.edu.cn;
leixia@uestc.edu.cn;
jg$\_$chen@std.uestc.edu.cn;
202221220122@std.uestc.edu.cn;
2385353188@qq.com).

C. Yuen is with the School of Electrical and Electronics Engineering, Nanyang Technological University, Singapore 639798 (email: chau.yuen@ntu.edu.sg).}
}

\IEEEtitleabstractindextext{
\begin{abstract}
Due to the broadcast nature of wireless communications, physical-layer security has attracted increasing attention from both academia and industry. Artificial noise (AN), as one of the promising physical-layer security techniques, leverages the spatial degree-of-freedom of channels to effectively enhance the security of wireless communications. In contrast to other physical-layer security techniques, the key distinguishing feature of AN lies in its ability to generate specific interfering signals according to channel characteristics, thereby increasing the secrecy capacity by reducing the wiretap channel capacity without affecting the legitimate channel capacity. Hence, this paper provides a comprehensive and up-to-date survey of AN, including its evolution, modeling, background, applications, and future trends. Initially, we introduce the development, fundamentals, and technical background of AN.
Subsequently, we review the current state of research across various AN-empowered scenarios and AN-combined technologies.
Finally, we discuss open challenges and outline potential future directions for AN-aided wireless security.
\end{abstract}

}

\maketitle

\section*{List of Abbreviations}

\begin{table}[!htp]
\begin{tabular}{ll}
2D&Two-dimensional\\
3D&Three-dimensional\\
5G&Fifth-generation\\
6G&Sixth-generation\\
AD&Alternating direction\\
AF&Amplify-and-forward\\
AFF&Artificial fast fading \\
AG&Air-ground\\
Alice&Legitimate transmitter\\
AN&Artificial noise\\
ANE&Artificial noise elimination\\
ANH&Artificial noise hopping\\
AN-SM&Artificial-noise-aided spatial modulation\\
ANSR&Artificial noise to signal ratio\\
ARQ&Automatic-repeat-request\\
AWGN&Additive white Gaussian noise \\
BCD&Block coordinate descent \\
BD&Block diagonalization \\
BER&Bit-error rate\\
Bob&Legitimate receiver\\
BS&Base station\\
CB&Cooperative beamforing\\
CBS&Cognitive base station\\
CJ&Cooperative jammer\\
CP&Cyclic prefix\\
CPS&Cyber physical system\\
CR&Cognitive radio\\
CRN&Cognitive radio network\\
CSI&Channel state information\\
CU&Cognitive user\\
D2D&Device-to-device\\
DCAN&Data carrying artificial noise\\
DF&Decode-and-forward\\
DFRC&Dual-functional Radar Communication\\
DM&Directional modulation\\
DNN&Deep neural network\\
\end{tabular}
\label{tbl}
\end{table}

\begin{table}[!htp]
\begin{tabular}{ll}
DoF&Degree-of-freedom\\
DQSM&Differential quadrature spatial modulation \\
DRFC&Dual-functional radar-communication\\
EH&Energy harvesting\\
ER&Energy receiver\\
ESR&Ergodic secrecy rate\\
Eve&Eavesdropper\\
FD&Full-duplex\\
FDMA&Frequency-division multiple access\\
FFT&Fast Fourier Transform\\
FH&Frequency hopping \\
GPI&General power iterative\\
GSM&Generalized spatial modulation\\
HD&Half-duplex\\
IA&Interference alignment\\
IFFT&Inverse fast Fourier transform\\
IoT&Internet of things\\
IOS&Intelligent omni surface\\
IQ&In-phase and quadrature \\
ISAC&Integrated sensing and communication\\
LED&Light-emitting diode\\
LoS&Line-of-sight\\
MAC&Medium-access control\\
MEC&Mobile-edge computing \\
MIMO&Multiple-input multiple-output\\
MIMOME&Multiple-input multiple-output multiple-antenna-eavesdropper\\
MIMOSE&Multiple-input multiple-output single-antenna-eavesdropper\\
MISO&Multiple-input single-output\\
MISOME&Multiple-input single-output multiple-antenna-eavesdropper\\
MISOSE&Multiple-input single-output single-antenna-eavesdropper\\
MLI&Main-lobe-integration \\
MM&Majorization-minimization\\
MMA&Multiple multiple-antenna\\
mMIMO&Massive multiple-input multiple-output\\
mmWave&Millimeter wave\\
MRC&Maximal ratio combining\\
MRT&Maximal ratio transmission\\
MSA&Multiple single-antenna\\
MSE&Mean squared error\\
MU&Multi-user\\
nLoS&Non-light-of-sight\\
NOMA&Non-orthogonal multiple access\\
NR&Non-regenerative\\
OFDM&Orthogonal frequency division multiplexing\\
OM&Oblique manifold\\
OSI&Open system interconnection\\
OSTBC&Orthogonal space-time block code\\
PAM&Pulse amplitude modulation\\
PAPR&Peak-to-average power ratio\\
PLS&Physical-layer security\\
PMAN&Power minimized artificial noise\\
\end{tabular}
\label{tbl}
\end{table}

\begin{table}[!htp]
\begin{tabular}{ll}
PU&Primary user\\
PLA&Physical layer authentication\\
PZ&Protected zone\\
QoMS&Quality of multicast service\\
QoS&Quality of service\\
QVP&Quality of violation probability\\
RCI&Regularized channel inversion\\
RF&Radio frequency\\
RIS&Reconfigurable intelligent surface\\
RLC&Rateless codes\\
SCA&Successive convex approximation \\
SDP&Semidefinite program\\
SDR&Semidefinite relaxation\\
SEE&Secrecy energy efficiency\\
SEEM&Secrecy energy efficiency maximization\\
SIC&Self-interference cancellation\\
SIMOME&Single-input multiple-output multiple-antenna-eavesdropper\\
SIMOSE&Single-input multiple-output single-antenna-eavesdropper\\
SER&Symbol error rate \\
SINR&Signal to interference plus noise ratio\\
SISO&Single-input single-output\\
SISOME&Single-input single-output multiple-antenna-eavesdropper\\
SISOSE&Single-input single-output single-antenna-eavesdropper\\
SKG&Secret key generation\\
SM&Spatial modulation\\
SMA&Single multiple-antenna\\
SNR&Signal to noise ratio\\
SOP&Secrecy outage probability\\
SRM&Secrecy rate maximization\\
SSA&Single single-antenna\\
SSR&Secrecy sum rate\\
STBC&Space time block code\\
STLC&Space time line code\\
STM&Secrecy throughput maximization\\
SU&Secondary user\\
SVD&Singular value decomposition\\
SWIPT&Simultaneous wireless information and power transfer\\
TAS&Transmit antenna selection\\
TDD&Time division duplexing\\
THP&Tomlinson-Harashima precoder\\
THz&Terahertz\\
TSPS&Time-switching power-splitting\\
UAV&Unmanned aerial vehicle \\
UC-SM&Unitary coded spatial modulation \\
UTWR&Untrusted two-way relay\\
VAA&Virtual antenna array\\
VLC&Visible light communication \\
WPT&Wireless power transfer\\
ZF&Zero-forcing\\

\end{tabular}
\label{tbl}
\end{table}

\section{Introduction}

\IEEEPARstart{D}{uring} the last decades, wireless communications have been proliferating with the maturity commercialization of the fifth-generation (5G) and the forthcoming development of the sixth-generation (6G) technologies \cite{b1}. 
Due to the nature of broadcast propagation, the data transmission in wireless communication suffers from a critical threat of information leakage, which calls for the paramount importance of improving wireless communications security\cite{b4}. As a widely used protocol architecture for wireless communications, the open system interconnection (OSI) model \cite{b5} consists of the application layer \cite{b6}, presentation layer \cite{b7}, session layer, transport layer \cite{b8}, network layer \cite{b9}, medium-access control (MAC) layer \cite{b10}, and physical layer \cite{PLSS1,PLSS2,PLSS3,PLSS4,PLSS5,PLSS6,PLSS7,PLSS8,PLSS9,PLSS10,PLSS11,PLSS12,PLSS13,PLSS14,PLSS15,PLSS16} from top to bottom. In order to comprehensively and effectively ensure the security of wireless communications, relevant security threats and vulnerabilities in each protocol layer are individually protected to meet the security requirements of authenticity, confidentiality, integrity, availability, and so on \cite{b13}. Generally above the network layer, the information security has been conventionally guaranteed by using pre-shared keys \cite{b14}. Although the cryptographic mechanism indeed enhances the secrecy performance of communications, it suffers from the redundant secret key distribution and management process, especially in high-speed and low-latency dynamic wireless communications. Hence, cryptography requires additional computational power and endures an increasing latency with the rapid growth of large-scale resource-constrained wireless devices. More fatally, all encryption measures are based on the premise that decrypting is impossible without the knowledge of key, which may be invalid with the relentless growth of computational power.

During recent decades, physical-layer security (PLS) has been heralded as a potential direction for information-theoretic security of wireless communications against eavesdropping, as a complement to upper-layer encryption mechanisms. The fundamental principle of PLS was initially studied by Wyner \cite{wyner1}, where the key principle is to exploit the inherent randomness of noise or channels to limit the amount of information intercepted by unauthorized receivers. In \cite{wyner1}, the gap of channel capacity between the legitimate user and unauthorized eavesdropper is defined as the secrecy capacity and it is proved that secure transmission can be achieved as long as the secrecy capacity is positive. Following this concept, the artificial noise (AN) technology has been widely investigated as a typical PLS technique due to its ability to utilize channel state information (CSI) to enhance the secrecy performance. The AN technique is usually implemented at the legitimate transmitter (Alice), leveraging the spatial degree-of-freedom (DoF) of channels to ensure secure communications. By invoking the AN into the null space\footnote{The null space is the set of all vectors that are mapped to the zero vector when the transformation is applied. In the context of AN, the null space is used to design the AN signal that has no influence on the information-bearing signal, allowing for secure communication while obscuring information from eavesdroppers (refer to Section II.C.).}of channels of legitimate receiver (Bob), it is able to further enhance the secrecy performance by reducing the channel capacity of eavesdropper (Eve) without seriously affecting that of Bob due to the channel differences between them.
Compared with the conventional cryptographic mechanism, AN has the following core features:

\textbf{Information-theoretic security:} AN realizes the PLS by exploiting the characteristic of channels, where the secrecy performance can be theoretically analyzed and information-theoretic security can be constructed.

\textbf{Acceptable power consumption:} Although the deployment of AN may occupy additional power, the performance loss of Bob is acceptable in contrast to that of Eve. Additionally, in some specific systems, the optimized AN may not invoke power consumption.

\textbf{Software programmable:} AN can realize real-time and dynamic regulation with the variation of channel response.

\textbf{Excellent compatibility:} As a beamforming technology for security, AN can be easily compatible with various systems in wireless propagation environments.

In a nutshell, AN has significant potential to become a crucial PLS technique for wireless communications. However, the state-of-art research reported in the literature has not conducted a comprehensive survey on AN, which is the original motivation for writing this paper. Although several noteworthy surveys have been published in the domain of PLS, there is still an insufficient focus on the AN and the coverage of its applications. As summarized in Table I, existing surveys mainly focus on the PLS rather than the AN technology, thus somewhat neglecting the introduction of AN principles, system models, related applications, and future directions. Additionally, due to the excessive number of technologies involved in PLS and different focuses, some applications have been omitted in the existing surveys. 

Against this backdrop, the main contributions of this paper are divided into two folds. On the one hand, we focus on the AN technology to fill the gap in existing surveys. On the other hand, we provide more comprehensive applications in terms of AN-empowered scenarios and AN-combined technologies. Through our efforts, readers are expected to quickly figure out the principles, research status, design features, and future challenges of AN in their interested topics.


\begin{table*}
\large
\centering
\caption{Summary of Surveys Related to AN }\label{tabI}
\renewcommand\arraystretch{1}
\resizebox*{\linewidth}{!}{
\begin{tabular}{|c|c|c|c|c|c|c|c|c|c|c|c|c|c|c|c|c|}
\hline
\textbf{Main Topics} & \multicolumn{8}{|c|}{\textbf{Related Scenarios}} & \multicolumn{7}{|c|}{\textbf{Related Technologies}} & \multirow{2}*{\textbf{Main Contributions}} \\ \cline{1-16}
\textbf{Reference (year)} & \textbf{HetNet$^1$} & \textbf{IoT}& \textbf{mmWave} & \textbf{MU} & \textbf{Relay} & \textbf{Satellite} & \textbf{UAV}& \textbf{VLC}& \textbf{Coding}& \textbf{DM}&\textbf{mMIMO}& \textbf{NOMA}& \textbf{OFDM}& \textbf{RIS}& \textbf{SM}& \\ \hline
\cite{PLSS1} (2014) & &  &  & \checkmark &  &  &  &  & \checkmark &  & & & & &  & PLS in MU networks\\ \hline
\cite{PLSS2} (2017)&\checkmark &  &  & \checkmark  & \checkmark &  &  &  &  &  & & & & &  & Multiple-antenna techniques for PLS \\ \hline
\cite{PLSS3} (2017)& &  &  & \checkmark &  & \checkmark &  &  &  &  &  & & & & & PLS in satellite systems \\ \hline
\cite{PLSS4} (2018) & \checkmark &  & \checkmark &  &  &  &  &  & \checkmark & & \checkmark & \checkmark & & & & PLS for 5G \\ \hline
\cite{PLSS5} (2018)& &  &  &  &  &  &  & \checkmark &  &  &  & & & & & PLS for optical wireless communications \\ \hline
\cite{PLSS6} (2019)&\checkmark & \checkmark & \checkmark &  & \checkmark &  & \checkmark & \checkmark & \checkmark & & & \checkmark & & & & Classifications and applications of PLS \\ \hline
\cite{PLSS7} (2019)& &  &  & \checkmark & \checkmark &  &  &  &  &  & & & & &  & Optimization approaches for PLS \\ \hline
\cite{PLSS8} (2019)& \checkmark &  &  &  & \checkmark &  &  &  & & & \checkmark & \checkmark & & & & Cooperative relaying and jamming for PLS \\ \hline
\cite{PLSS9} (2019)& \checkmark&  & \checkmark &  &  &  &  &  & & & \checkmark & & & &  & PLS for 5G \\ \hline
\cite{PLSS10} (2020)&\checkmark & \checkmark &  &  &  & \checkmark &  & & & & &  &  &  &  & PLS in space information networks \\ \hline
\cite{PLSS11} (2020)& &  & \checkmark &  &  &  &  & \checkmark & & & & &  &  &  & PLS for VLC \\ \hline
\cite{PLSS12} (2021)& & \checkmark  &  &  &  &  &  &  &  &  & & & & &  & PLS for IoT \\ \hline
\cite{PLSS13} (2022)& &  &  &  &  &  & \checkmark &  &  &  & & & & &  & PLS for UAV \\ \hline
\cite{PLSS14} (2022)& & \checkmark &  &  & \checkmark &  &  & & & & &  &  &  &  & PLS in satellite networks \\ \hline
\cite{PLSS15} (2023)& &  &  &  &  &  & \checkmark & & & & &  &  &  &  & PLS for UAV \\ \hline
\cite{PLSS16} (2024)& &  &  &  &  &  &  &  &  &  & & & & \checkmark & & RIS-assisted PLS \\ \hline
This survey & \checkmark & \checkmark & \checkmark & \checkmark & \checkmark & \checkmark & \checkmark & \checkmark & \checkmark & \checkmark & \checkmark & \checkmark & \checkmark & \checkmark &  \checkmark & AN-assisted PLS with wider coverage \\ \hline
\multicolumn{14}{l}{\small{$^1$Heterogeneous networks, including CRN, D2D, SWIPT, etc.}}
\end{tabular}
}
\end{table*}

To this end, the remainder of this paper is organized as follows.
\begin{itemize}
\item In Section II, the evolution and fundamentals of AN are introduced, along with common design guidelines.
\item In Section III, a comprehensive survey of the current state of research on AN-empowered scenarios is presented.
\item In Section IV, a comprehensive survey of the current state of research on AN-combined technologies is provided.
\item In Section V, the most significant challenges worth future tackling are discussed.
\item In Section VI, a brief conclusion is drawn.
\end{itemize}

\section{AN Evolution and Modeling}

\subsection{AN Evolution}

The theoretical framework of AN was academically mentioned by Goel and Negi in 2005 \cite{Goel1} and 2008 \cite{Goel2} to guarantee the secrecy performance of multiple-input single-output (MISO)/ multiple-input multiple-output (MIMO) systems by mixing the information-bearing and AN signals at Alice\footnote{AN and friendly jamming are both techniques designed to improve communication security by thwarting eavesdropping. The AN seeks to obscure the intended signal by introducing random noise that primarily impacts potential Eves, whereas friendly jamming actively disrupts an Eve's ability to intercept communications by creating intentional interference. Although both strategies aim to protect legitimate communication, they are distinct technologies with different foundational principles and implementation methods. {AN is a technique that introduces randomness at the transmitter, whereas friendly jamming requires cooperative nodes to create the interference \cite{PLSS8,FJ1,FJ2}.}}. Since the AN lies in the null space of Bob's channel, it has no influence on the signal detection of Bob. However, due to the channel difference between Bob and Eve, the received signal at Eve will be severely affected by the AN, which can be regarded as the degradation of Eve's channel. In addition, the existence of null space requires satisfying the constraint that the number of Alice's antennas is larger than that of Bob, which can be seen as using the spatial DoF to generate AN \cite{DOF1}.

Later on, some investigation on the performance analysis of AN sprang up, with the purpose of illustrating the secrecy performance of AN \cite{analysis2,analysis3,analysis4,analysis5}.
For example, in \cite{analysis2}, the closed-form expressions of the connection and secrecy outage probabilities (SOPs) were derived for multi-antenna small-cell networks, where the results show that in a low cell-load case, deploying more base stations (BSs) will improve the connection and secrecy outage performance, and deploying more transmit antennas at each BS will only improve the connection outage performance. In \cite{analysis3}, the closed-form expression for the ergodic secrecy sum rate (SSR) was derived for the large system in the multiuser downlink, and the power allocation between information-bearing signals and the AN was optimized by maximizing the SSR. It shows that more power needs to be used for AN when Eve has more antennas and when the system serves fewer users. In \cite{analysis4}, the upper and lower bounds of the leakage rate to the eavesdropper in the high signal to noise ratio (SNR) regime were represented by a single compact expression as a function of the number of Eve's antennas, the dimensionality of signal space, and the channel coherence time, which offers useful insights in exploiting the secrecy potential of the AN-assisted massive multiple-input multiple-output (mMIMO) systems. In \cite{analysis5}, the SOP and mean secrecy rate were derived for the large-scale broadcast channels. Analytical results show that i) when the AN is adopted, the SOP exponentially decays with the number of transmission antennas, ii) in most of the power allocation cases, the per-user secrecy rate can be improved significantly, e.g., there is an almost 2.7 times of improvement for the particular transmission antennas number.


In order to further expand the usage scope of AN, in 2012, the authors of \cite{GAN3} first studied a new form of AN by removing the limitation of null space, which is also known as generalized AN \cite{GAN4}. The authors of \cite{GAN3} and \cite{GAN4} investigated the secrecy rate optimization problems of generalized AN in multiple-input multiple-output single-antenna-eavesdropper (MIMOSE) and multiple-input single-output single-antenna-eavesdropper (MISOSE) systems, respectively, where the conventional AN based on null space is demonstrated to be strictly sub-optimal. Then the authors of \cite{GAN1} addressed the generalized AN to the discriminatory channel estimation for multiple-input multiple-output multiple-antenna-eavesdropper (MIMOME) systems, in which the AN covariance matrix, the pilot signal power, and the linear estimator are jointly determined to minimize the channel estimation error. Moreover, the authors of \cite{GAN5} focused on the generalized AN generation in MIMOME Rician channels using the complex non-central Wishart distribution, while the authors of \cite{GAN6} paid attention to the generalized AN with the CSI of the eavesdropping channel for MIMOME systems. Recently, the authors of \cite{GAN2} pointed out that the optimality of generalized AN in \cite{GAN4} is valid only under some ideal assumptions such as perfect channel estimation and spatially uncorrelated channels. To break through this limitation, the authors of \cite{GAN2} first exploited a deep neural network (DNN) to jointly design and optimize the precoders for the information signal and the AN, which is called the deep AN scheme. 

In 2014, the researchers of \cite{OFDM9} proposed a time-domain AN generation technique to orthogonal frequency division multiplexing (OFDM) systems when the null space in the spatial domain may not exist, and showed that this technique requires a longer cyclic prefix (CP).
In 2016, the combination of AN and secret key generation (SKG) was first named data carrying artificial noise (DCAN) in \cite{SKG1} to summarize the case that Bob uses the pre-shared knowledge of AN and SKG to mitigate the influence.
On one hand, finding a balanced trade-off for power allocation between the information-bearing signal and the AN has attracted wide attention. For example, the authors of \cite{PA3} investigated the power allocation by maximizing a lower bound of ergodic secrecy rate (ESR) for MIMOME in 2018. On the other hand, the optimization for a new form of AN is desired to reduce the power consumption and enhance the jamming effect of AN. Specifically, in 2020, an Euclidean distance optimized AN scheme was proposed by minimizing the Euclidean distance between the transmit signal and a random jamming signal, which introduces no waste of transmit power and provides a stronger jamming intensity to eavesdroppers in contrast to conventional AN scheme using complex Gaussian random matrix \cite{SM5}.

In general, the AN has been progressively evolving towards broader application scenarios, enhanced energy efficiency, and improved jamming effectiveness.

\subsection{Modeling of AN-assisted Wireless Communications}

\begin{figure*}[!htp]
\centering
\includegraphics[width=1\textwidth]{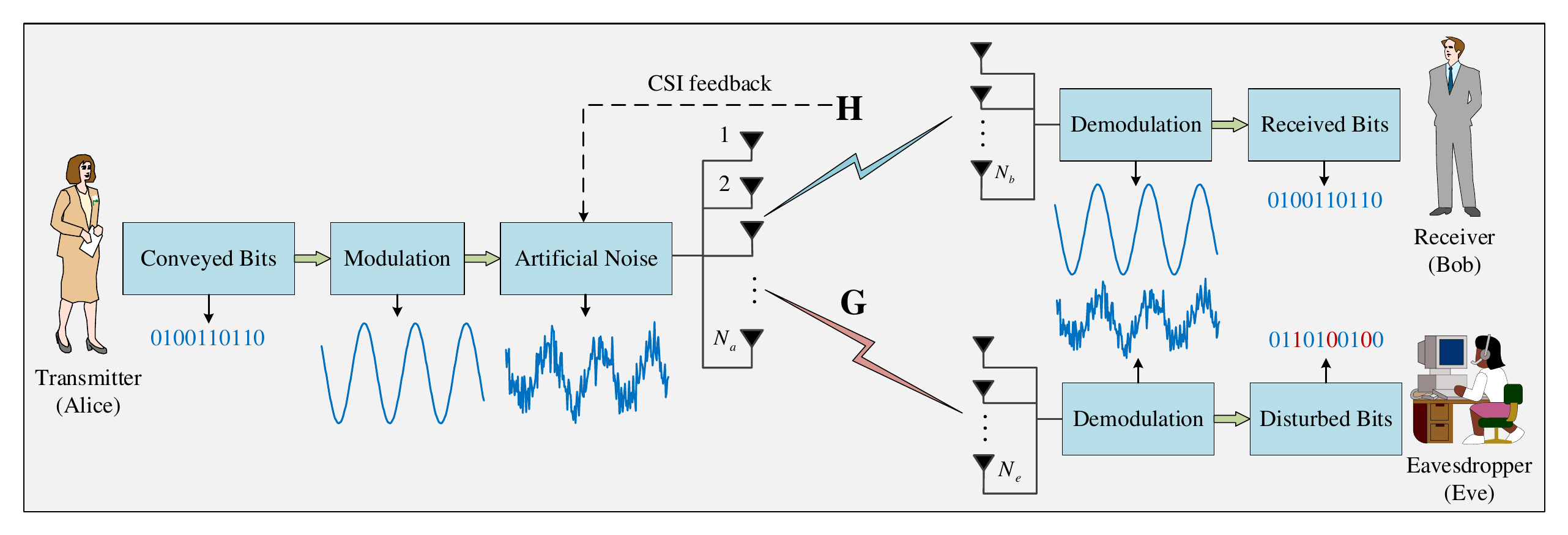}
\caption{Framework of AN-assisted wireless communications. }
\label{fig_2}
\end{figure*}

As depicted in Fig. \ref{fig_2}, most wireless secrecy communication systems consist of an Alice with ${N_a}$ antennas, a Bob with ${N_b}$ antennas, and an unauthorized Eve with ${N_e}$ antennas, wherein Alice wishes to convey a private message to Bob under the wiretapping of a passive Eve. While transmitting the information-bearing signal, Alice is able to generate specific interfering signals termed AN so that only Eve is adversely affected by the interfering signals, while Bob remains slightly unaffected.
The transmit signal at Alice can be modeled as
\begin{equation}\label{x:1}
{\bf{x}} = {\bf{Ws}} + {\bf{Vr}},
\end{equation}
where ${\bf{Ws}}$ and ${\bf{Vr}}$ denote the information-bearing signal and AN, respectively. In practice, the precoding/beamforming matrix ${\bf{W}}{\in ^{{N_a} \times {N_s}}}$ attempts to enhance the transmission quality of the information-bearing vector ${\mathbf{s}} \in {\mathbb{C}^{{N_s}}}$ and ${{N_s}}$ denotes the number of data stream, while the AN matrix ${\mathbf{V}} \in {\mathbb{C}^{{N_a} \times \left( {{N_a} - {N_b}} \right)}}$ tries to reduce the impact of AN vector ${\mathbf{r}} \in {\mathbb{C}^{\left( {{N_a} - {N_b}} \right)}}$ for Bob. The transmit signal is conveyed to Bob via an MIMO channel ${\mathbf{H}} \in {\mathbb{C}^{{N_b} \times {N_a}}}$ and also received by Eve via a wiretap channel ${\mathbf{G}} \in {\mathbb{C}^{{N_e} \times {N_a}}}$. Hence, the received signals at Bob and Eve can be expressed as
\begin{equation}\label{A:2}
{\mathbf{y}} = {\mathbf{HWs}} + {\mathbf{HVr}} + {\mathbf{u}},
\end{equation}
\begin{equation}\label{A:1}
{\mathbf{z}} = {\mathbf{GWs}} + {\mathbf{GVr}} + {\mathbf{v}},
\end{equation}
where ${\mathbf{u}} \in {\mathbb{C}^{{N_b}}}$ and ${\mathbf{v}} \in {\mathbb{C}^{{N_e}}}$ stand for complex additive white Gaussian noise (AWGN) with elements obeying i.i.d. $\mathcal{C}\mathcal{N}(0 ,{\sigma _u^2})$ and $\mathcal{C}\mathcal{N}(0 ,{\sigma _v^2})$, respectively.

\subsubsection{Orthogonal AN}
With the CSI of Bob, an orthogonal AN can be conducted using null-space projection. Specifically, assuming ${\mathbf{V}}$ is the null space of ${\mathbf{H}}$, i.e., ${\mathbf{HV}} = {\mathbf{0}}$, the influence of AN can be mitigated at Bob as
\begin{equation}\label{A:3}
{\mathbf{y}} = {\mathbf{HWs}} + {\mathbf{u}}.
\end{equation}

It is worth mentioning that the null space ${\mathbf{V}}$ can be obtained by the singular value decomposition (SVD) of ${\mathbf{H}}$ as
{\begin{equation}\label{orthogonal:1}
{\mathbf{H}} = {\mathbf{U}}\left[ {\begin{array}{*{20}{c}}
  {\mathbf{D}}&0
\end{array}} \right]{\left[ {\begin{array}{*{20}{c}}
  {{{\mathbf{V}}_1}}&{\mathbf{V}}
\end{array}} \right]^H},
\end{equation}
where ${\mathbf{U}} \in {\mathbb{C}^{{N_b} \times {N_b}}}$ denotes a unitary matrix composed of left singular vectors of ${{\mathbf{H}}}$, ${\mathbf{D}} \in {\mathbb{C}^{{N_b} \times {N_b}}}$ represents a diagonal matrix whose entries are the singular values of ${{\mathbf{H}}}$, ${{\mathbf{V}}_1} \in {\mathbb{C}^{{N_a} \times {N_b}}}$ consists of the first ${N_b}$ right singular vectors of ${{\mathbf{H}}}$, spanning the subspace corresponding to the information-bearing, and ${{\mathbf{V}}} \in {\mathbb{C}^{{N_a} \times \left( {{N_a} - {N_b}} \right)}}$ contains the remaining ${N_a}-{N_b}$ right singular vectors, spanning the null space associated with the noise component.}
Another more straightforward way to obtain the null space of $\mathbf{H}$ is 
\begin{equation}\label{orthogonal:2}
\mathbf{V} = \mathbf{I} - \mathbf{H}^{\text{H}} (\mathbf{H}\mathbf{H}^{\text{H}})^{-1} \mathbf{H}.
\end{equation}

\subsubsection{Non-orthogonal AN}
Orthogonal AN may not be optimal when the CSI of Eve is known. An intuitive explanation is that one can further interfere with Eve by sacrificing part of Bob's communication quality. Mathematically, a generalized optimization problem can be cast to design the beamforming matrix and the AN to improve the secrecy performance, which can be given by
\begin{subequations}\label{nonorthogonal:1}
\begin{align}
  &\max /\min f\left( {{\mathbf{W}},{\mathbf{V}},{\mathbf{r}}} \right) \hfill \\
  &{\text{s}}{\text{.t}}{\text{. }}{\left\| {\mathbf{x}} \right\|^2} \leqslant {\text{1,}} \hfill \\
  &\;\;\;\;\;\;\;{\mathbb{P}_{out}} \leqslant {\mathbb{P}_\gamma }, \hfill \\
  &\;\;\;\;\;\;\;{R_b} \geqslant {\gamma _b}, \hfill \\
  &\;\;\;\;\;\;\;{R_e} \leqslant {\gamma _e}, \hfill
\end{align}%
\end{subequations}%
{where $f$ denotes a specific secrecy-related objective function such as secrecy rate, secrecy capacity, or SOP. For instance, a maximization problem can be formulated as $\max {R_s}$, which aims to maximize the secrecy rate by enhancing Bob’s achievable rate while suppressing Eve’s. Conversely, a minimization problem may take the form $\min {\mathbb{P}_{out}}$, which attempts to minimize the SOP under the given constraints.} Moreover, ${\left\| {\mathbf{x}} \right\|^2} \leqslant {\text{1}}$ represents the power constraint triggering the issue of power allocation between information-bearing signal and AN, ${\mathbb{P}_{out}} \leqslant {\mathbb{P}_\gamma }$ means that the SOP is restricted below a certain threshold, ${R_b} \geqslant {\gamma _b}$ indicates that the quality of service (QoS) of Bob is desired to exceed a certain threshold, while ${R_e} \leqslant {\gamma _e}$ implies that of Eve is lower than a certain threshold.

\subsubsection{AN through Symbol-level Precoding}
\begin{figure}
    \centering
    \includegraphics[scale=0.75]{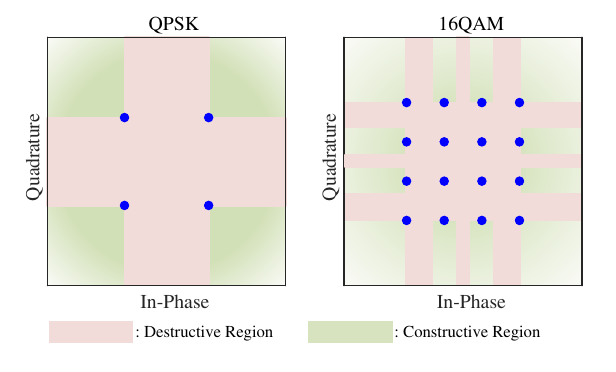}
    \caption{Constructive and destructive interference illustration for QPSK and 16QAM constellations.}
    \label{symbol-level}
\end{figure}

Different from the aforementioned orthogonal and non-orthogonal AN design at the channel fading rate, AN can also be conducted at the symbol rate with symbol-level precoding techniques \cite{SLP1,SLP2,SLP3,SLP4,SM6}. As shown in Fig. \ref{symbol-level}, the effect of AN can be divided into constructive and destructive types. Thus, AN can be designed as constructive interference towards Bob for transmit power saving. Meanwhile, when the CSI of Eve is available, AN can be designed as destructive interference towards Eve. A general optimization for AN design in multi-user MISO $\mathcal{M}$-PSK transmissions through symbol-level precoding can be cast as
\begin{subequations}
\begin{align}
&\mathop {\min }\limits_{\mathbf{x}} \left\| {\mathbf{x}} \right\|_2^2\\
&\text{s.t.} \; \left[\Re\left(\lambda_{n_{\text{b}}}\right) \sqrt{\Gamma_{n_{\text{b}}} \sigma_u^2}\right]\tan\theta_\mathrm{th}\geq\left|\Im\left(\lambda_{n_\text{b}}\right)\right|, \\
&\;\;\;\;\;\left[\Re\left(\lambda_{n_{\text{e}}}\right)-\sqrt{\Gamma_{n_{\text{e}}}\sigma_v^{2}}\right]\tan\theta_{\mathrm{th}}\leq |\Im\left(\lambda_{n_{\text{e}}}\right)|, \\
&\;\;\;\;\;1 \leq {n_{\text{b}}} \leq N_{\text{b}},\; 1 \leq {n_\text{e}} \leq N_{\text{e}},
\end{align}
\end{subequations}
{where we have $\mathbf{h}_{n_\text{b}}^\text{T} \mathbf{x} = \lambda_{n_{\text{b}}} s_{n_{\text{b}}}$ and $\mathbf{h}_{n_\text{e}}^\text{T} \mathbf{x} = \lambda_{n_{\text{e}}} s_{n_{\text{e}}}$. $s_{n_{\text{b}}}$ and $s_{n_{\text{e}}}$ denote the expected receive PSK signals at the $n_{\text{b}}$-th Bob and $n_{\text{e}}$-th Eve, respectively, $\lambda_{n_{\text{b}}}$ and $\lambda_{n_{\text{e}}}$ represent the scalar factor for the $n_{\text{b}}$-th Bob and $n_{\text{e}}$-th Eve, respectively, $\mathbf{h}_{n_\text{b}} $ and $\mathbf{h}_{n_\text{e}} $ are the CSI for the $n_{\text{b}}$-th Bob and $n_{\text{e}}$-th Eve, respectively, $\Gamma_{n_{\text{b}}}$ and $\Gamma_{n_{\text{e}}}$ denote the SINR of the $n_{\text{b}}$-th Bob and $n_{\text{e}}$-th Eve, respectively, ${\Re \left(  \cdot  \right)}$ and $\Im \left(  \cdot  \right)$ represent the operations on taking the real and imaginary parts of complex numbers, respectively, and $\theta_{\text{th}} = \pi/\mathcal{M}$ stands for the phase deviation between adjacent PSK symbols.}

From the perspective of secrecy performance, the non-orthogonal AN can provide a higher secrecy rate in contrast to the orthogonal one due to the intuitive objective function of the optimization problem. However, from the implementation perspective, the non-orthogonal AN requires Eve's CSI, which is unavailable as the passive receiver has no cooperation with Alice to perform the channel feedback. Although relevant papers may relax this restriction to only know the location range of Eve, the essence of requiring the CSI of Eve is unchanged. By contrast, the orthogonal AN, as a security technique that only requires the CSI of Bob, benefits from better feasibility. Furthermore, the orthogonal AN has no additional interference to Bob in comparison with the non-orthogonal one, hence maintaining the hardware requirements and detection technology for Bob. Finally, AN based on symbol-level precoding can improve secure transmission performance by designing the AN as constructive interference for Bob and destructive interference for Eve. For example, the authors of \cite{SM6} investigated such AN in generalized spatial modulation (GSM) systems, resulting in both power saving and security enhancement.

\subsection{Technical Background}
In order to provide researchers with a smoother reading experience, we first address some common issues, such as why AN technology is effective, what is the required workflow, and what are the minimum hardware requirements.

\textbf{Why AN Is Effective?}

AN is tailored to the characteristics of the legitimate channel, enabling it to disrupt Eve while allowing Bob to effectively extract the information. By masking the information-bearing signal with customized AN, the distinction between the legitimate channel and the wiretap channel prevents Eve from successfully eavesdropping. Specifically, the presence of random AN lowers the SNR at Eve, complicating the extraction of the information and increasing the likelihood of decoding errors. When AN is designed to align with the null space of the legitimate channel, it minimizes the impact on Bob's signal detection.

\textbf{What Is the Complete Workflow?}

The complete AN transmission workflow consists of the following three main steps.
\subsubsection{Channel Assessment:} Evaluate the CSI of the legitimate channel via channel estimation and channel feedback.

\subsubsection{AN Design and Integration:} Design the AN to match the specific characteristics of the legitimate channel, ensuring it effectively obscures the signal for Eves without significantly impacting Bob. Combine the AN with the information-bearing signal before transmission.

\subsubsection{Transmission and Reception:} Transmit the combined signal over the legitimate channel to Bob and over the wiretap channel to Eve. Since the AN is tailored to the legitimate channel, Bob can successfully demodulate the information, while the detection of Eve is hindered by AN.

From the above steps, the prior information that Alice and Bob need to share includes the pilot signals, the CSI feedback mechanism, and the synchronization parameters.

Note that a capable Eve may estimate the CSI of the wiretap channel based on the pilot signal sent by Alice. However, due to the inherent difference between the wiretap channel and the legitimate channel, the random AN continues to interfere with Eve's reception. In practice, the differences between the wiretap channel and the legitimate channel are common due to the distinct placements of Bob and Eve, allowing Alice to design the AN solely based on the characteristics of the legitimate channel.

\textbf{What Are the Minimum Hardware Requirements?}

The minimum hardware requirements for implementing AN consist of signal processing units, modulators, demodulators, channel estimation tools, synchronization components, and power control modules. These components collaborate to generate, transmit, and process the AN effectively. Furthermore, the minimum number of RF chains varies depending on specific cases.

In scenarios where the deployment of AN has no influence on Bob, Alice are required to have more RF chains than Bob. This spatial-domain requirement typically applies to MISO and MIMO systems. However, in SISO systems, this restriction can be relaxed by sacrificing some frequency or time domain resources. For instance, in OFDM systems, the CP matrix offers time-domain DoF, which can reduce the number of RF chains needed by Alice. Similarly, in time-domain multiplexing systems, increasing the number of time slots also lessens Alice's RF chain requirements. Additionally, a mutual retransmission mechanism between Alice and Bob can also decrease the RF chain requirements.

In scenarios where the deployment of AN impacts Bob, the requirement for the number of RF chains becomes irrelevant. Here, effective AN design necessitates that the interference to Eve exceeds the interference to Bob. This requirement is often represented in research through quantitative metrics, such as secrecy capacity and secrecy rate, etc. However, these metrics rely on the channel CSI of the wiretap channel, challenging their implementation.

In a nutshell, the AN design generally requires Alice to have more RF chains than Bob. However, this requirement can be relaxed by sacrificing some frequency/time-domain resources or establishing a retransmission mechanism in advance between Alice and Bob.

Subsequently, we explain the meaning of some basic technical concepts.

\textbf{Null Space:}
Given a channel matrix ${\mathbf{H}}$, its null space ${\mathbf{V}}$ is the combination of all orthogonal basis vectors such that
\begin{equation}
{\mathbf{HV}} = {\mathbf{0}}.
\end{equation}
By deploying the AN in the null space of the legitimate channel, the legitimate receiver will not be interfered with, thus it is also called orthogonal AN. Although non-orthogonal AN is also mentioned in many papers, it is not as well-known as orthogonal AN as it may require the CSI of Eve. In technical implementation, the null space can be generated by performing the SVD or orthogonal projection on channel matrix ${\mathbf{H}}$ as shown in (\ref{orthogonal:1}) or (\ref{orthogonal:2}).

\textbf{Spatial DoF:}
In general, the DoF refers to the number of independent parameters or variables that can vary in a system without violating any constraints. It is a measure of the number of independent directions or dimensions in which a system can move or be manipulated.

In AN technology, the elements of the random vector ${\mathbf{r}}$ are such independent parameters. The reason is that by multiplying the random vector ${\mathbf{r}}$ at the right side of the null space matrix , the AN ${\mathbf{Vr}}$ will not affect Bob no matter how the random vector ${\mathbf{r}}$ changes. The dimension of the random vector ${\mathbf{r}} \in {\mathbb{C}^{\left( {{N_a} - {N_b}} \right)}}$ implies that the existence of spatial DoF requires Alice to have more antennas than Bob.

\textbf{Secrecy Capacity:}
\begin{equation}
{C_s} \triangleq \mathop {\max }\limits_{p\left( {\mathbf{s}} \right)} \left\{{I\left( {{\mathbf{s}};{\mathbf{y}}} \right) - I\left( {{\mathbf{s}};{\mathbf{z}}} \right)} \right\}.
\end{equation}

The definition of secrecy capacity was firstly proposed in \cite{wyner1} as the gap of channel capacity between Bob and Eve. In \cite{Goel1} and \cite{Goel2}, the expression for secrecy capacity under the AN technology was derived. Since the closed-form expressions for secrecy capacity are not always available \cite{LF6}, the following secrecy rate is often considered.

\textbf{Secrecy Rate:}
\begin{equation}
{R_s} \triangleq I\left( {{\mathbf{s}};{\mathbf{y}}} \right) - I\left( {{\mathbf{s}};{\mathbf{z}}} \right).
\end{equation}

The definition of secrecy rate is the gap of achievable rate between Bob and Eve, whose theoretical upper bound is the secrecy capacity when the distribution of ${\mathbf{s}}$ satisfies certain conditions. Note that in some literature, the secrecy rate is equivalent to the secrecy capacity because the authors may assume the perfect distribution of the source.

In extensive papers, the optimization problem was formulated to maximize the secrecy capacity or secrecy rate under specific constraints (see \cite{MISO4,VLC1,CRN4,indicator1}). Moreover, the objective function can be transformed to some similar indicators. 

For example, the ESR is the average of the secrecy rate over different channel realizations. This average is taken over the distribution of possible channel conditions, reflecting long-term or statistical performance as \cite{PA3,MIMO3}
  \begin{equation}
R_s^{erg} = {\mathbb{E}_{{\mathbf{H}},{\mathbf{G}}}}\left[ {{R_s}} \right].
\end{equation}

The SSR is the sum of the secrecy rates for all users in the system, which provides an aggregate measure of the secure transmission capacity across multiple users as \cite{ssr1}
  \begin{equation}
R_s^{sum} = \sum\limits_{k = 1}^K {R_s^k},
\end{equation}
where $k$ denotes the index of $k$-th user.

Combing the definitions of ESR and SSR, the ergodic SSR becomes a more comprehensive metric that takes into account the sum of secrecy rate for all users over multiple channel realizations as \cite{analysis3}
\begin{equation}
R_{sum}^{erg} = {\mathbb{E}_{{\mathbf{H}},{\mathbf{G}}}}\left[ {\sum\limits_{k = 1}^K {R_s^k} } \right].
\end{equation}

 When Eve's position appears in a certain area and obeys a certain probability distribution, the mean secrecy rate can be calculated by averaging the secrecy rate with respect to Eve's position ${l_e}$ as \cite{analysis5,msr1}
 \begin{equation}
R_s^m = {\mathbb{E}_{{l_e}}}\left[ {{R_s}} \right].
\end{equation}

Furthermore, some approximations of the secrecy rate, such as its upper bounds \cite{analysis4} or lower bounds \cite{LF6,PA2,PA8}, were also applied to obtain results with better form.

\textbf{SOP:}
\begin{equation}\label{SOP:11}
{\mathbb{P}_{out}} \triangleq \mathbb{P}\left\{{{R_s} < {R_t}} \right\}.
\end{equation}

The definition of SOP was firstly proposed in \cite{indicator2} as the probability that the instantaneous secrecy rate is below a target secrecy rate threshold ${{R_t}}$. Commonly, it quantifies the probability of unsafe transmission and can be minimized to guarantee the security of transmission \cite{SOP1}, or it can be used as a constraint to optimize other parameters \cite{SOP3}.

\begin{figure*}[!htp]
\centering
\includegraphics[width=1\textwidth]{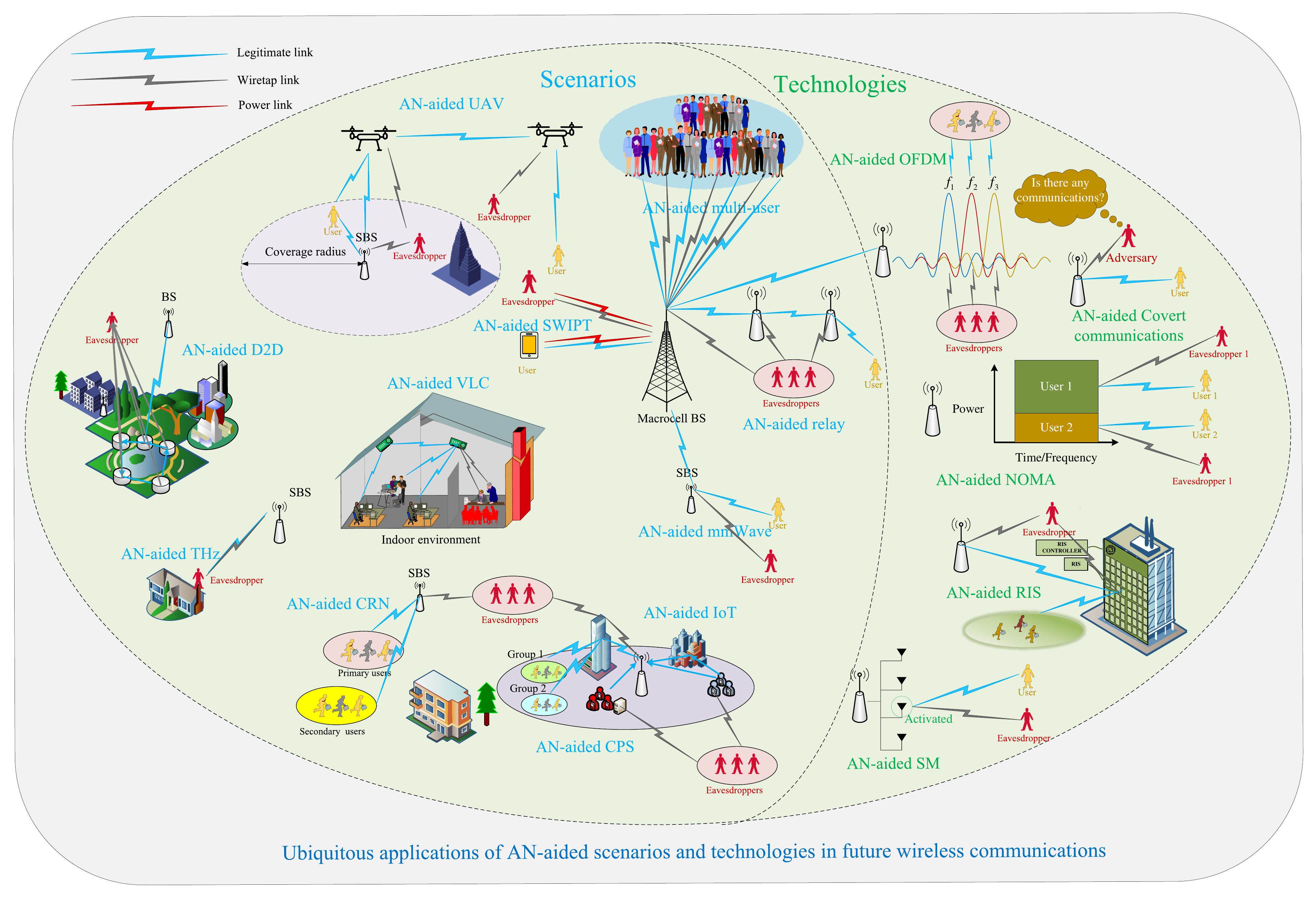}
\caption{The applications of AN in future wireless communications.}
\label{fig_3}
\end{figure*}

\textbf{Power Allocation:}
\begin{equation}
\theta  = {{{\left\| {{\mathbf{Ws}}} \right\|}^2}}/ {{{\left\| {\mathbf{x}} \right\|}^2}}.
\end{equation}

Power allocation between the information-bearing signal and the AN is an important design parameter. Allocating more power to AN will lead to a decrease in the throughput of information-bearing signals, while less AN power will reduce the security of systems. Hence, the problem of finding a balanced trade-off for power allocation has attracted lots of attention. In \cite{PA3}, the power allocation was derived by maximizing a lower bound of ESR for MIMOME without the CSI of Eve. In \cite{PA2}, the power allocation was given by the water-filling algorithm to maximize a lower bound of secrecy rate for MIMOME without the CSI of Eve. In \cite{PA8}, the power allocation was designed to maximize a closed-form lower bound of secrecy capacity and showed that equal power allocation is a simple and generic strategy to achieve near-optimal capacity performance. In \cite{PA1}, the power allocation was optimized by maximizing the SSR for multi-user (MU) downlink without the CSI of Eve. In \cite{PA4}, the power allocation was obtained with constraints for reliability and secrecy level and the corresponding secrecy transmission rate in the presence of network interference for the worst scenario. In \cite{PA5}, the power allocation was studied by minimizing the SOP under a target secrecy rate or maximizing the secrecy rate under an SOP constraint, respectively. In \cite{PA6}, the power allocation was investigated by minimizing the total transmission power in multiple-input single-output multiple-antenna-eavesdropper (MISOME) systems. In \cite{PA7}, the power allocation was designed to ensure a target secrecy probability under QoS constraints at Bob and Eve in MISOME systems without Eve's CSI. In \cite{PA9}, the power allocation was derived as a closed-form solution by maximizing the achievable secrecy rate for MIMOME, where the derived results show that equal power and water-filling power allocations lead to similar solutions and rate loss. In \cite{PA10}, the power allocation was considered to maximize an analytical closed-form expression of an achievable secrecy rate for MISOME.

\textbf{QoS:}
\begin{equation}
{\text{SIN}}{{\text{R}}_b} = {\left\| {{\mathbf{HWs}}} \right\|^2}/\left( {{{\left\| {{\mathbf{HVr}}} \right\|}^2} + \sigma _b^2} \right),
\end{equation}
\begin{equation}
{\text{SIN}}{{\text{R}}_e} = {\left\| {{\mathbf{GWs}}} \right\|^2}/\left( {{{\left\| {{\mathbf{GVr}}} \right\|}^2} + \sigma _e^2} \right).
\end{equation}

QoS means the attainable signal quality at Bob or Eve, which can be seen as the requirement of SNR or signal to interference plus noise ratio (SINR). QoS is always used as a constraint to jointly optimize beamforming and AN by minimizing the transmit power \cite{QoS1}, \cite{QoS2}.

\textbf{BER:}
{
\begin{equation}
{P_b} = {\mathbb{E}_{{\bf{H}}}}\left( {{n_{b}}/n} \right),
\end{equation}
\begin{equation}
{P_e} = {\mathbb{E}_{{\bf{H}},{\bf{G}}}}\left( {{n_{e}}/n} \right).
\end{equation}

The bit-error rate (BER) is defined as the ratio of the number of error bits--denoted as ${n_{b}}$ for Bob and ${n_{e}}$ for Eve--to the number of the total transmitted bits $n$.} The AN technology aims to maintain acceptable BER loss for Bob, while sharply increasing the BER of Eve even when SNR goes to infinity \cite{SPSM1}. Specifically, the notion of practical secrecy was proposed in \cite{indicator3} as a new design criterion based on the behavior of the BER of Eve, as its SNR goes to infinity.

\section{Overview of AN-empowered Scenarios}


As shown in Fig. \ref{fig_3}, AN is expected to pervade future wireless networks. Existing works fall into two categories: (i) AN-empowered scenarios; (ii) AN-integrated technologies. This section reviews key scenarios, with Table II summarizing AN types, models, node roles, CSI availability, and contributions.

\begin{table*}[!htp]
\begin{center}
\center
\caption{Overview of AN-empowered Scenarios}
  \begin{tabular}{cccccccc}
    \toprule[1.5pt]
    \makebox[0.01\textwidth][c]{References} & \makebox[0.01\textwidth][c]{Scenarios} & \makebox[0.01\textwidth][c]{AN Types}
                                     & \makebox[0.01\textwidth][c]{System Models} & \makebox[0.05\textwidth][c]{Bob types} &\makebox[0.01\textwidth][c]{Eve types} & \makebox[0.01\textwidth][c]{CSI of Bob/ Eve}   & \makebox[0.01\textwidth][c]{Involved Contributions}           \\
    \midrule[1pt]
    \cite{CPS1}&CPS&AN&MIMOME& SMA  &SMA  &Imperfect/No&  Virtual noise space  \\
    \cite{CPS2}&CPS&AN&MIMOME& SMA &SMA &Either/Either &  AN Optimization for full-duplex \\
    \cite{CRN2}&CRN&AN&MISOSE& MSA & MSA &Perfect/Either&   Two SEEM schemes \\
    \cite{CRN3}&CRN&AN&MIMOME& SMA &MMA  &Perfect/Perfect&  Convergent solution of SRM problem \\
    \cite{CRN6}&CRN&AN&MISOSE& MSA & SSA/MSA &Statistical/Statistical&  SRM  \\
    \cite{CRN7}&CRN&AN&MISOSE& SSA & SSA &Statistical/Statistical&  SRM \\
    \cite{CRN5}&CRN&AN&MISOME& SSA & SMA &Either/Either&  SEEM in OFDM-CRNs\\
    \cite{D2D1}&D2D&AN&MISOSE& MSA & SSA &Statistical/Statistical  &  Three schemes of precoding matrix  \\
    \cite{D2D2}&D2D&AN&MISOME& MSA & SMA & Perfect/ No &  Secrecy sum rate maximization  \\
    \cite{IoT1}&IoT&AN&MISOSE&SSA  & SSA & Either/Either &   Cooperative beamforming \\
    \cite{IoT2}&IoT&AN&MISOSE& SSA & SSA & Perfect/No & A dual-user two-phase system\\
    \cite{mmWave3}&mmWave&Hybrid ANH&MIMOME& SMA & SMA/MMA& Perfect/ No &Minimal hardware complexity \\
    \cite{mmWave1}&mmWave&Two-stage AN&MISOME& MSA & MMA & Perfect/ No &Hybrid beamforming for AF relay \\
    \cite{mmWave2}&mmWave&Hybrid AN&MISOSE& SSA & MSA & Perfect/ Partial &A secrecy rate lower bound \\
    \cite{mmWave4}&mmWave&AN&MISOSE& SSA & SSA & Perfect/ Statistical & Secrecy throughput maximization \\
    \cite{mmWave5}&mmWave&AN&MISOSE& SSA & SSA & Perfect/ Partial & SOP minimization \\

    \cite{multiuser1}&MU&AN&MISOSE& MSA & MSA & Perfect/ Perfect &  AN in service integration\\
    \cite{multiuser2}&MU&Robust AN&MISOSE& MSA & MSA & Imperfect/ Imperfect &  Imperfect CSI\\
    \cite{multiuser3}&MU&AN&MISOSE& MSA & MSA & Either/ Either &   Secrecy rate region maximization\\
    \cite{multiuser4}&MU&Jammer AN&MISOME& MSA & SMA & Perfect/ No &  Opportunistic scheduling OSTBC and AN\\
    \cite{multiuser5}&MU&AN with THP&MISOME& MSA & SMA & Either/ No &  Nonlinear precoding\\
    \cite{multiuser6}&MU&AN&MISOME& MSA & SMA & Imperfect/ No &  Closed-form expression of ESR\\
    \cite{multiuser7}&MU&AN&MISOSE& MSA & MSA & Perfect/ Either &  SRM in joint optimization framework\\
    \cite{multiuser8}&MU&Jammer AN&MISOSE& SSA & MSA & Perfect/ Perfect &  Optimal multiuser diversity\\
    \cite{multiuser9}&MU&Jammer AN&SIMOME& SMA & SMA & Perfect/ Statistical &  SOP minimization\\
    \cite{multiuser10}&MU&AN&MIMOME& SMA & SMA & Perfect/ No &  Joint MU constellations and AN alignment\\
    \cite{multiuser11}&MU&Time-domain AN&MISOSE& SSA & SSA & Perfect/ No &  AN and secret key aided schemes\\

    \cite{PZ2}&PZ&AN&MISOSE& SSA & SSA & Perfect/ No & Optimization of the protected zone size\\
    \cite{PZ1}&PZ&AN&SISOSE& SSA & MSA & Statistical/ No & Optimal use of AN with a protected zone\\
    \cite{relay3}&AF relay&AN&MIMOME& SMA & SMA & Perfect/ No &  SRM\\
    \cite{relay4}&AF relay&AN&MIMOME& SMA & SMA & Perfect/ No &  AN for IA-based multipair relay networks\\
    \cite{relay7}&AF relay&AN&MIMOME& SMA & SMA & Perfect/ No &  Antenna selection on relay with AN\\
    \cite{relay8}& AF relay&Bob AN&SISOSE& SSA & SSA relay & Perfect/ Perfect &  AN from Bob for untrusted relay\\
    \cite{relay10}& AF relay&Relay AN&SISOSE& SSA & MSA & Perfect/ Imperfect & SINR maximization\\
    \cite{relay2}&AF relays&Relay AN&SISOSE& SSA & MSA & Perfect/ No &  AN generation from cooperative relays\\
    \cite{relay9}& AF relays&Relay AN&SISOSE& SSA & MSA & Perfect/ Perfect &  SSR maximization through SDR and SCA\\
    \cite{relay12}& AF relays&Relay AN&SISOSE& SSA & MSA & Perfect/ Perfect & SRM via two-level optimization and SDR\\
    \cite{relay15}& AF relays&Relay AN&SISOME& SSA & MMA & Perfect/ Imperfect & SRM for MMA relays \\
    \cite{relay5}&DF relay&Jammer AN&MIMOME& SMA & SMA & Perfect/ No &  AN from a cooperative FD jammer\\
    \cite{relay6}&DF relay&AN and relay AN&MIMOME& SMA & MMA & Perfect/ No &  STM under an SOP constraint\\
    \cite{relay11}& DF relay&Relay PR AN&MISOSE& MSA & MSA & Perfect/ Statistical & Joint NOMA and AN-aided FD relay\\
    \cite{relay16}& DF relay&Relay AN&SISOSE& SSA & SSA/MSA & Perfect/ No & NOMA-based two-way relay network \\
    \cite{relay13}& DF relays&Relay AN&SISOSE& MSA & MSA & Perfect/ No & Joint AN and relay selection for CRNs\\
    \cite{relay17}& DF relays&Relay AN&SISOSE& SSA & SSA & Imperfect/ No & AN-aided two-way relay selection \\
    \cite{relay18}& DF relays&AN&MISOME& SSA & SMA & Perfect/ Statistical & Joint user and relay selection with AN \\
    \cite{relay14}& NR relay&Relay AN&SISOSE& SSA & SSA & Imperfect/ Imperfect & Two energy harvesting strategies\\
    \cite{relay1}&UTWR&AN&MIMOME& SMA & SMA relay & Either/ Either &  Maximum SSR for untrusted relay\\

    \cite{LEO1}&Satellite&DCAN&MIMOME&SMA  & SMA & Imperfect/Imperfect & Power and time slot allocation\\
    \cite{LEO2}&Satellite&AN&MISOSE&MSA  & MSA & Perfect/Perfect & Transmit power minimization\\

     \cite{SLP4}&SWIPT&AN&MISOSE& MSA & MSA & Either/ Either & Total transmit power minimization \\
    \cite{SWIPT2}&SWIPT&AN&MIMOME& SMA & MMA & Perfect/ No & SEEM for SWIPT  \\
    \cite{SWIPT1}&SWIPT&AN&SISOSE& SSA & SSA & Statistical/ No & AN-aided hybrid TSPS scheme  \\
    \cite{SWIPT3}&SWIPT&AN&MISOSE& MSA & MSA & Perfect / Either & Power minimization for NOMA\\
    \cite{SWIPT4}&SWIPT&AN&MIMOME& MMA & SMA & Statistical/ No & AN assisted IA scheme with WPT   \\
    \cite{SWIPT5}&SWIPT&AN&MISOSE&MSA  & MSA & Either/Either&   Multi-cell coordinated beamforming \\
    \cite{SWIPT6}&SWIPT&AN&MISOME&MSA  & MMA & Imperfect/ Imperfect& Cooperative jamming \\
    \cite{SWIPT7}&SWIPT&AN&MIMOSE&SMA  & MSA & Perfect/Statistical & SOP analysis with secondary terminals \\

    \cite{THZ1,THZ2}&THz&SIC-free AN&MIMOME& SMA & SMA/MMA & Statistical/ No & DNN-powered SIC-free receiver AN \\
    \cite{UAV3}&UAV&AN&MISOSE& SSA & MSA & Statistical/ Statistical & Max-min secrecy, trajectory design  \\
    \cite{UAV2}&UAV&AN&SISOSE& SSA & SSA & Perfect/ No &  Joint optimization  \\
    \cite{UAV1}&UAV&AN&MISOSE& MSA & SSA & Imperfect/No & SEEM for UAV-based NOMA \\
    \cite{VLC3}&VLC&AN&MISOSE& SSA & SSA & Perfect/ Imperfect & Closed-form expression for secrecy rate\\
    \cite{VLC5}&VLC&AN&MISOSE& SSA & MSA & Perfect/ Either &  Two sub-optimal low-complexity schemes\\
    \cite{VLC4}&VLC&AN&MISOSE& MSA & SSA & Statistical/ Either & ZF design\\
    \cite{VLC6}&VLC&AN&MISOSE& MSA & SSA & Imperfect/ Imperfect & Power minimization for NOMA \\
    \cite{VLC2}&VLC&Time-domain AN&SISOSE& SSA & SSA & Statistical/ Statistical & Time-domain AN for restricting PAPR \\

    \bottomrule[1.5pt]
  \end{tabular}
\end{center}
\end{table*}

\subsection{AN-aided CPS}
Cyber-physical systems (CPSs) integrate the cyber world, physical processes, and control to enable real-time sensing and dynamic control in large-scale systems. Wireless communication is essential in CPSs, especially in mobile applications like autonomous vehicles, medical monitoring, and robotics, where security remains critical due to the broadcast nature \cite{CPS1,CPS2}.

As pointed out in \cite{CPS1}, the main challenge of adding AN to the CPSs is that the AN direction matrix is given in advance as the excitation signals, thus optimizing the AN direction matrix will lead to the re-excitation of CPSs. To solve this problem, the authors of \cite{CPS1} introduced a searchable virtual noise space to design the AN injection direction as 
\begin{equation}
{S_{\mathbf{V}}} = \left\{{{{\mathbf{V}}_ - } \in {\mathbb{R}^{h \times N}}\left| {{\text{rank}}\left( {{{\mathbf{V}}_ - }} \right)} \right. = h} \right\},
\end{equation}
{where ${{\mathbf{V}}_ - }$ denotes a specific virtual noise matrix in the searchable virtual noise space ${S_{\mathbf{V}}}$. For any ${{\mathbf{V}}_ - } \in {S_{\mathbf{V}}}$, there exists a direction matrix ${\mathbf{E}} \in {\mathbb{R}^{h \times h}}$ such that ${\mathbf{V}} = {\mathbf{E}}{{\mathbf{V}}_ - }$ with $h$ denoting the dimension of the virtual noise space and $N$ representing the number of the state data samples.} Driven by this searchable virtual noise space, the influence of AN is not amplified for Bob in the CPS.

The authors of \cite{CPS2} focused on the full-duplex communication mode in CPS, which is challenging as the AN is unable to remain in the null space in both forward and reverse channels. Therefore, the optimization problem aims to jointly optimize the AN and the transmit filter with respect to the QoS constraint, which is formulated as 
{\begin{subequations}
\begin{align}
  \mathop {\min }\limits_{{\mathbf{W}},{\mathbf{V}}} P_t^A + P_t^B, \hfill \\
  {\text{s}}{\text{.t}}{\text{. SINR}} \geqslant \gamma,  \hfill
\end{align} 
\end{subequations}
where $P_t^A$ and $P_t^B$ denote the transmit power for forward and reverse communications, respectively.}
Additionally, the research has delved into the impact of channel estimation errors and self-interference on the overall system performance, providing a comprehensive explanation of these effects.

AN-aided CPSs face unique challenges in effectively integrating AN to enhance both communication and sensing functionalities without degrading system performance. Key challenges include optimization for performance improvements. Future trends are likely to focus on developing adaptive noise management strategies to dynamically balance performance across diverse operational conditions. These trends will leverage advanced machine learning for real-time adjustments and anomaly detection while enhancing security measures to protect against vulnerabilities introduced by AN. Additionally, innovations will aim to improve energy efficiency and system robustness to support increasingly complex CPS applications.

\subsection{AN-aided CRN}

\begin{figure}[t]
\centering
\includegraphics[width=3.5in]{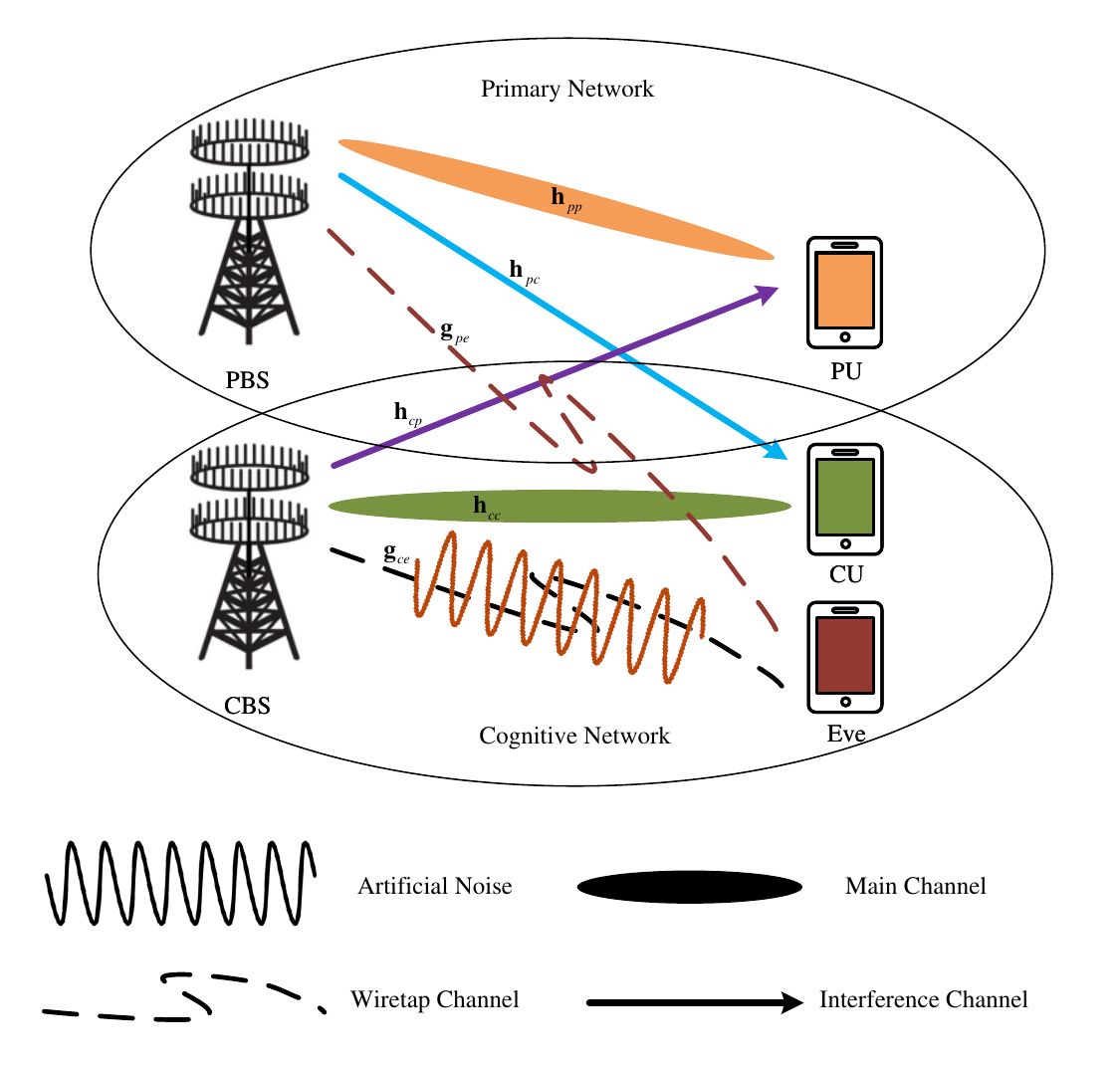}
\caption{Framework of the AN-aided CRN wireless secure communications.}
\label{fig_CRN}
\vspace{-0em}
\end{figure}

Over the past two decades, cognitive radio networks (CRNs) have rapidly evolved to address spectrum scarcity by allowing cognitive users (CUs) to share licensed spectrum with primary users (PUs) under QoS constraints, as shown in Fig. \ref{fig_CRN}. To enhance the PLS of CRN, extensive analytical studies have been conducted \cite{CRN2,CRN3,CRN6,CRN7,CRN5}.

In order to address multiple Eves in the network and to improve energy efficiency, the authors of \cite{CRN2} proposed a new optimization framework to maximize the secrecy energy efficiency (SEE) at the cognitive base station (CBS). This was achieved by optimizing the beamforming under the constraints of the secrecy rate for both PU and CU, limitation of PUs interference from CBS, and adhering to the total transmit power constraint for the CBS. The SEE metric is given by
\begin{equation}
\eta_{SEE}={R_{c,\sec}}/{P_{tot}},
\end{equation}
where $R_{c,\sec}$ represents the secrecy rate and $P_{tot}$ is the total power consumption. Moreover, a secrecy energy efficiency maximization (SEEM) scheme was proposed by exploiting the instantaneous CSI of Eves. In addition, when the instantaneous CSI is unavailable, the authors proposed an alternative SEEM scheme by exploiting the statistical CSI of Eves. Numerical results demonstrated that the proposed method can achieve a trade-off between schemes that focus solely on secrecy rate and those that focus solely on energy efficiency.

In \cite{CRN4}, the authors studied the AN-aided precoding scheme for MIMOME CRNs by solving a secrecy rate maximization problem. Later, the authors of \cite{CRN3} identified and corrected gaps in \cite{CRN4}. To handle the non-convex capacity constraint, the capacity function was reformulated as
\begin{equation}
C_k(\mathbf{W},\mathbf{V})=\varphi_k(\mathbf{W},\mathbf{V})-\phi_k(\mathbf{V}),
\end{equation}
where $\varphi_k(\mathbf{W},\mathbf{V})$ and $\phi_k(\mathbf{V})$ are auxiliary functions. Subsequently, this problem was iteratively solved using successive convex approximation (SCA), with simulations confirming convergence and satisfactory performance.

In \cite{CRN6}, an AN-aided cognitive transmit strategy was proposed to maximize the joint secrecy rate of primary and cognitive links under power and primary QoS constraints. In the first slot, the cognitive transmitter listens while an inactive user jams with AN; in the second slot, it relays the primary signal while sending its own data and AN. Specifically, the secrecy rate is formulated as
\begin{equation}
R_{\mathrm{secrecy}}=\left[R_{p}+R_{s}^{\mathrm{sum}}-R_{e}\right]^{+},
\end{equation}
where $R_{p}$ denotes the achievable rate at the primary receiver,  $R_{s}^{\mathrm{sum}}$ represents the sum rate of cognitive network, and $R_{e}$ stands for the achievable rate at Eve. Furthermore, the authors proposed a computationally efficient approximation method combining semidefinite relaxation (SDR) with a two-step alternating optimization to obtain a local optimum.

Further in \cite{CRN7}, an alternative optimization method combined with one-dimensional linear searching was employed to design the optimal beamforming. Furthermore, by confining the AN to the null space of the legitimate channel, a low-complexity secure beamforming scheme was also presented.

The authors of \cite{CRN5} proposed a SEE optimization scheme for OFDM-based cognitive radio downlink transmissions. This was achieved by optimizing the power allocation between information-bearing and AN signals across different OFDM subcarriers subject to the total transmit power constraints for the PBS and CBS while guaranteeing a required secrecy rate for the CU and PU. Additionally, two schemes were proposed for scenarios where the instantaneous CSI or statistical CSI of Eve was available. Since there are no closed-form solutions for the proposed problems, new two-tier algorithms were proposed to achieve the optimal power allocation solutions to the formulated problems.

In general, AN is generated within the CRN to secure communications of both PUs and CUs. However, as the CRN inherently interferes with the PU, optimal power allocation between the information signal and AN, along with system energy efficiency, emerges as a key challenge, further complicating the optimization problem.

\subsection{AN-aided D2D}
\begin{figure}[t]
\centering
\includegraphics[scale=0.4]{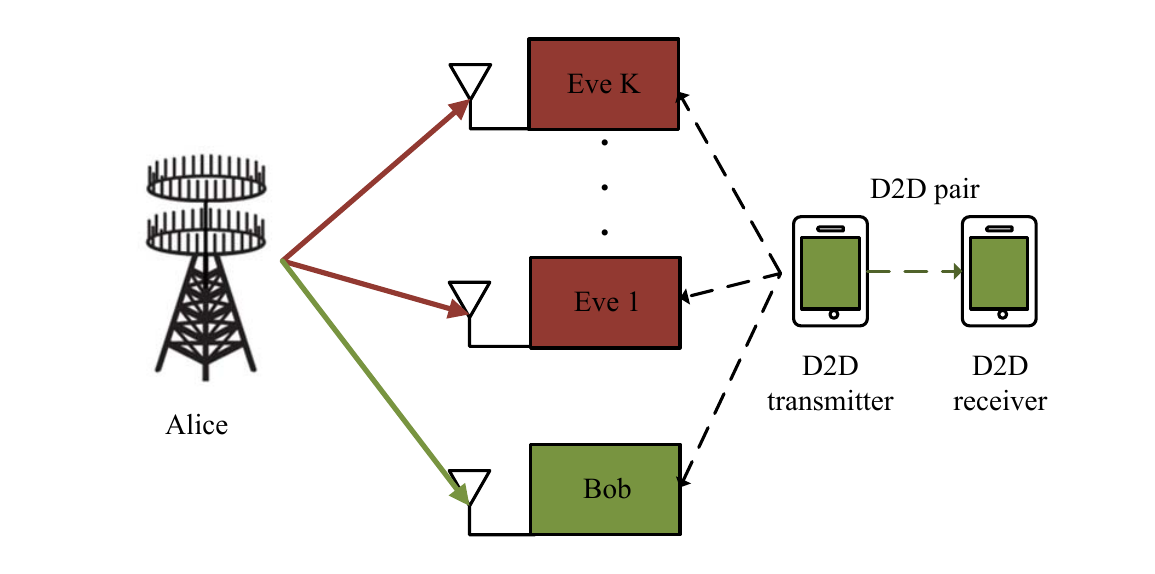}
\caption{Framework of the AN-aided D2D wireless secure communications.}
\label{fig_5}
\vspace{-0em}
\end{figure}

Device-to-device (D2D) communication, enabling direct transmission without BS involvement, is a key technology in 5G networks. As illustrated in Fig. \ref{fig_5}, scenarios involving multiple Eves targeting both cellular and D2D links have drawn increasing research interest \cite{D2D1,D2D2}.


The authors of \cite{D2D1} devised a secure D2D communication strategy using AN, jamming-aided precoding, and signal precoding to maximize secrecy capacity while sharing downlink resources with cellular users. AN was placed in the null space of interference channels. Three precoding schemes and an optimal power allocation algorithm were developed, all showing strong security improvements in simulations.

In \cite{D2D2}, the authors aimed to enhance the SSR for both D2D and cellular users. AN was precoded using the null space of CSI from the BS to both user types, and power was allocated via a one-dimensional search. A pricing-based pairing and power control algorithm was also introduced to match D2D pairs with cellular users, maximizing secrecy throughput.

Due to device limitations, AN is typically generated by the BS over the downlink, making power allocation in D2D communications difficult. In addition, how to pair D2D devices for security improvement is also a great challenge.

\subsection{AN-aided IoT}
\begin{figure}[t]
\centering
\includegraphics[scale=0.54]{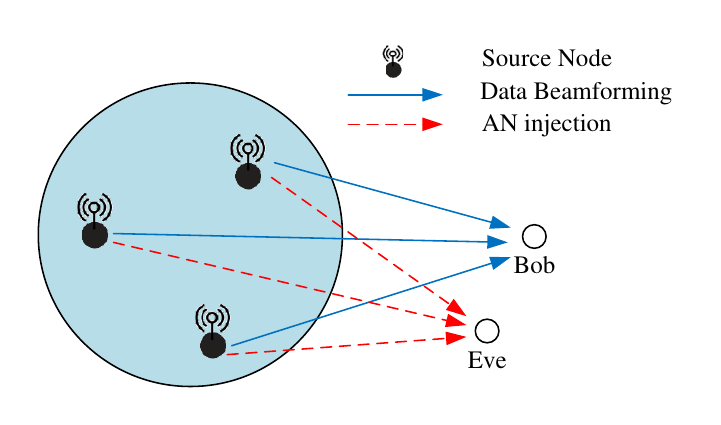}
\caption{Framework of the AN-aided IoT wireless secure communications.}
\label{fig_IoT}
\vspace{-0em}
\end{figure}

With the appearance of the Internet of Things (IoT), human-computer interaction technology is more important in our lives. Due to the sharp growth in the number of connected devices, conventional cryptography mechanisms may suffer from the difficulties caused by the key distribution of multiple devices. As shown in Fig. \ref{fig_IoT}, to overcome such critical issues for IoT, AN techniques can be generated cooperatively \cite{IoT1,IoT2}.

To reduce overhead in large-scale IoT networks, the authors of \cite{IoT1} proposed an AN-aided cooperative beamforming (CB) scheme using a virtual antenna array (VAA). The method exploits the statistical properties of CB links to avoid CSI collection and precoding weight sharing. Closed-form secrecy rate expressions were derived with and without CSI errors, and the impact of system parameters like VAA size was analyzed for secure CB system design.

In \cite{IoT2}, the authors proposed an AN-aided secure transmission scheme for cooperative non-orthogonal multiple access (NOMA)-based IoT systems with multi-antenna nodes. The BS transmits signals with AN, and strong nodes relay them to weak nodes. Exact SOP expressions were derived, and secrecy diversity analysis showed that system performance is limited by the weaker node’s channel quality.

In conclusion, AN in IoT needs to be generated collaboratively by multiple nodes, which indicates that additional coordination and synchronization issues between multiple nodes need to be addressed.

\subsection{AN-aided mmWave}
The millimeter wave (mmWave) frequency band is a promising candidate for 5G wireless networks. The narrow and directional beams in mmWave communications have significantly enhanced transmission security. However, eavesdropping remains a potential threat in both indoor and outdoor environments \cite{mmWave3,mmWave1,mmWave2,mmWave4,mmWave5}.

In \cite{mmWave3}, a practical secure transmission scheme called artificial noise hopping (ANH) was proposed for MIMO mmWave systems with single or multiple colluding Eves. ANH avoids symbol-level beam steering and requires only two RF chains-one for data and another one for ANH. Considering LoS-dominant channels and MRC, the scheme achieves low overhead and minimal hardware complexity, with both simulations and experiments confirming its practicality.

The authors of \cite{mmWave1} proposed an AN-aided two-stage hybrid beamforming scheme for MU-MISO mmWave relay systems with two passive Eves. In phase I, RF analog beamformers and relay combiners are designed to maximize signal power and suppress interference using codebook-based beam training. In phase II, digital baseband beamforming uses zero-forcing to reduce complexity. The same approach applies to both relay phases, avoiding joint optimization and minimizing feedback.

In \cite{mmWave2}, a low-complexity AN-aided hybrid analog–digital precoding scheme was proposed for mmWave MISO systems. Using the joint moment generating function, a lower bound of secrecy rate ${{\tilde R}_{{\rm{Sec}}}}$ were derived and optimized as
\begin{subequations}
\begin{align}
&\mathop {\max }\limits_{\phi ,{{\bf{F}}_{{\rm{RF}}}},{{\rm{f}}_{{\rm{BB}}}},{{\bf{U}}_{{\rm{BB}}}}} {{\tilde R}_{{\rm{Sec}}}},{\rm{}}\\
&\quad\quad\text {s.t. }  {{\bf{F}}_{{\rm{RF}}}} \in {{\cal F}_{{\rm{RF}}}},\\
&\;\;\;\;\;\;\;\;\;\;\;\;\;{\left\| {{{\rm{f}}_{{\rm{BB}}}}} \right\|^2} = 1,\\
&\;\;\;\;\;\;\;\;\;\;\;\;\;\left\| {{{\bf{U}}_{{\rm{BB}}}}} \right\|_{\rm{F}}^2 = 1,\\
&\;\;\;\;\;\;\;\;\;\;\;\;\;0 \le \phi  \le 1
\end{align}
\end{subequations}
where $\mathbf{F}_{\text{RF}}$ denotes the analog beamforming matrix, $\mathbf{f}_{\text{BB}}$ represents the digital beamforming vector and AN digital baseband precoder, and $\phi$ is the power allocation factor. The problem was decomposed into optimizing $\phi$ and solving a hybrid precoder via eigen-decomposition and projection methods.

The authors of \cite{mmWave4} proposed a dynamic parameter transmission scheme for mmWave systems over multipath slow fading channels, aiming to maximize secrecy throughput under an SOP constraint. The CDF of Eve’s SINR was derived for SOP analysis, enabling optimal codeword rate and power allocation design. AN beamforming was based on the array responses of Bob and Eve, and the study revealed that secrecy performance heavily depends on the spatial path overlap between them.

The same system model with partial CSI of Eve at Alice was considered in \cite{mmWave5} under an on-off transmission scheme. An optimization problem was formulated to maximize the secrecy rate subject to codeword rate and SOP constraints as
\begin{subequations}
\begin{align}
&\mathop {\max }\limits_{\eta, R_t}  R_s\left(\gamma^{\circ}\right) \\
&\text {s.t. }  0<R_s\left(\gamma^{\circ}\right)<R_t\left(\gamma^{\circ}\right) \leq C_d, \\
&\;\;\;\;\; \mathcal{P}_{s o}\left(\gamma^{\circ}\right) \leq \epsilon, \\
&\;\;\;\;\; 0 \leq \eta \leq 1,
\end{align}
\end{subequations}
where $\gamma^0 \in {\Upsilon}$ represents overall channel gain satisfying the transmission constraint, $R_s$ and $R_t$ are defined as the secrecy rate and codeword rate. $\eta$ is the power allocation factor between useful signal and AN. Using the CDF of Eve's SINR from \cite{mmWave4}, a closed-form SOP and optimal power allocation $\eta^{\star}$ were derived. Results highlighted that the secrecy outage is more likely to occur when common paths are weak or Eve is close to Alice.

Overall, AN design in mmWave systems needs to consider unique channel characteristics like array response, codebook structure, orthogonality, and sparsity. Despite signal processing challenges, hybrid architectures are essential due to the high overhead of fully digital massive MIMO.

\subsection{AN-aided MU}

\begin{figure}[t]
\centering
\includegraphics[width=3.5in,height=2.0in]{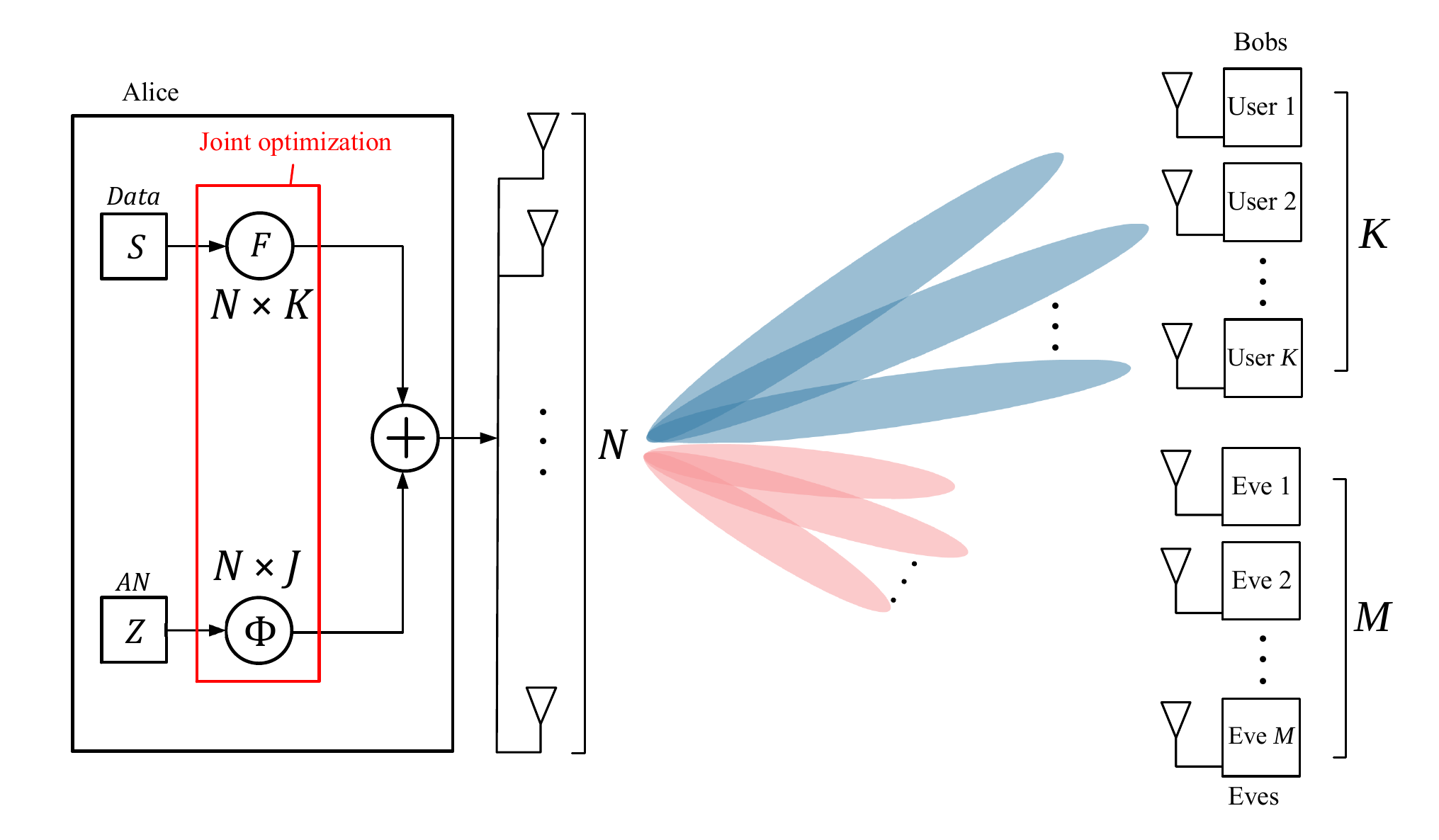}
\caption{AN-aided MU wireless secure communications.}
\label{fig_MU}
\vspace{-0em}
\end{figure}

The application of AN can be easily extended to MU networks with multiple receivers or transmitters. Depending on the characteristics of receivers and transmitters in actual application scenarios, MU can be divided into downlink and uplink. As displayed in Fig. \ref{fig_MU}, in the downlink, a powerful BS serves multiple users, necessitating high spectral efficiency and secure communication. A heuristic approach is to integrate coexisting services, typically, multicast service and confidential service, into one integral service for one-time transmission, where establishing the security of the confidential service while not compromising the public service becomes a crucial problem. The authors of \cite{multiuser1,multiuser2,multiuser3} designed the optimal input of multicast message, confidential message, and AN jointly in this scenario. Furthermore, an appropriate user scheduling strategy can enhance security, as demonstrated by \cite{multiuser4}. In the downlink scenario, feedback from the user end is limited. The authors of \cite{multiuser5} and \cite{multiuser6} addressed this issue by using quantified CSI to complete the design of the corresponding transceiver and transmission framework. In addition, Ref. \cite{multiuser7} studied the joint design of precoding and AN, and proposed a new secure precoding algorithm. In the uplink, the performance of the user transmission is limited, making the utilization of the MU feature an important topic. The authors of \cite{multiuser8} and \cite{multiuser9} used MU diversity to make non-scheduled users in the network generate AN, thereby improving  security performance. Ref. \cite{multiuser10} studied the power allocation problem of MU constellations and AN in the mMIMO scenario, and provided a new NOMA solution. In addition, time-domain AN can also be combined with the secret key in single-carrier frequency-division multiple access (FDMA) systems to obtain enhanced security \cite{multiuser11}.

The authors of \cite{multiuser1} studied optimal AN-aided transmit design for MU-MISO systems with integrated multicast, confidential messages, and AN, focusing on quality of multicast service (QoMs). They aimed to maximize secrecy rate under QoMs constraints and defined the secrecy rate region boundary. Results showed this beamforming approach effectively enhances security performance.

In \cite{multiuser2}, a secure MU-MISO transmission design with AN was proposed to serve multicast and confidential messages under imperfect CSI modeled as $\mathbf{h}=\tilde{\mathbf{h}}+\mathbf{e}$ with bounded error. The goal was worst-case secrecy rate maximization (SRM) under sum power and QoS constraints. The problem’s hidden convexity was revealed and reformulated into a sequence of semidefinite programs (SDPs).

Building on \cite{multiuser2}, the authors of \cite{multiuser3} showed that transmit beamforming with well-designed AN outperforms existing schemes for confidential message transmission under both perfect and imperfect CSI. They analyzed computational complexity and proposed two suboptimal schemes, including one with a power splitting factor $\rho$ to decouple multicast and confidential transmissions: 
\begin{equation}
\mathrm{Tr}(\mathbf{Q}_{c}+\mathbf{Q}_{a})=\rho P, \quad
\mathrm{Tr}(\mathbf{Q}_{0})=(1-\rho)P,
\end{equation}
where $\mathbf{Q}_{c}$, $\mathbf{Q}_{a}$, and $\mathbf{Q}_{0}$ denote covariance of $\mathrm{x}_c$, $\mathrm{x}_a$ ,and $\mathrm{x}_0$, respectively, while $P$ represents the total transmission power budget. Numerical results showed AN effectively enlarges secrecy rate regions, and suboptimal methods achieve near-optimal performance with less complexity.

To enhance MU-MISO network security, the authors of \cite{multiuser4} proposed an AN relay strategy using orthogonal space-time block code (OSTBC) at the BS and AN transmission from a cooperative relay that avoids the selected legitimate user. Two user scheduling schemes were used for user selection. Exact closed-form expressions for SOP and secrecy throughput were derived, and simulations confirmed the scheme’s effectiveness and the impact of varying antenna numbers.

In \cite{multiuser5}, an AN-aided nonlinear low-complexity transceiver using Tomlinson-Harashima precoding (THP) was proposed for secure MU-MISO communications with limited feedback. The authors derived analytical ESR approximations and near-optimal power allocation via numerical methods.

A secure MU multi-antenna transmission framework using AN-aided linear ZF beamforming with limited CSI feedback from distributed Bobs was proposed in \cite{multiuser6}. Users quantize the channel direction as
\begin{equation}
\hat{\mathbf{h}}=\arg\max_{\mathbf{c}\in\mathcal{C}}\left|\tilde{\mathbf{h}}\mathbf{c}^H\right|,
\end{equation}
where $\mathbf{c}$ is the codeword for feedback. The authors derived a tractable lower bound on ESR and optimized power allocation to maximize it. They analyzed the effects of transmit power and feedback bits, identifying minimum requirements for both to ensure desired performance.

In \cite{multiuser7}, a joint optimization framework was proposed to maximize the sum secrecy rate in MU-MIMO wiretap channels with multiple Eves by jointly optimizing secure precoders and AN. The non-convex, non-smooth SRM problem was reformulated into a smooth one, defining a joint vector
\begin{equation}
\overline{\mathbf{v}}=\begin{bmatrix}\mathbf{f}_1^\top,\mathbf{f}_2^\top,\cdots,\mathbf{f}_K^\top,{\mathbf{a}}_1^\top,{\mathbf{a}}_2^\top,\cdots,{\mathbf{a}}_J^\top\end{bmatrix}^\top,
\end{equation}
with the channel rate expressed as
\begin{equation}
R_k=\log_2\left(\frac{\bar{\mathbf{v}}^\mathsf{H}\mathbf{A}_k\bar{\mathbf{v}}}{\bar{\mathbf{v}}^\mathsf{H}\mathbf{B}_k\bar{\mathbf{v}}}\right).
\end{equation} 
Algorithms were developed for both perfect and imperfect CSI, including a null-space projection AN design. The approach achieved faster convergence, required only a small AN subspace, and outperformed existing methods in secrecy rate.

The authors of \cite{multiuser8} proposed a user scheduling scheme with AN to enhance security in an MU uplink wiretap network with one desired receiver and multiple Eves. One user transmits data while others generate AN, with the received signal at the desired user given by
\begin{equation}
y=h_nx_n+\sum_{s=1}^Sh_s\delta_s+z,
\end{equation}
where $x_n$  denotes the desired data symbol for the $n$-th user and $\delta_s$ represents an independent jamming signal generated by the $s$-th user. By using threshold-based feedback, the scheme achieves optimal multiuser diversity gain $\log N$ and reduces signaling overhead. Simulations show it outperforms conventional methods with higher average secrecy rates.

Similar to \cite{multiuser8}, the authors of \cite{multiuser9} proposed a secure user scheduling scheme with AN for an MU SIMOME wiretap network, selecting one user for data and another for AN. The scheme requires only Eve’s channel statistics, not instantaneous CSI. Closed-form SOP expressions with MRC were derived, showing that AN effectively lowers SOP by degrading Eve’s link and is robust to changes in Eve’s antenna number.

To secure an energy-based mMIMO system, a joint MU constellation alignment and AN design was proposed in \cite{multiuser10} to manage multiuser interference with finite-alphabet inputs and optimize error performance. With large antenna arrays at the BS and passive Eve, and multi-antenna Bobs, the authors designed an energy-efficient pulse amplitude modulation (PAM) constellation and a stepped water-filling algorithm to generate AN vectors unknown to Eve. Simulations showed Bob’s error rate vanishes with more antennas, while Eve’s error rate remains bounded, highlighting improved security.

In \cite{multiuser11}, a time-domain AN-aided scheme was proposed for single-carrier-FDMA with eavesdropping, where AN is projected onto the null space of the Alice-Bob channel:
\begin{equation}
\mathbf{R}^{\mathrm{cp}}\mathbf{H}_t^{(k)}\mathbf{V}^{(k)}=0,
\end{equation}
where $\mathbf{R}^{\mathrm{cp}}$ is cyclic prefix removal matrix. Analysis showed robustness even when Eve used machine learning detectors, with secrecy rate growing linearly with AN streams. The study also combined AN with a channel-based secret key to further enhance the PLS.

In summary, AN in downlink scenarios are mainly utilized to enhance the quality of confidential service, whereas in uplink scenarios AN are often co-generated by users and employed to secure communications. Multi-user scenarios bring multiple variables that complicate the problem and increase the algorithmic complexity. Consequently, the development of efficient algorithms with minimal performance loss should be a primary focus in current research.

\subsection{AN-aided Protected Zone}

The secrecy-protected zone (PZ) is an Eve-free area formed inherently or intentionally, preventing eavesdroppers from accessing it. Installing Alice or Bob in restricted or isolated locations, like high communication towers, ensures such zones. Beyond physical barriers, defining a protected zone near Alice helps prevent Eve’s channel from outperforming Bob’s, allowing AN to achieve higher secrecy rates \cite{PZ2,PZ1}.

The authors of \cite{PZ2} studied the impact of a nearby Eve by defining a PZ around Alice. A weighted normalized cost function was introduced as
\begin{equation}
\mathrm{CF}(a,b,r_s)=\kappa_1\frac{a+b}{P_{max}}+\kappa_2\frac{r_s}{r_{s_{max}}},
\end{equation}
where $P_{max}$ and $r_{s_{max}}$ denote the maximum available transmission power and the maximum allowable radius of the PZ, respectively. $a+b$ represents the total transmitted power and $r_s$ stands for the size of PZ, while $\kappa_1$ and $\kappa_2$ are the weights, respectively. The study proposed an optimization strategy to smartly allocate power and PZ size, effectively maximizing the probability of achieving a target secrecy rate.

Furthermore, the authors of \cite{PZ1} enhanced secrecy by combining AN and a PZ, considering unknown Eves and jammers. Without CSI, a fixed secrecy rate 
$R_S$ is set, and the secrecy transmission rate is defined as
\begin{equation}
\mathcal{R}\triangleq(1-\Pr[\mathbb{E}_1])(1-\Pr[\mathbb{E}_2])R_\mathrm{S},
\end{equation}
where $\mathbb{E}_1$ represents for Bob's inability to decode the codeword at the specified rate and $\mathbb{E}_2$ denotes for the capacity of at least one Eve is higher than the limit. They analyzed how AN, PZ radius, and jammer/Eve density affect performance and derived optimal power allocation in interference-limited networks to maximize $\mathcal{R}$.

The advantage of using the AN for secure communication in the protected zone is that as long as Eve cannot approach Alice, more resources could be allocated to Bob to ensure robust security. Therefore, resource allocation in this scenario becomes a critical issue, which fundamentally explains the impact of Eve channel quality on the PLS boundary.

\subsection{AN-aided Relay}

To enhance the PLS of relay networks, various relaying strategies have been explored in the literature, as evidenced in citations like \cite{relay3,relay4,relay7,relay8,relay10,relay2,relay9,relay12,relay15,relay5,relay6,relay11,relay16,relay13,relay17,relay18,relay14,relay1}. The research on AN-aided relay techniques can be categorized into different scenarios, including single amplify-and-forward (AF) relay \cite{relay3,relay4,relay7,relay8,relay10}; multiple AF relays \cite{relay2,relay9,relay12,relay15}; single decode-and-forward (DF) relay \cite{relay5,relay6,relay11,relay16}; multiple DF relays \cite{relay13,relay17,relay18}; non-regenerative (NR) relay \cite{relay14}; untrusted two-way relay (UTWR) \cite{relay1}.
These studies aim to devise effective strategies to bolster the security and performance of relay-assisted communication networks.

In \cite{relay3}, an AN-aided secure MIMOME system with a multi-antenna AF relay was studied under unknown Eve CSI and location. The goal was to maximize the expected secrecy rate, where
a penalty method with stochastic decomposition was used. Results showed that secrecy rate improves with more antennas at Bob, and higher AN power becomes optimal as SNR increases. Furthermore, two novel AN algorithms were introduced in \cite{relay4} for interference alignment (IA)-based multiuser communication systems with a multi-antenna half-duplex (HD) relay and Eve. The goal was to minimize transmit power while ensuring each user’s SNR exceeds a threshold.
An iterative algorithm with SDP was proposed to solve this problem, leading to four AN-based transmission models balancing sum and secrecy rates.

In \cite{relay7}, three AN-aided MIMOME transmission schemes using a two-way multi-antenna relay were studied with different security–complexity trade-offs. First, secure transmission conditions were derived for relay antenna selection without AN. Then, Alice's AN injection was analyzed under various eavesdropping strategies. Finally, a low-complexity joint antenna selection and AN scheme was proposed.
Results showed that joint AN and antenna selection improves secrecy at the cost of increased complexity.

In \cite{relay8}, the concept of AN was applied in an untrusted relay network for single-input single-output (SISO) systems. In the first time slot, Alice sends the information-bearing signal while Bob transmits AN $r$, resulting in the relay receiving
\begin{equation}
{\tilde y_R} = {h_{ar}}s + {h_{br}}r + \tilde n,
\end{equation}
{where ${h_{ar}}$ denotes the CSI between Alice and the relay, ${h_{br}}$ represents the CSI between Bob and the relay, and $\tilde n$ is the AWGN.}
In the second time slot, the relay amplifies and forwards the combined signal to Bob, who can cancel the AN using prior knowledge. The SER was derived under AWGN, and the optimal AN phase and power allocation were obtained by maximizing the SER. Notably, the study revealed that Gaussian-distributed AN is not always optimal.

The AN was injected at the trusted multi-antenna AF relay for SISO in the presence of multiple single-antenna (MSA) Eves in \cite{relay10}. Assuming imperfect knowledge of the Eves' channels, the power of Alice, the AF relaying matrix, and the covariance of the AN transmitted by the relay were jointly optimized to maximize the received SINR of Bob under robust secrecy constraints. 
Furthermore, a penalized difference-of-convex algorithm was proposed to efficiently solve the non-convex SDP with guaranteed convergence.

In \cite{relay2}, the authors employed AN from multiple cooperative relays using MU diversity in SISO systems to combat MSA eavesdroppers. They derived the maximum number of uniformly distributed Eves that can be tolerated under secrecy constraints. For identical path-loss Eves, an exponential number (in system nodes) is tolerable, while for uniformly distributed Eves, sub-linear growth in Eves is supported.

For two-way AF relay networks, a joint design of cooperative beamforming and AN was proposed as \cite{relay9}
\begin{subequations}
\begin{align}
&\mathop {\min }\limits_{{\mathbf{w}},{\mathbf{\Sigma }}} \mathop {\max }\limits_{l \in {\cal L}}  {- \left( {{R_{B,1}} + {R_{B,2}} - {R_{E,l}}} \right)}  \\
 & {\text{s.t.}} \; {\mathbf{e}}_k^T\left( {{\mathbf{Cw}}{{\mathbf{w}}^H} + {\mathbf{\Sigma }}} \right){{\mathbf{e}}_k} \leqslant {P_{R,k}},\forall k,
\end{align}
\end{subequations} 
where ${P_{R,k}}$ is the power budget of the $k$-th relay, ${{\mathbf{e}}_k}$ denotes a unit vector, ${\mathbf{C}}$ represents the covariance matrix of the information-bearing signal, and ${\mathbf{\Sigma }}$ stands for AN covariance matrix.
Based on the SSR maximization criterion, a stationary point was efficiently developed by utilizing SDR and SCA techniques. The proof was also given to show that SDR is always tight and yields a rank-one solution.

In \cite{relay12}, the joint cooperative beamforming and AN strategy was employed for a set of single-antenna AF relays in the presence of a single-antenna Alice, a single-antenna Bob, and MSA Eves. Under both the total and individual relay power constraints, an optimal joint cooperative beamforming and AN design based on SRM was proposed as
\begin{subequations}
\begin{align}
& \mathop {\max }\limits_{{\mathbf{w}},{\mathbf{\Sigma }}}  {R_s} \\
  &{\text{s.t.}}  {{\mathbf{w}}^H}{\mathbf{Cw}} + {\text{Tr}}\left( {\mathbf{\Sigma }} \right) \leqslant P,{\mathbf{\Sigma }}\underset{\raise0.3em\hbox{$\smash{\scriptscriptstyle-}$}}{\succ } {\mathbf{0}}, \\
  & \;\;\;\;\; {\mathbf{e}}_k^T\left( {{\mathbf{Cw}}{{\mathbf{w}}^H} + {\mathbf{\Sigma }}} \right){{\mathbf{e}}_k} \leqslant {P_{R,k}},\forall k,
\end{align}
\end{subequations}
{where ${{\mathbf{e}}_k}$ denotes a unit vector with the $k$-th entry being one.}
The SRM problem was treated by a combination of two-level optimization and SDR. For the latter, it was proved that SDR always has a rank-one solution for the SRM problem, which identifies that SDR is optimal.

The authors of \cite{relay15} tackled an optimization problem in a two-hop relay single-input single-output multiple-antenna-eavesdropper (SISOME) system, wherein multiple multiple-antenna (MMA) AF relays collaboratively transmit information-bearing and AN signals from a single-antenna Alice to a single-antenna Bob in the presence of MMA Eves. More specifically, under the the spherical uncertainty model of wiretap channel $\left\| {{{\mathbf{g}}_m} - {{{\mathbf{\bar g}}}_m}} \right\| \leqslant {\varepsilon _m}$, the worst achievable rate of Eve is expressed as 
\begin{equation}
\mathop {\max }\limits_{{{\mathbf{g}}_m}} {C_{e,m}} = \log \left( {1 + {P_s}{{\left( {\left\| {{{{\mathbf{\bar g}}}_m}} \right\| + {\varepsilon _m}} \right)}^2} + {{\mathbf{\Lambda }}_m}} \right),
\end{equation}
where ${{\varepsilon _m}}$ reflects the accuracy for the knowledge of Eves' CSI and ${{{\mathbf{\Lambda }}_m}}$ represents the SINR under the accurate CSI. The SRM problem was solved via a two-level polynomial-time approach using SDR, which was proven optimal through reformulation and Karush-Kuhn-Tucke analysis. AN played a key role in ensuring the SDR solution’s optimality.

In \cite{relay5}, the AN was generated by a cooperative FD jammer for MIMOME with a multi-antenna FD relay, where the cooperative jammer (CJ) was assumed to be an FD node that is solely charged by the ambient RF transmissions. Without the CSI of Eve, the authors of \cite{relay5} investigated the self-energy recycling at the CJ and derived a sufficient condition for the CJ to operate as a node, expressed as
\begin{equation}
{B_J} \geqslant \left( {{P_J}{\text{+ }}{P_p}} \right)T,
\end{equation}
where ${B_J}$ denotes the battery energy of the CJ, ${P_J}$ represents the signal power transmitting information-bearing signal and AN, ${P_p}$ stands for the signal processing power consumption for DF, and $T$ is the activation duration. It reveals that the battery state increases with time. Finally, the numerical results demonstrated that the introduction of AN-aided CJ achieves a significant average secrecy rate improvement.

The authors of \cite{relay6} designed a relay-aided secure transmission scheme for MIMOME. By assuming that both Alice and the relay transmit AN alongside information-bearing signals, a closed-form expression for the transmission outage probability and SOP were derived as
\begin{equation}
\begin{gathered}
  {P_{to}} = 1 - \left( {1 - {F_{{\gamma _{sr}}}}\left( {{\tau _b}} \right)} \right)\left( {1 - {F_{{\gamma _{rd}}}}\left( {{\tau _b}} \right)} \right), \hfill \\
  {P_{so}} = 1 - \exp \left( {- 2\lambda \left( {{J_1} + {J_2} - {J_3}} \right)} \right), \hfill \\ 
\end{gathered} 
\end{equation}
where ${{\gamma _{sr}}}$ and ${{\gamma _{rd}}}$ stand for the SINR from S to R and from R to D, respectively, ${F_\gamma }\left( \tau  \right)$ denotes the CDF of $\gamma$, while ${J_1}$, ${J_2}$, and ${J_3}$ are calculated by the CDF of Eve's SINR. Asymptotic analysis with a large number of Alice's antennas led to the secrecy throughput expression
\begin{equation}
{T_s} = \left( {{R_b} - {R_e}} \right)\left( {1 - {P_{to}}} \right).
\end{equation}
By solving the secrecy throughput maximization (STM) problem under the SOP constraint, the system and channel parameters were determined for practical deployment.

Based on a pseudo-random sequence known to Bob, an AN scheme was investigated in \cite{relay11} for NOMA FD relay networks. In this scheme, the optimal power allocation between the information-bearing and the AN signals was determined to minimize the SOP
\begin{equation}
{\theta ^ * } = \mathop {\arg \min }\limits_{0 \leqslant \theta  < 1} {P_{so}}.
\end{equation}
A closed-form SOP expression showed that combining FD and AN at relays significantly improves physical layer secrecy in NOMA cooperative networks.

In \cite{relay16}, a NOMA-based two-way relay network was developed where a trusted relay assists two users in exchanging messages while continuously transmitting AN to impair SSA/MSA Eves. The relay’s signals in phases 1 and 2 are given by
\begin{equation}
\begin{gathered}
  {\mathbf{x}}_R^{\left( 1 \right)} = {\mathbf{V}}\sqrt {\frac{{{P_R}}}{{\left( {{N_R} - 2} \right)}}} {{\mathbf{r}}^{(1)}}, \hfill \\
  {\mathbf{x}}_R^{\left( 2 \right)} = {\mathbf{W}}\sqrt {\frac{{\theta {P_R}}}{2}} \left[ {\begin{array}{*{20}{c}}
  {{s_B}} \\ 
  {{s_A}} 
\end{array}} \right] + {\mathbf{V}}\sqrt {{P_R}/\left( {{N_R} - 2} \right)} {{\mathbf{r}}^{(2)}}, \hfill \\ 
\end{gathered} 
\end{equation}
where ${\mathbf{x}}_R^{\left( 1 \right)}$ and ${{\mathbf{r}}^{(1)}}$ denote the transmit signal and random vector during the phase 1, while ${\mathbf{x}}_R^{\left( 2 \right)}$ and ${{\mathbf{r}}^{(2)}}$ represent that during phase 2. The relay and users employ FD to enhance efficiency and secrecy. Successive interference cancellation-based decoding schemes were proposed, and closed-form ESRs were derived under both SSA and MSA Eve scenarios.

The authors of \cite{relay13} proposed a novel cooperative AN scheme for underlay cognitive DF relay networks, where a secondary transmitter communicates with an SU via multiple cognitive DF relays while MSA passive Eves attempt interception. Without Eves' CSI, two relay selection schemes were introduced as
\begin{equation}
\begin{gathered}
  {k^*} = \arg \mathop {\max }\limits_{k \in K} {\left| {{h_{kB}}} \right|^2}, \hfill \\
  {k^o} = \arg \mathop {\max }\limits_{k \in K} {R_s}, \hfill \\ 
\end{gathered} 
\end{equation}
{where ${{h_{kB}}}$ denotes the CSI between the $k$-th relay and the SU, while ${k^*}$ and ${k^o}$ represent the indices of selected relays.}
The conventional scheme uses only main channel CSI, while the best relay scheme also considers interference CSI. Closed-form SOP expressions for both schemes were derived.

To secure source nodes communicating via multiple two-way DF relays in the presence of an Eve, an AN-aided two-way opportunistic relay selection scheme was conceived as \cite{relay17}
\begin{equation}
o = \arg \mathop {\max }\limits_{k \in K} \left[ {\min \left( {{{\left| {{h_{kA}}} \right|}^2},{{\left| {{h_{kB}}} \right|}^2}} \right)} \right]
\end{equation}
where $o$ denotes the selected relay, while ${{h_{k{A}}}}$ and ${{h_{k{B}}}}$ are the channels between the relay $k$ to the users $A$ and $B$, respectively. The scheme’s outage and intercept probabilities were analyzed to characterize security and reliability, with a security-reliability trade-off compared against direct transmission and one-way relaying.

A joint user and relay selection scheme was proposed to enhance PLS in an MU multi-relay network, where the best pair of the user and relay maximizing the user-to-destination SINR is jointly selected in \cite{relay18} as
\begin{equation}
\left( {{m^ * },{n^ * }} \right) = \arg \mathop {\max }\limits_{m,n} \min \left( {{\gamma _{m,n}},{\gamma _n}} \right),
\end{equation}
where ${\gamma _{m,n}}$ and ${\gamma _n}$ are the SINR of the relay and destination, respectively.
In addition, the authors analytically examined the SOP of this scheme, based on which the optimal power allocation between the information-bearing signal and AN at the users was determined to minimize the SOP.
To avoid the high complexity of the joint selection, a separate user and relay selection scheme was also proposed, where a relay is first selected to maximize the relay-to-destination SINR, and a user is then selected to maximize the SINR from the user to the selected relay as
\begin{equation}
\begin{gathered}
  {n^ * } = \arg \mathop {\max }\limits_n {\gamma _n}, \hfill \\
  {m^ * } = \arg \mathop {\max }\limits_m {\gamma _{m,{n^ * }}}. \hfill \\ 
\end{gathered} 
\end{equation}
Moreover, the authors also derived the SOP and optimal power allocation for the separate selection scheme. While the joint user-relay selection outperforms the separate scheme, the low-complexity separate scheme achieves similar secrecy performance under certain conditions, such as when the number of users far exceeds relays or when the user-to-relay SNR is much higher than relay-to-destination SNR.

The authors of \cite{relay14} investigated a non-regenerative relay network supporting simultaneous wireless information and power transfer, in which the energy harvesting relay is powered by RF signals from Alice. Assuming imperfect CSI from the Alice/relay to the Bob/Eve, the authors investigated the AN-aided secure robust beamforming that minimizes the transmission power at the relay, while guaranteeing the secrecy rate constraint and the transmit power constraint at the relay.
In particular, two energy harvesting strategies, namely power splitting and time switching, were studied. Numerical results showed power-splitting outperforms time switching.

For a UTWR, Ref. \cite{relay1} aimed to maximize the secrecy sum rate by jointly designing the sources' signal precoder ${{{\mathbf{W}}_i}}$, AN matrix ${{{\mathbf{V}}_i}}$, and the relay's precoder ${\mathbf{F}}$ as
\begin{subequations}
\begin{align}
  &\mathop {\max }\limits_{{\mathbf{F}},{{\mathbf{W}}_i},{{\mathbf{V}}_i}} {\left( {{R_1} + {R_2} - {R_r}} \right)^ + } \\
  &{\text{s.t.}} \; {\left\| {{{\mathbf{W}}_i}} \right\|^2} + {\left\| {{{\mathbf{V}}_i}} \right\|^2} \leqslant {P_i},i = 1,2, \\
  & \sum\limits_{i = 1}^2 {\left( {{{\left\| {{\mathbf{F}}{{\mathbf{H}}_i}{{\mathbf{W}}_i}} \right\|}^2} + {{\left\| {{\mathbf{F}}{{\mathbf{H}}_i}{{\mathbf{V}}_i}} \right\|}^2}} \right) + \sigma _r^2{{\left\| {\mathbf{F}} \right\|}^2}}  \leqslant {P_R},
\end{align}
\end{subequations} 
where ${{R_1}}$, ${{R_2}}$, and ${{R_r}}$ denote the achievable rates at source 1, source 2, and UTWR, respectively. Under perfect CSI, the problem was reformulated as a difference-of-convex program and solved iteratively for a local optimum, with asymptotic analysis guiding precoder design at high relay power. Under imperfect CSI, channel uncertainties were handled via a worst-case model, and a weighted minimum mean square error approach was developed for robust secure precoding.

In a nutshell, applying the AN in relay systems faces several challenges, including optimizing noise power allocation, balancing security with signal quality, and managing increased computational complexity. Ensuring adaptability to dynamic conditions and mitigating interference are also critical concerns. Future directions involve developing advanced optimization and adaptive algorithms, integrating emerging technologies like 5G, enhancing interference mitigation and energy efficiency, and improving robustness and real-time implementation. Addressing these challenges through innovative research and technology can enhance the effectiveness of AN in improving security and performance in relay networks.

\subsection{AN-aided Satellite Communications}

As a result of extended communication range, wide coverage, and adaptable networking capabilities, satellite communication systems offer extensive utility in both military and civilian domains, encompassing broadcasting, navigation, and emergency communication, each benefiting from their distinctive attributes. To enhance the PLS of satellite communications, some AN methods were put forth in \cite{LEO1,LEO2}.

A power and time slot allocation method was proposed in \cite{LEO1} to secure satellite transmission. The authors first proposed a DCAN method based on weighted fractional Fourier transform (WFRFT), outperforming conventional AN methods under imperfect CSI. An optimized allocation method using fractional and convex differential programming addressed secrecy degradation with more signals. Simulations showed improved performance, especially with imperfect CSI, and maximized total secrecy capacity.

In \cite{LEO2}, a secure AN-assisted beamforming scheme was proposed for cognitive satellite-ground networks, where the satellite shares spectrum with multi-cell terrestrial networks amid multiple unauthorized Eves.
The goal is to minimize the total transmit power of the satellite and BSs while meeting secrecy rate constraints $C_p$ for satellite users and SNR constraints ${\rm SINR}_{n,m}$ for terrestrial users as
\begin{subequations}
\begin{align}
&\mathop {\min }\limits_{{{\mathbf{w}}_0},{{\mathbf{w}}_{n,m}}} {\left\| {{{\mathbf{w}}_0}} \right\|^2} + \sum\limits_{n = 1}^N {\sum\limits_{m = 1}^{{M_n}} {{{\left\| {{{\mathbf{w}}_{n,m}}} \right\|}^2}} } \\
&{\text{s.t.}} \; {C_p} \geqslant R,\\
&\;\;\;\;\;\; {\text{SIN}}{{\text{R}}_{n,m}} \geqslant {\gamma _{n,m}},
\end{align}
\end{subequations}
where a low-complexity ZF method was employed to obtain a suboptimal solution.

In summary, the precision of CSI and power constraints are crucial concerns in implementing AN for satellite communication. Inaccurate CSI may lead to the leakage of AN into the main channel, affecting the intended receiver. Furthermore, integrating AN into transmitted signals demands extra power, which is a scarce resource on satellites.

\subsection{AN-aided SWIPT}

The simultaneous wireless information and power transfer (SWIPT) has garnered significant interest in recent years due to its potential to enhance the energy efficiency of networks. It is particularly appealing in scenarios with low energy demands, where it can power wireless devices with limited battery capacity, often in situations where battery replacement or recharging is impractical. However, SWIPT is susceptible to eavesdropping, primarily because energy receivers (ERs) can not only harvest energy but also intercept information-bearing messages intended for Bobs, as noted in \cite{SLP4}. To mitigate this vulnerability, the introduction of AN can further enhance the PLS in SWIPT, as discussed in \cite{SWIPT2,SWIPT1,SWIPT3,SWIPT4,SWIPT5,SWIPT6,SWIPT7}.

 The authors of \cite{SWIPT2} focused on an SEEM problem for a SWIPT-based MIMOME wiretap channel. Concentrating the optimization of the transmit and AN covariance matrix ${{{\bf{Q}}_c},{{\bf{Q}}_a}}$, an SEEM problem was cast and sloved by
applying fractional programming to simplify the objective, followed by exploiting its primal decomposability.

The authors of \cite{SWIPT1} proposed an AN-aided time-switching power-splitting (TSPS) scheme for AN-assisted SWIPT in OFDM systems. This scheme exploits the temporal degrees of freedom from the CP structure to degrade Eve’s signal while providing power to Bob. The AN precoder is designed to cancel interference at Bob. An SRM problem was formulated over CP length, time switching, and power splitting parameters, subject to energy harvesting constraints. Numerical results demonstrated that the TSPS scheme achieves higher average secrecy gain than pure power splitting.

In \cite{SWIPT3}, a downlink MISO NOMA CRNs network with SWIPT was studied, where a primary BS sends confidential signals to PUs across clusters, and NOMA enhances power efficiency in the secondary network. The authors formulated security and energy harvesting constraints for power minimization and designed beamforming under both perfect and bounded CSI error models using SDR and cost-function algorithms. Simulations showed the AN-aided cooperative scheme reduces transmission power in MISO-NOMA SWIPT systems.

In \cite{SWIPT4}, the authors proposed an AN-assisted interference alignment (IA) scheme with wireless power transfer for interference networks with $K$ users. Each user’s transmission involves unitary precoding/decoding matrices to manage signals and AN, while conventional systems require canceling interference and AN at legitimate receivers through specific zero-interference conditions.

The authors of \cite{SWIPT5} treat AN and interference as redundant energy rather than interference to be removed. Using power splitting, received power is divided between information decoding and energy harvesting. Their precoding and decoding algorithm aligns and cancels AN and interference at legitimate receivers. To enhance performance, they maximize total AN transmit power by jointly optimizing information transmit power and power-splitting coefficient.
Since the coupled effect between SINR and EH variables $P_{t}^{[k]}$ and $\rho^{[k]}$, the constraint is non-convex. Due to the coupling between SINR and EH, this non-convex problem was reformulated into a convex form (as in \cite{SWIPT4}) and solved by CVX, with a low-complexity suboptimal closed-form solution also derived.

The authors of \cite{SWIPT5} studied AN-aided multi-cell coordinated beamforming for SWIPT under a non-linear energy harvesting (EH) model. They formulated power-minimization problems to design transmit beamforming vectors and AN covariance matrices at all BSs, ensuring SINR and harvested energy thresholds for authorized users while limiting eavesdroppers' SINR. Key variables include beamforming vectors, power splitting factors between information decoding and EH, and AN covariance matrices. Under perfect CSI, the non-robust design was solved via SDR under imperfect CSI, including worst-case and statistical robust designs by S-Procedure and Bernstein-type inequality. Additionally, a distributed ADMM-based AN-aided multi-cell beamforming framework was proposed, enabling each BS to optimize using only local CSI, greatly reducing overhead compared to centralized designs.

To enhance secrecy robustness, \cite{SWIPT6} incorporated cooperative jamming into AN-SWIPT. They jointly optimized the transmit beamformer, AN and CJ covariance matrices, and power splitting ratio under imperfect CSI with norm-bounded uncertainty. The semi-infinite problem was tackled in two steps: reformulated into SDPs using the S-Procedure and Schur complement, then reduced to a single-variable optimization solved via one-dimensional line search.

In \cite{SWIPT7}, the authors analyzed secrecy outage in AN-aided cognitive radio SWIPT systems with randomly located non-colluding Eves. They proposed a MIMO CR-SWIPT system with AN over Rayleigh fading and selection combining at Bob. A closed-form SOP expression was derived using Gauss-Laguerre quadrature, and asymptotic analysis revealed the secrecy diversity order and array gain at high SNR.
 
Overall, due to the additional constraints brought by EH, the design of AN in SWIPT systems is more complex compared to conventional systems. Furthermore, in SWIPT systems, AN can simultaneously serve as an energy signal for EH devices, which is a feature not present in traditional AN systems.

\subsection{AN-aided THz}

Due to the enticing data rate brought by the abundant bandwidth resource, the potential of PLS of Terahertz (THz) communications needs in-depth analysis. Despite the narrow beam of THz communications, the PLS in the THz band is challenging when eavesdroppers are inside the beam radiation sector \cite{THZ1,THZ2}.

To enhance THz secure communications, \cite{THZ1} and \cite{THZ2} proposed a self-interference-cancellation (SIC)-free AN-assisted receiver scheme leveraging the distinct temporal broadening of AN at Eve. By introducing a time delay in Bob’s AN pulses, Eve experiences pulse overlap, degrading her reception. This eliminates the need for complex SIC at Bob. The authors further optimized system parameters via a DNN to maximize secrecy rate. Results showed the proposed method achieves 4 bps/Hz with lower hardware and computational complexity than SIC-based designs.

In summary, due to the narrow beamwidth in THz scenarios, traditional spatial AN is no longer suitable for interfering with Eves within the beam. Therefore, temporal broadening effect has become the new choice for AN in terahertz scenarios.

\subsection{AN-aided UAV}

Recently, unmanned aerial vehicle (UAV)-aided wireless communications have attracted significant research interest. UAVs offer unique attributes, including flexible deployment, dominant line-of-sight (LoS) and air-ground (AG) channels, and controlled mobility in three-dimensional (3D) space. Consequently, UAVs can serve as flying BSs, aerial radio access points, or aerial relays, establishing AG links to extend coverage, ensure seamless connectivity, and support high-rate communications. However, the open nature of AG links inevitably renders such systems vulnerable to eavesdropping attacks, and hence, safeguarding UAV communication is of significant importance \cite{UAV3}. 

Currently, the form of AN employed in UAV resembles a shared jamming signal at both Alice and Bob, which can also be seen as a pre-shared key in cryptography mechanisms and differs from space-domain AN \cite{UAV3,UAV2,UAV1}. In UAV-aided wireless networks, the joint optimization of power allocation and trajectory for AN-aided UAV has emerged as a new challenge \cite{UAV3,UAV2}. Furthermore, the joint optimization of resources, trajectory, and AN was explored in  \cite{UAV1} for a dual-UAV-based NOMA scenario with imperfect CSI.

In \cite{UAV3}, a secure UAV-based data dissemination system was proposed, where the UAV transmits both information and pre-known AN. To maximize the average minimum secrecy rate $\eta$, the user scheduling, power splitting, and UAV trajectory were jointly optimized. The resulting non-convex problem was tackled using an iterative algorithm based on alternating and successive convex optimization.

For UAV-aided wireless networks, a secure two-phase transmission protocol was introduced in \cite{UAV2}, where Alice shares AN with Bob beforehand for cancellation during data transmission. The UAV’s trajectory, transmit power, and AN power allocation were jointly optimized to enhance secrecy rate. To handle the non-convex problem, the authors decomposed it into four subproblems and solved it using a successive convex approximation-based iterative algorithm.

In \cite{UAV1}, joint optimization of resource allocation, trajectory, and AN was studied for a dual-UAV NOMA system under imperfect CSI. Probabilistic constraints were handled using the Markov inequality and Marcum Q-function, enabling a reformulation into deterministic constraints. The communication UAV’s trajectory was optimized using slack variables and Taylor expansion.

 In UAV secure transmission systems, it is important to jointly optimize the power allocation of AN, other resource allocation, and the UAV's trajectory. This typically requires complex optimisation methods, and finding a low-complexity algorithm remains a significant challenge.

\subsection{AN-aided VLC}

The visible light communication (VLC), primarily based on light-emitting diodes (LEDs) and photo diodes, has been regarded as one of the most promising wireless communication technologies due to its impressive data rate and unregulated wide bandwidth. 
By combining communication and illumination, VLC is well-suited for future indoor access scenarios, which, however requires security, particularly in the presence of potential wiretapping threats. To fulfill secure VLC communications, AN has been introduced into the MISO VLC wiretap channels to protect against a single Eve \cite{VLC3} or multiple Eves \cite{VLC5}. In addition, AN-aided precoding design for multi-user VLC channels was addressed in \cite{VLC4}, and this concept was extended to NOMA in \cite{VLC6}. Furthermore, it found application in indoor SISO direct-current-biased optical OFDM systems in \cite{VLC2}. In general, when adding the AN ${\mathbf{v}}$ to the VLC, an additional concern is the influence of the direct current bias ${I_D}$ in the transmit signal
\begin{equation}
{\mathbf{x}} = \sum\limits_{k = 1}^K {{{\mathbf{W}}_k}{{\mathbf{s}}_k}}  + {\mathbf{v}} + {I_D}{{\mathbf{1}}_N},
\end{equation}
where ${{\mathbf{W}}_k}$ and ${{\mathbf{s}}_k}$ denote the precoding matrix and transmit signal towards the $k$-th user.


Due to the unique nonlinear transfer characteristic of the LEDs, the power of transmit signal must be constrained within a certain range, which inspires new requirements regarding the power allocation of AN
\begin{equation}\label{delta:1}
\sum\limits_{k = 1}^K {{{\left\| {{{\mathbf{W}}_k}} \right\|}_1}}  + \rho {\left\| {\mathbf{V}} \right\|_1} \leqslant {\Delta _n},
\end{equation}
where ${\Delta _n} = \min \left( {I_n^{DC} - {I_{\min }},{I_{\max }} - I_n^{DC}} \right)$ is the direct current limitation with $\left[ {{I_{\min }},{I_{\max }}} \right]$ bounding the allowable range of current.

For the MISO VLC wiretap channel, the authors of \cite{VLC3} employed an AN-based beamforming approach to maximize an upper bound of secrecy rate. The upper bound of secrecy rate is expressed as
\begin{equation}
{R_u}\left( {{{\mathbf{h}}_E}} \right) = \frac{1}{2}\log \left[ {\frac{{1 + N{A^2}{{\left( {{\mathbf{h}}_B^T{\mathbf{e}}\left( {{{\mathbf{h}}_E}} \right)} \right)}^2}/\sigma _B^2}}{{1 + N{A^2}{{\left( {{\mathbf{h}}_B^T{\mathbf{e}}\left( {{{\mathbf{h}}_E}} \right)} \right)}^2}/\sigma _E^2}}} \right],
\end{equation}
where $N$ denotes the number of LED fixtures, $A$ represents the peak-power constraint for transmit signals, and ${\mathbf{e}}\left( {{{\mathbf{h}}_E}} \right)$ is the only one active orthonormal eigenvector of the matrix ${{\mathbf{h}}_B}{\mathbf{h}}_B^T/\sigma _B^2 - {{\mathbf{h}}_E}{\mathbf{h}}_E^T/\sigma _E^2$.

In \cite{VLC5}, the authors explored the design of PLS in VLC systems with multiple Eves. They employed AN-assisted precoding in scenarios where the CSI of Eves was both known and unknown. The primary objective of the design is to minimize the total transmission power, subject to specific constraints on the SINR for both Bob and Eves as
\begin{subequations}
\begin{align}
  &\mathop {\min }\limits_{{\mathbf{v}},{\mathbf{W}}} \left\| {\mathbf{W}} \right\|_2^2 + \left\| {\mathbf{v}} \right\|_2^2 \hfill \\
  &{\text{s}}{\text{.t}}{\text{.\; SINR}} \geqslant \gamma , \hfill \\
  &\;\;\;\;\;\;\left| {{{\left[ {\mathbf{v}} \right]}_n}} \right| + {\left\| {{{\left[ {\mathbf{W}} \right]}_{n,:}}} \right\|_1} \leqslant {\Delta _n}, \hfill 
\end{align}
\end{subequations}
where $n$ denotes the index of $n$-th Eve and ${\Delta _n}$ is defined in (\ref{delta:1}).
In the case of unknown Eve's CSI, the AN was strategically placed in the null space of Bob's channel, simplifying the design problem. When Eve's CSI is known, the problem became non-convex due to constraints imposed by the Eves' SINR. To address this, the authors investigated two different sub-optimal but low-complexity approaches: a concave-convex process and SDR.

The AN-aided precoding designs with respect to Bobs' and Eves' SINR performances for multi-user VLC wiretap channels were discussed in \cite{VLC4}. In the case of passive Eves, the AN was strategically designed to occupy the null space of the Bobs' aggregate channel matrix. When dealing with active Eves, the design objective is to limit the Eves' SINR to be below a certain threshold, thereby enhancing security. Aside from the general precoding design, the authors explored a specific design of ${\mathbf{W}}$ as
\begin{equation}
{\mathbf{W}} = \left( {{{\mathbf{H}}^\dag } + \left( {{\mathbf{I}} - {{\mathbf{H}}^\dag }{\mathbf{H}}} \right){\mathbf{R}}} \right){\text{diag}}\left\{{\sqrt {\mathbf{p}} } \right\},
\end{equation}
which employs the ZF technique as the fundamental precoding scheme for Bobs, effectively decoupling the multi-user channel into multiple subchannels. This decoupling facilitated confidentiality among Bobs and contributed to securing the communication in multi-user VLC scenarios.

For a MISO VLC system with NOMA, a robust AN-aided secure beamforming design was optimized in \cite{VLC6}, formulated by 
\begin{subequations}
\begin{align}
  &\mathop {\min }\limits_{{{\mathbf{W}}_k},{\mathbf{V}}} {\text{Tr}}\left( {\sum\limits_{k = 1}^K {{{\mathbf{W}}_k} + {\mathbf{V}}} } \right)\hfill \\
  &{\text{s}}{\text{.t}}{\text{.  }}R_{B,k}^L \geqslant {\gamma _b}, \hfill \\
  &\;\;\;\;\;\;R_{E,k}^U \leqslant {\gamma _E}, \hfill \\
  &\;\;\;\;\;\;{\text{Tr}}\left( {\sum\limits_{k = 1}^K {{{\mathbf{W}}_k}{{\mathbf{e}}_n}{\mathbf{e}}_n^T} } \right) \leqslant \Delta _n^2, \hfill \\
  &\;\;\;\;\;\;{{\mathbf{W}}_k}\underset{\raise0.3em\hbox{$\smash{\scriptscriptstyle-}$}}{\succ } {\mathbf{0}},{\mathbf{V}}\underset{\raise0.3em\hbox{$\smash{\scriptscriptstyle-}$}}{\succ } {\mathbf{0}}, \hfill 
\end{align}
\end{subequations}
where ${{\mathbf{W}}_k}$ and ${\mathbf{V}}$ denote the precoding matrix for $k$-th user and the AN matrix, respectively. ${{\mathbf{e}}_n}$ represents the $n$-th column of the identity matrix and ${\Delta _n}$ is defined in (\ref{delta:1}). The optimization leveraged SDP relaxation to minimize transmit power under QoS and Eve’s rate constraints. Simulations showed that the AN-aided beamforming improved security and efficiency, though CSI imperfections led to conservative resource use and limited performance gains.

In \cite{VLC2}, the authors focused on indoor SISO direct-current-biased optical OFDM systems. They proposed a precoding scheme using time-domain AN to exploit the DoF provided by the CP of OFDM, i.e.,
\begin{equation}
{{\mathbf{R}}^{cp}}{\mathbf{HV}} = {\mathbf{0}}.
\end{equation}
The authors proposed a convex optimization approach to limit PAPR while maximizing secrecy rates. Results showed improved secrecy and reduced PAPR. For further details on time-domain AN, refer to the AN-OFDM chapter.

In a nutshell, the use of AN in VLC needs to fully comply with the requirements of the LED constrained power range, and the influence of direct current bias should be paid attention to by researchers.

\subsection{AN-aided ISAC}

Integrated sensing and communications (ISAC), a key 6G feature, faces risks of information leakage from both communication and sensing. To address this, AN has been widely studied as a promising PLS technique for secure ISAC transmission.\cite{ISAC1,ISAC2_1,ISAC2_2,ISAC3}.

In \cite{ISAC1}, a secure beamforming scheme for dual-functional radar-communication (DFRC) systems was proposed, where radar beams act as AN to suppress eavesdropping. The system jointly optimizes communication and radar beamformers to maximize the sum secrecy rate while maintaining a desired radar beampattern and limiting Eve’s SINR. To solve the non-convex problem, a ZF-based strategy and SDR with eigenvalue decomposition were used. Additionally, a jamming power maximization problem was formulated to reduce complexity. Simulations confirmed that both SRM and JRM approaches enhance security without requiring Eve’s CSI.

Refs. \cite{ISAC2_1} and \cite{ISAC2_2} introduced AN in DFRC systems to enhance secure transmission. The transmitter, equipped with a uniform linear array, serves multiple Bobs and detects a single-antenna Eve. A joint design of transmit beamforming and AN was proposed to optimize the radar beampattern while minimizing Eve’s SNR and ensuring Bobs' SINR constraints. The problem was solved using SDR, Dinkelbach’s method, and quadratic transform, with extensions for angle uncertainty and imperfect CSI. Additionally, from a symbol-level precoding view, AN was shown to act as constructive interference for Bobs, achieving superior radar SINR.

In \cite{ISAC3}, the authors studied an ISAC system combining an MU-MISO downlink and a colocated MIMO radar. The BS with a ULA serves multiple single-antenna users in the presence of eavesdroppers, while the radar detects a target. A joint design of BS and radar beamforming, along with AN, was proposed to minimize Eve's SINR while meeting communication and radar SINR constraints. The optimization was solved via BCD, fractional programming, and SDR. AN was also incorporated from both BS and radar to further degrade Eve's reception. Simulations confirmed that the proposed schemes significantly reduced Eve’s SINR versus baselines.

In a nutshell, AN-aided ISAC must balance tradeoffs among communication throughput, sensing accuracy, and AN design complexity. Future work will explore adaptive algorithms, ML-enhanced signal processing, energy-efficient designs, and application-specific, secure solutions.

\section{Overview of AN-combined Technologies}

\begin{table*}[!htp]
\begin{center}
\center
\caption{Overview of AN-combined Technologies}
  \begin{tabular}{cccccccc}
    \toprule[1.5pt]
    \makebox[0.00\textwidth][c]{References} & \makebox[0.00\textwidth][c]{Technologies} & \makebox[0.00\textwidth][c]{AN Types}
                                     & \makebox[0.00\textwidth][c]{System Models} & \makebox[0.00\textwidth][c]{Bob types} &\makebox[0.00\textwidth][c]{Eve types} & \makebox[0.00\textwidth][c]{CSI of Bob/ Eve}   & \makebox[0.00\textwidth][c]{Involved Contributions}           \\
    \midrule[1pt]
    \cite{code4}&STBC&AN&MIMOME& MMA & SMA & Imperfect/ No & Aligned AN symbols\\
    \cite{code2}&STLC&AN&MIMOME& SMA & SMA & Perfect/ No & Lower bound of SSR\\
    \cite{code3}&RLC&Jammer AN&MISOME& SSA & SMA & Perfect/ No & TAS for reducing QVP\\
    \cite{code1}&Polar codes&AN&SISOSE& SSA & SSA & Perfect/ Perfect & Jamming positions selecting\\
    \cite{covert2}&Covert&AN&SISOSE& SSA & MSA & Statistical/ Statistical &  AN from the closest friendly node\\
    \cite{covert3}&Covert&AN&SISOSE& SSA & MSA & Statistical/ Statistical &  Increased covert transmission bits \\
    \cite{covert5}&Covert&Jammer AN&SISOSE& SSA & SSA & Statistical/ Statistical &  AN from a friendly jammer\\
    \cite{covert1}&Covert&AN&MISOSE& MSA & MSA & Perfect/ Statistical &  AN for D2D cellular network\\
    \cite{covert4}&Covert&AN&MISOSE& MSA & MSA & Statistical/ Statistical & Stackelberg game formulation for IoT\\
    \cite{DM1}&DM&Analog AN&MISOSE&MSA&SSA&Perfect/No&Analog AN far-field beampattern design\\
    \cite{DM2,DM3}&DM&Analog AN&MISOSE&MSA&SSA&Perfect/No&Simplified AN design for dynamic DM\\
    \cite{DM4}&DM&AN&MISOSE&MSA&SSA&Perfect/No&GPI-based precoding for SRM\\
    \cite{DM5}&DM&AN&MISOME&MSA&SSA&Either/Either&Robust multi-beam for broadcasting\\
    \cite{DM6}&DM&AN&MISOME&MSA&SSA&Either/Either&MLI-based SINR maximization for MU\\
    \cite{FH3}&FH&AN&SISOSE& SSA & SSA & No/ No &  Pre-stored AN for FH\\
    \cite{FH1}&FH&AN&SISOSE& SSA & MSA & No/ No &  Storage AN for IQ balances\\
    \cite{FH2}&FH&AN&SISOSE& SSA & MSA & No/ No &  Secrecy analysis and power allocation\\
    \cite{MISO2}&MISO&AN&MISOSE& SSA & SSA & Perfect/ No & Channel correlation \\
    \cite{MISO11}&MISO&Robust AN&MISOSE& SSA & SSA & Partial/ Partial & Imperfect CSI \\
    \cite{MISO13}&MISO&Generalized AN&MISOSE& SSA & SSA & Perfect/ Statistical & SRM \\
    \cite{MISO3}&MISO& AN&MISOSE& SSA & SSA & Perfect/ Statistical & On-off and adaptive schemes \\
    \cite{MISO8}&MISO& AN&MISOME& SSA & SMA & Perfect/ Statistical & SOP minimization \\
    \cite{MISO5}&MISO&Generalized AN&MISOME& SSA & SMA & Perfect/ Statistical & SRM \\
    \cite{MISO6}&MISO& AN&MISOME& MSA & SMA & Perfect/ No & Three AN schemes for CRNs\\
    \cite{MISO10}&MISO& AN&MISOME& SSA & SMA & Perfect/ Statistical & On-off and adaptive schemes \\
    \cite{MISO14}&MISO& AN-AFF&MISOME& SSA & SMA & Perfect/ No & Hybrid AN-AFF scheme\\
    \cite{MISO1}&MISO&Robust AN&MISOSE& SSA & MSA & Perfect/ Imperfect & SRM\\
    \cite{MISO12}&MISO&Generalized AN&MISOSE& SSA & MSA & Perfect/ Imperfect & A safe convex approximation\\
    \cite{MISO7}&MISO&AN&MISOSE& SSA & MSA & Perfect/ No & SOP-based security region\\
    \cite{MISO9}&MISO&Generalized AN&MISOSE& SSA & MSA & Perfect/ Statistical & Semiadaptive scheme for STM\\
    \cite{MISO15}&MISO&Generalized AN&MISOME& SSA & MMA & Perfect/ Either & SRM \\
    \cite{MIMO1}&mMIMO&AN&MISOME& MSA & SMA & Either/ Either & SRM under different CSI accessibility\\
    \cite{MIMO2}&mMIMO&AN&MISOME& MSA & SMA & Imperfect/ No & Three AN precoders and analysis\\
    \cite{MIMO4}&mMIMO&AN&MISOME& MSA & SMA & Imperfect/ No & Two AN schemes in two phases\\
    \cite{MIMO5}&mMIMO&AN&MISOME& MSA & SMA & Statistical/ No & Low-complexity AN over Ricean channel\\
    \cite{MIMO6}&mMIMO&AN&MISOSE& MSA & SSA & Imperfect/ No &  SSR maximization for NOMA\\
    \cite{NOMA1}&NOMA&Two-phase AN&SIMOSE& SMA/SSA & SSA & Perfect/ No &  Two-phase AN and relay selection\\
    \cite{NOMA3}&NOMA&Bob AN&SISOSE& MSA & SSA & No/ No &  Pseudo random AN from FD Bobs\\
    \cite{NOMA2}&NOMA&AN&MISOSE& MSA & SSA & Perfect/ Perfect &  SRM via AO for RIS-NOMA\\
    \cite{NOMA4}&NOMA&AN&MISOME& MSA & SMA & Perfect/ No &  Closed-form SOP for MISO-NOMA\\
    \cite{OFDM8}&OFDM&Temporal AN&SISOSE& SSA & SSA & Perfect/ Either &  Temporal AN for OFDM\\
    \cite{OFDM7}&OFDM&Time-domain AN&SISOSE& SSA & SSA & Perfect/ Perfect &  Time-domain AN with discrete inputs\\
    \cite{OFDM6}&OFDM&Time-domain AN&MISOME& MSA & SMA & Perfect/ Perfect &  Time-domain AN for multiuser\\
    \cite{OFDM1}&OFDM&Time-domain AN&SISOSE& SSA & SSA & Perfect/ Perfect &  Time-domain AN for AF relay\\
    \cite{OFDM5}&OFDM&Hybrid AN&MIMOME& SMA & SMA & Perfect/ No &  Closed-form expressions for secrecy rate\\
    \cite{OFDM4}&OFDM&Hybrid AN&MIMOME& SMA & MMA & Perfect/ Perfect &  Hybrid spatial and temporal AN\\
    \cite{OFDM3}&OFDM&Frequency AN&SISOSE& MSA & MSA & Perfect/ Perfect &  Frequency-domain AN for SWIPT\\
    \cite{OFDM2}&OFDM&Bob AN&SISOSE& SSA & MSA & Perfect/ No &  Hybrid parallel power-line/wireless OFDM\\
    \cite{Au1}&PLA&AN&MIMOME& SMA & SMA & Imperfect/ No & AN fingerprint embedding authentication\\
    \cite{Au2}&PLA&AN&SISOSE& SSA & SSA & No/ No & AN-aided message authentication codes\\
    \cite{Au3}&PLA&Tikhonov AN&SISOSE& SSA & SSA & Perfect/ No & OFDM challenge-response authentication\\
    \cite{Au4}&PLA&Two frame AN&SISOSE& SSA & SSA & Perfect/ No & Authentication over time-variant channels\\
    \cite{RIS4}&RIS&AN&MISOSE& SSA & MSA & Perfect/ Perfect &  Performance validation of AN-RIS\\
    \cite{RIS1}&RIS&AN&MIMOME& SMA/MMA & SMA & Perfect/ Perfect &  BCD for RIS-MIMO\\
    \cite{RIS2}&IOS&AN&MIMOME& SMA & SMA & Perfect/ Perfect &  Complexity reduction for refractive RIS\\
    \cite{RIS3}&RIS&AN&MISOSE& MSA & MSA & Perfect/ Perfect &  RIS and AN assisted MEC systems\\
    \cite{RIS6}&RIS&AN&MISOSE& SSA & SSA & Perfect/ No &  Efficient OM and MM algorithms\\
    \cite{RIS5}&RIS&AN&MISOSE& SSA & SSA & Perfect/ No &  Computational complexity reduction\\
    \cite{SISO2}&SISO&Time-domain AN&Multi-path& SSA & SSA& Perfect/ No &Inter-symbol interference compensation\\
    \cite{SISO4}&SISO&Adaptive AN&SISOSE-ARQ& SSA & SSA & Perfect/ No &PAPR and out-of-band emission reduction \\
    \cite{SISO3}&SISO&Time-domain AN&TDD SISOSE& SSA & SSA & Perfect/ No & Repetition coding in two time slots \\
    \cite{SISO1}&SISO&Two-phase AN&HD SISOSE& SSA & SSA & No/ No &Broadcasting and forwarding of AN\\
    \cite{SISO5}&SISO&Reference-free AN&SISOSE& SSA & SSA & No/ No & AN reconstruction from received signals \\
    \cite{SM1}&SM&AN&MISOSE& SSA & SSA & Perfect/ No &  Single RF chain\\
    \cite{SM2}&SM&AN&MISOSE& SSA & SSA & Imperfect/ Imperfect &  Closed-form expression of ESR\\
    \cite{SM3}&DQSM&Jammer AN&MISOSE& SSA & SSA & Perfect/ Either &  AN from a CJ\\
    \cite{SM4}&SM&AN&MISOSE& SSA & SSA & Perfect/ No &  Antenna selection for two RF chains\\
    \cite{SM6}&GSM&AN&MIMOME& SMA & SMA & Perfect/ No &  PMAN and BER analysis\\
    \cite{SM7}&GSM&AN&MIMOME& MMA & SMA & Perfect/ No &  Multi-user GSM with AN and BD\\
    \bottomrule[1.5pt]
  \end{tabular}
\end{center}
\end{table*}

In this section, an overview of the AN-combined technologies is provided and the technical contributions of related papers are summarized in Table III.

\subsection{AN-aided Coding}

Over the past few decades, several coding technologies have been extensively explored to improve communication quality. The property of channel characteristic enables them to combine with AN for PLS, which includes but is not limited to the AN-aided space time block code (STBC) \cite{code4}, space time line code (STLC) \cite{code2}, rateless codes (RLC) \cite{code3}, and polar code \cite{code1}.

In \cite{code4}, a secure STBC scheme was presented for MIMO systems when Alice has the CSI of Bob but without that of Eve. To bolster security, the authors introduced AN symbols ${v_1}$ and ${v_2}$ at the first slot, which are carefully aligned with each other, yielding
\begin{equation}
\begin{gathered}
  {v_1} = \eta \left[ {{{\bar h}_1}\left( {c_1^ *  - c_2^ * } \right) + {{\bar h}_2}\left( {c_3^ *  - c_4^ * } \right)} \right], \hfill \\
  {v_2} = \eta \left[ {{{\bar h}_3}\left( {c_1^ *  - c_2^ * } \right) + {{\bar h}_4}\left( {c_3^ *  - c_4^ * } \right)} \right], \hfill \\ 
\end{gathered} 
\end{equation}
where ${{{\bar h}_1}}$, ${{{\bar h}_2}}$, ${{{\bar h}_3}}$, and ${{{\bar h}_4}}$ are equivalent parameters composed of CSI, while $\eta$ is a defined power constraint parameter.

In \cite{code2}, the STLC with AN was proposed for enhancing the PLS in a time division duplex (TDD) mode. Note that the AN symbols are injected in two time slots, i.e.,
\begin{equation}
\begin{gathered}
  \left[ {\begin{array}{*{20}{c}}
  {v_{1,1}^ * } \\ 
  {v_{2,1}^ * } 
\end{array}} \right] = \frac{{\left\| {{{\mathbf{h}}_2}} \right\|}}{{\left\| {{{\mathbf{h}}_1}} \right\|}}\left[ {\begin{array}{*{20}{c}}
  {{h_{1,1}}}&{h_{2,1}^ * } \\ 
  {{h_{2,1}}}&{- h_{1,1}^ * } 
\end{array}} \right]{\mathbf{r}}, \hfill \\
  \left[ {\begin{array}{*{20}{c}}
  {v_{1,2}^ * } \\ 
  {v_{2,2}^ * } 
\end{array}} \right] = \frac{{- \left\| {{{\mathbf{h}}_1}} \right\|}}{{\left\| {{{\mathbf{h}}_2}} \right\|}}\left[ {\begin{array}{*{20}{c}}
  {{h_{1,2}}}&{h_{2,2}^ * } \\ 
  {{h_{2,2}}}&{- h_{1,2}^ * } 
\end{array}} \right]{\mathbf{r}}. \hfill \\ 
\end{gathered} 
\end{equation}
Moreover, the authors derived the lower bound for the SSR under various scenarios: without AN, with AN, and with AN in the presence of imperfect CSI.

Ref. \cite{code3} investigated a secure transmission scheme using RLC in a delay-constrained system. By using RLC, secrecy is achieved if Bob can accumulate the required number of packets before Eve does. To achieve this, the authors employed transmit antenna selection (TAS) at Alice to improve the main channel quality as
\begin{equation}
{h_S} = \mathop {\max }\limits_{k \in \left\{{1,2, \ldots ,{N_a}} \right\}} \left| {{h_k}} \right|,
\end{equation}
and the AN based on the null space was generated to degrade Eve's channel.

Ref. \cite{code1} proposed an AN-aided polar coding algorithm to improve the secrecy rate. Specifically, a secure coding model based on AN was presented, where the AN is pre-shared by transceivers and injected into the current codeword
\begin{equation}
{\mathbf{x}} = {\mathbf{s}} \oplus {\mathbf{v}},
\end{equation}
where $ \oplus $ is the module-2 sum. With the knowledge of AN ${\mathbf{v}}$, Bob can mitigate its impact and detecting codewords.

In general, the combination of AN and coding is practical and feasible. However, the difficulty that researchers need to show solicitude for is how to utilize the characteristics of different coding methods.

\subsection{AN-aided Covert Communications}

\begin{figure}[t]
\centering
\includegraphics[width=3.5in,height=4.5in]{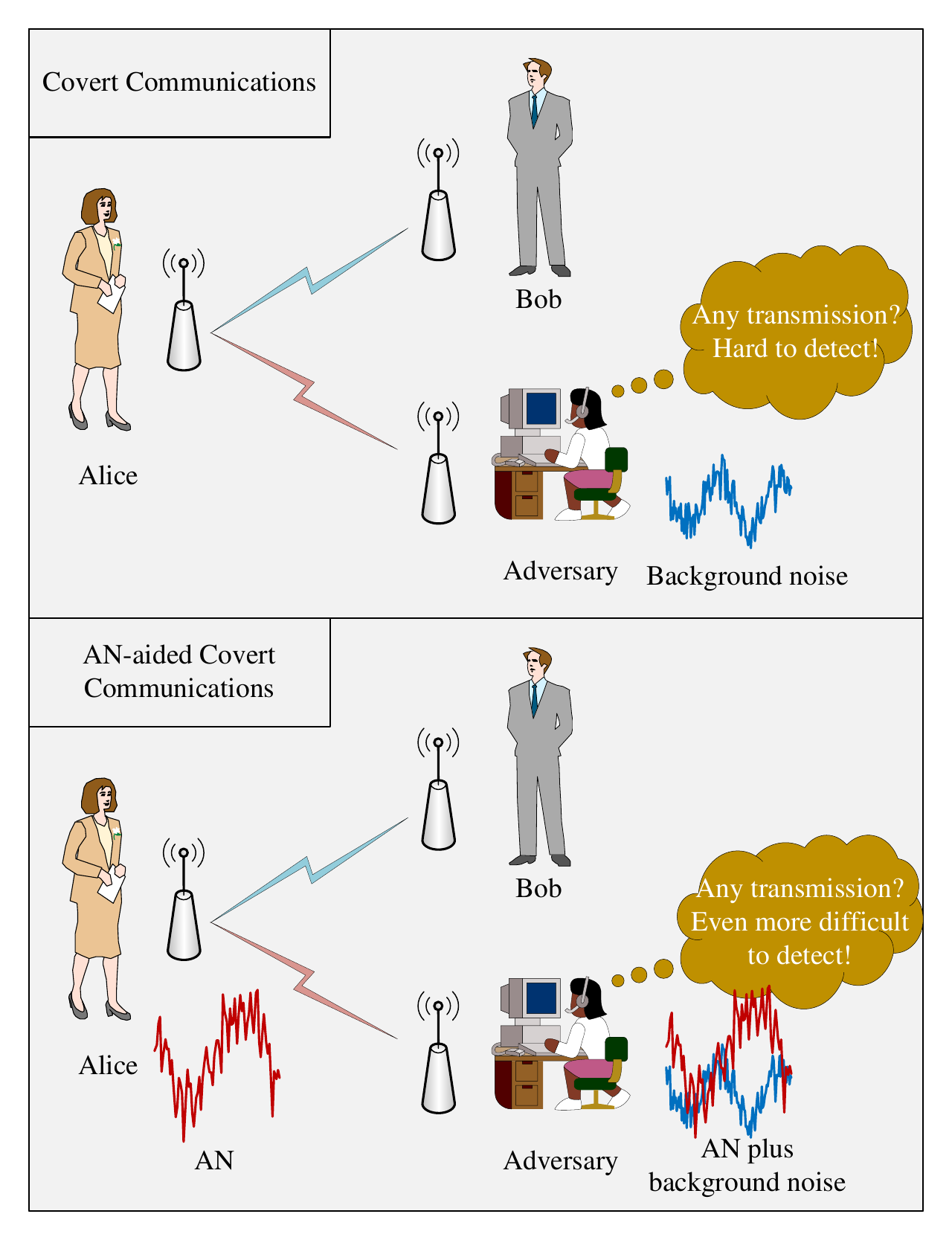}
\caption{The effect of AN in covert communications.}
\label{fig_covert}
\vspace{-0em}
\end{figure}

Due to its ability to create interference towards the adversary, the AN technology can be considered as an implementation method for covert communications, which is also known as low probability of detection (LPD) communications. \cite{covert2,covert3,covert5,covert1,covert4}. As depicted in Fig. \ref{fig_covert}, the main concept of covert communications is to confuse the judgment of the adversary, making it hard to determine whether Alice and Bob are communicating. With the aid of AN, this goal can be achieved more easily, making it more difficult for the adversary to detect the conduct of communications.

The authors of \cite{covert2} explored network scenarios with friendly nodes generating AN to enable covert communication, suggesting that the node nearest the adversary should do so to satisfy the error probability constraint
\begin{equation}
{\mathbb{P}_{FA}} + {\mathbb{P}_{MD}} \geqslant 1 - \varepsilon ,\forall \varepsilon,
\end{equation}
where ${\mathbb{P}_{FA}}$ and ${\mathbb{P}_{MD}}$ denote false alarm and 
miss detection probabilities, respectively, and $\varepsilon$ is small positive constant indicating the maximum deviation from perfect covertness.

The authors of \cite{covert3} focused on the scenario where other friendly nodes were distributed according to a 2D Poisson point process with a specific density. They proposed an AN generation strategy from the friendly node closest to Eve, and demonstrated that compared to the conventional work that can only transmit $\mathcal{O}\left( {\sqrt n } \right)$ bits in $n$ channel uses, the method in \cite{covert3} allows Alice to reliably send increased $\mathcal{O}\left( {\min \left\{{n,{m^{\gamma /2}}\sqrt n } \right\}} \right)$ bits to Bob.

The authors of \cite{covert5} studied a wireless covert communication system, where a friendly jammer generates intermittent AN to confuse Eve. The transmission probabilities of information and AN were jointly optimized to maximize the communication covertness under a throughput requirement. Then, closed-form optimal solutions are obtained by simplifying the original problem to a one-dimensional problem.

To confuse Eve in a D2D underlying cellular network, the AN-aided covert signals were used at Alice in \cite{covert1}. To evaluate performance, the achievable D2D communication rate is defined as the product of the covert signal desirable communication rate ${R_d}$, the cellular link connection probability $1 - {\mathbb{P}_{co}}$, and the D2D link connection probability $1 - {\mathbb{P}_{do}}$. By maximizing the achievable D2D communication rate under the constraints of error probabilities (including false alarm and mis-detection probability), the optimization problem was formulated as 
\begin{subequations}
\begin{align}
  &\mathop {\max }\limits_{{P_t}} {R_d}\left( {1 - {\mathbb{P}_{co}}} \right)\left( {1 - {\mathbb{P}_{do}}} \right) \hfill \\
  &\;\;{\text{s}}{\text{.t}}{\text{.  }} \mathbb{P}_{FA}^k + \mathbb{P}_{MD}^k \geqslant 1 - \varepsilon . \hfill
\end{align}
\end{subequations}

By applying uplink power control and AN scheme, a covert IoT system was studied in \cite{covert4}. Specifically, the AN is transmitted by in-band FD IoT gateways as a jamming operation to intends to hide the legitimate transmission from the observant Eves. Moreover, a Stackelberg game was formulated to study the interaction between the Eves and Bobs as
\begin{subequations}
\begin{align}
  &\mathop {\max }\limits_{p,q} \mathbb{P}\left( {{\text{SIN}}{{\text{R}}_b} \geqslant {\gamma _b}\left| {{\mathcal{L}_1}} \right.} \right){u_b} \hfill \\
   &  \;\;{\text{s}}{\text{.t}}{\text{. }} - \mathbb{P}\left( {{\text{SIN}}{{\text{R}}_e} \geqslant {\gamma _e}\left| {{\mathcal{L}_1}} \right.} \right){u_e} - {v_D}p \hfill \\
    &\;\;\;\;\;\;\;\;\;\mathbb{P}_{FA}^k + \mathbb{P}_{MD}^k \geqslant 1 - \varepsilon , \hfill \\
    &\;\;\;\;\;\;\;\;\;{p^L} \leqslant p \leqslant {p^U},{q^L} \leqslant q \leqslant {q^U}, \hfill
\end{align}
\end{subequations}
where $p$ and $q$ are the power of information-bearing signal and AN bounded by $\left[ {{p^L},{p^U}} \right]$ and $\left[ {{q^L},{q^U}} \right]$, respectively, ${u_b}$ and ${u_b}$ are the reward of guaranteeing the transmission reliability, while ${v_D}p$ is the power cost for the IoT device.

Objectively speaking, the AN significantly enhances the performance of covert communications. However, the main concern that researchers need to be aware of is that the calculation of error probabilities for the adversary, i.e., ${\mathbb{P}_{FA}}$ and ${\mathbb{P}_{MD}}$, includes the knowledge of its CSI, which may contradict the fact that Alice and adversary cannot collaborate to perform channel estimation in the practical scenarios.

\subsection{AN-aided DM}

\begin{figure}[t]
\centering
\includegraphics[width=3.5in,height=2.4in]{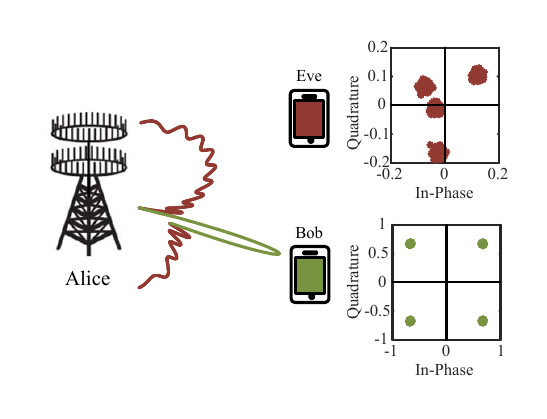}
\caption{Framework of the AN-aided DM wireless secure communications.}
\label{fig_7}
\vspace{-0em}
\end{figure}

As shown in Fig. \ref{fig_7}, directional modulation (DM) is capable of conducting beamforming towards the legitimate user while projecting AN to eavesdroppers in other directions, resulting in a standard constellation appearing only in the desired direction of legitimate users. As a multi-antenna-based PLS technique, DM extensively incorporates the notion of AN \cite{DMSurvey,DM1,DM2,DM3,DM4,DM5,DM6}.

The earliest utilization of AN in DM was proposed in \cite{DM1}, where the interference pattern for introducing AN was designed separately from information patterns by the far-field pattern null steering method. Specifically, the information beampattern towards the direction of Bob was first constructed with the array excitation $\mathbf{w}$, after which the orthogonal null-space interference beampattern towards the $i$-th sidelobe $\mathbf{v}_i$ was constructed using an orthogonal projection matrix of the array response of Bob $\mathbf{h}$. Therefore, the transmission can be synthesized as
\begin{equation}
{{\bf{x}}_m} = {\bf{w}}{s_m} + \sum\limits_{n = 1}^{N - 1} {\left( {{{\bf{v}}_i}{r_i}} \right)},
\end{equation}
where the interference beampattern is given by
\begin{equation}
{{\bf{v}}_i} = \left[ {{{\bf{I}}_N} - {{\bf{h}}^{- 1}}{\bf{h}}} \right]{\bf{w}}_i^{{\rm{sl}}},
\end{equation}
with ${\bf{w}}_i^{{\rm{sl}}}$ being the excitation of beampattern towards the $i$th sidelobe. Although this method was operated at the analog RF level, the main idea is consistent with the digital AN.

\begin{figure}[t]
\centering
\includegraphics[scale = 0.7]{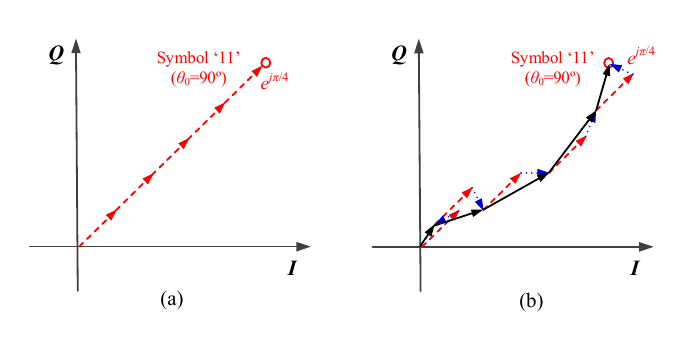}
\caption{Vector paths along the boresight for one QPSK symbol: (a) without AN; (b) with AN \cite{DM1}.}
\label{fig_DM2}
\end{figure}

This approach underwent refinement in \cite{DM2} and \cite{DM3}, where an innovative orthogonal vector representation for AN was employed for the so-called dynamic DM. The authors intuitively explained the diversity of AN through the geometric process of vector synthesis as shown in Fig. \ref{fig_DM2}. 

It can be observed that the different vector paths form the same constellation points in the boresight direction, which means that these vector paths have the same transmission effect for the legitimate receiver. On the contrary, for eavesdroppers in other directions, these different vector paths have different degrees of influence on the transmission effect.

To maximize the secrecy rate in a single-user MISO-DM system, the authors of \cite{DM4} proposed a general power iterative (GPI)-based DM precoding, which optimizes the beamforming vector and AN projection matrix iteratively in an alternating manner. With the given AN projection matrix $\mathbf{V}$, the beamforming optimization can be cast as
\begin{subequations}\label{SUDMOpt}
\begin{align}
&\mathop {\max }\limits_{\mathbf{w}}\frac{{{{\mathbf{w}}^{\text{H}}}\left( {{\mathbf{H}} + {A_{\text{B}}}{{\mathbf{I}}_N}} \right){\mathbf{w}}}}{{{{\mathbf{w}}^{\text{H}}}\left( {{\mathbf{G}} + {A_{\text{E}}}{{\mathbf{I}}_N}} \right){\mathbf{w}}}}\\
&\;\;{\rm{s}}{\rm{.t}}{\rm{.}}{{\bf{w}}^{\rm{H}}}{\bf{w}} = 1,
\end{align}
\end{subequations}
where $\mathbf{w}$ denotes the beamforming vector, while $\mathbf{H} = \mathbf{h}\mathbf{h}^{\text H}$ and $\mathbf{G} = \mathbf{g}\mathbf{g}^{\text H}$ represent the channel covariance matrices for Bob and Eve, respectively.

For broadcasting MU-MISO systems, a robust multi-beam DM synthesis was explored in \cite{DM5}. The beamforming vector was obtained by maximizing signal-to-leakage-noise ratio.

For performance improvement, the authors of \cite{DM6} expanded the algorithm in \cite{DM5} to an MU-MISO scenario. Specifically, a main-lobe-integration (MLI)-based SINR maximization problem was cast and solved by the generalized Rayleigh-Ritz theorem to obtain the beamforming vector for the MU-MISO-DM system in both perfect and imperfect knowledge of desired directions scenarios. Recently, an analog DM precoding was proposed in \cite{DMNF} for near-field secure transmission in both angle and distance dimensions, significantly improving the secure capbility of AN.

From the perspective of maximizing the secrecy rate, AN has been applied comprehensively in DM systems, including orthogonal and non-orthogonal AN. However, these methods tend to require high computational complexity, which may not be suitable for resource-limited DM systems. Additionally, the current application of AN is limited to projection in the angular domain, restricting DM to secure transmission solely in the directional dimension. Expanding AN to other domains is crucial for enhancing the security capabilities of DM.

\subsection{AN-aided FH}

Through changing the carrier frequencies, the frequency hopping (FH) technique has surged to avoid electromagnetic interference. To address both electromagnetic interference and wiretapping, one appealing research idea is to jointly utilize FH and AN techniques to provide rigorous security. Song et al. discovered that the wide hopping bandwidth will raise a non-negligible in-phase and quadrature (IQ) imbalance at transceiver oscillators, yielding significant signal distortions during IQ mixing \cite{FH1,FH2,FH3}.

To further protect FH systems against severe wiretapping, an architecture of AN-shielded FH systems was proposed in \cite{FH3}, wherein AN cancellation architecture is utilized at Bob in the presence of Eve. 
Specifically, AN reconstruction and cancellation were employed to depict the residual AN, where the time delay and channel fading were supposed to be perfectly estimated. In the focus of the impact of frequency deviation, the estimated normalized frequency offset was assumed to be ${\hat F_{{d_r}}} = {F_{{d_r}}} - \Delta {F_{{d_r}}}$, where $\Delta {F_{{d_r}}}$ denotes the frequency deviation after synchronization. Under this estimation, the AN can be reconstructed as
\begin{equation}
    \hat{c}_{\mathrm{ref}}(n)=\tilde{h}_{r}c(n-D_{r})\mathrm{e}^{\mathrm{j}2\pi n\hat{F}_{d_{r}}}g(n-D_{r}-iL),
\end{equation}
where $h_r$, $c(n)$, $D_r$, and $g(n)$ denote the complex channel gain, AN signal, normalized propagation delay, and shaping signal. Theoretical and simulation results confirmed that frequency deviation in practical FH systems will degrade both AN cancellation and system secrecy performance, and shorter hop length or proper power allocation for secret data and AN can mitigate the negative impact of frequency deviation, which in turn guides the practical transceiver design.

To alleviate signal distortion from IQ imbalances, the authors modeled signal distortions for AN-shielded FH systems \cite{FH1}. In order to measure the quality of the obtained signal after AN cancellation, the signal-to-distortion-plus-noise ratio for Bob’s residual signal and $n$-th Eve's composite signal are presented. Subsequently, the secrecy capacity was derived to evaluate system secrecy under multiple Eves as
\begin{equation}
C_{s}= 1/2\log_{2}\left(1+\Lambda_{r}\right) - \max_{n}\left\{1/2\log_{2}\left(1+\Lambda_{e}^{n}\right)\right\},
\label{SRFH}
\end{equation}
where $\Lambda_{r}$ and $\Lambda_{e}^{n}$ are the signal to AN
plus noise ratio at Bob and $n$-th Eve.

To protect military broadcasting against both hostile interference and eavesdropping, an AN-sheltered FH broadcasting architecture was proposed in \cite{FH2}, and the secrecy capacity in (\ref{SRFH}) was employed to measure its secrecy performance. Besides, the optimal transmitting power allocation for the information-bearing signal and AN was provided in a closed-form to maximize secrecy capacity in the presence of frequency mismatch. Specificialy, the power allocation problem was first cast as
\begin{subequations}
    \begin{align}\max_{\alpha}&\;C_{s}=\frac{1}{2}\log_{2}\frac{(a\alpha+b)(c\alpha+1)}{(a\alpha+1)(c\alpha+c)}\\
    &\mathrm{s.t.}\;\gamma_{r}>\gamma_{e}\\
    &\;\;\;\;\;\;\;\alpha\geq 0,
    \end{align}
\end{subequations}
where $\alpha$ denotes the power allocation factor between the confidential information signal and AN. $a$, $b$, $c$, and $d$ are four constant variables with respect to channel, frequency mismatch, and normalized power budget. The closed-form optimal solution for this problem is given by 
\begin{equation}
    \alpha^*=\left\{\begin{array}{ll}\alpha_1,&\text{if}\;ac+c<a+bc\; , \;ab+b\leq a+bc,\\0,&\text{if}\;a+bc<ab+b\; , \;b>c,\\\phi,&\text{if}\;ac+c\geq a+bc\; , \;b\leq c,\end{array}\right.
\end{equation}
where $\alpha_1$ is expressed as
\begin{equation}
    \alpha_1=\frac{a(b-c)-\sqrt{a(b-1)(c-1)(a-b)(a-c)}}{a(ac+c-a-bc)}.
\end{equation}

Conclusively, AN-aided FH systems can avoid electromagnetic interference and guarantee security simultaneously. However, imperfect analog components are major challenges in such systems since the legitimate receivers need to perform FH synchronization, AN reconstruction and cancellation operations. This process is sensitive to analog component issues, such as IQ imbalance, power amplifier nonlinearity, phase noise, and frequency deviation. Considering these issues, the performance analysis and transmission design of AN-aided FH systems will be more accurate and meaningful.

\subsection{AN-aided MISO}

\begin{figure}[t]
\centering
\includegraphics[width=3.5in,height=2.4in]{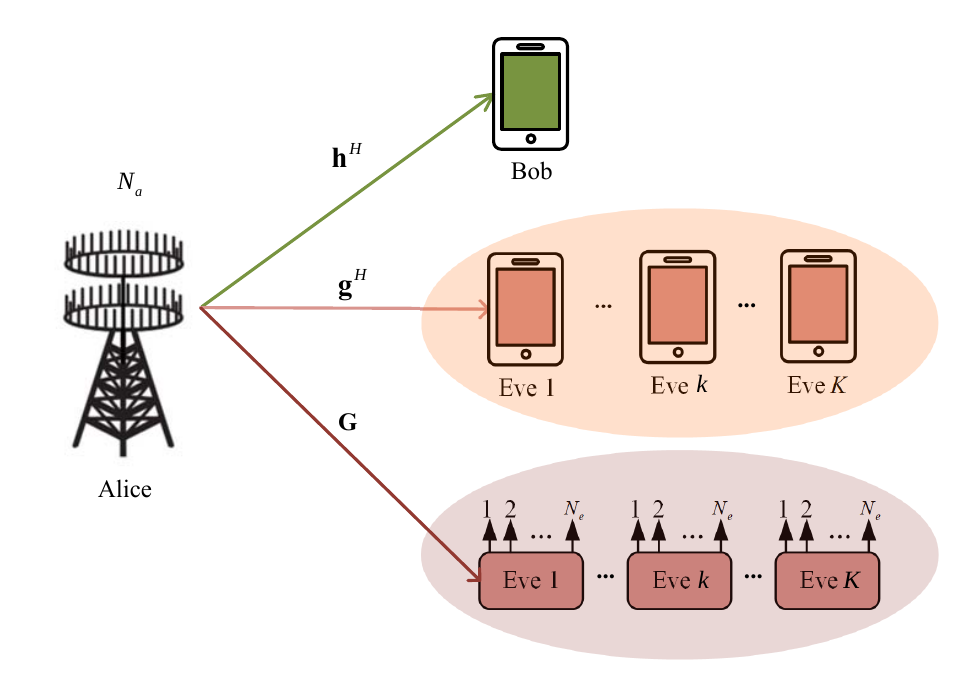}
\vspace{+3mm}
\caption{Framework of the AN-aided MISO wireless secure communications.}
\label{fig_MISO}
\vspace{-0em}
\end{figure}

Since the null space of the channel involved in the AN technology exists under the condition that the number of transmit antennas exceeds the number of receive antennas, the MISO channel naturally aligns with this condition, making them compatible. As depicted in Fig. \ref{fig_MISO}, the framework of AN-aided MISO wireless communications consists of a multiple-antenna Alice and a single-antenna Bob, which communicate in the presence of single or multiple Eves. For a fair comparison, most of the papers assume SSA Eve \cite{MISO2,MISO11,MISO13,MISO3}. In order to achieve a stronger secrecy performance, some papers focused on the scenarios involving single multiple-antenna (SMA) Eve \cite{MISO8,MISO5,MISO6,MISO10,MISO14}, MSA Eves \cite{MISO1,MISO12,MISO7,MISO9}, or MMA Eves \cite{MISO15}.

For the MISOSE system, the authors of \cite{MISO2} studied how the correlation of the wiretap channel affects the secrecy rate of the AN scheme. Specifically, a simple but tight lower bound on the ESR of the AN scheme was developed as
\begin{equation}
\begin{aligned}
R_{s}& \geq \log\left(1+\phi P(N_{A}-1)\right) \\
&-\left[\log\left(\sigma_e^2+\rho\phi PN_A+(1-\rho)P\right)\right] \\
&-\log\left(\sigma_e^2+(1-\rho)(1-\phi)P\frac{N_A-2}{N_A-1}\right)\bigg].
\end{aligned}
\end{equation}
Utilizing the derived bound, the influence of correlation on the secrecy rate was investigated.

Under two uncertainty models that Alice only has the imperfect CSI of Bob and Eve, the authors of \cite{MISO11} designed a robust AN for MISOSE. For the deterministic uncertainty model, the channel errors are assumed to be bounded by
\begin{equation}\label{Uncertain:1}
\begin{aligned}
\left\| {{\mathbf{\Delta h}}} \right\|\leq{\epsilon_b},\left\| {{\mathbf{\Delta g}}} \right\|\leq{\epsilon_e}.
\end{aligned}
\end{equation}
For the stochastic uncertainty model, the channel errors are zero-mean Gaussian random variables
\begin{equation}
\begin{aligned}
{\mathbf{\Delta h}} \sim \mathcal{C}\mathcal{N}\left( {{\mathbf{0}},\sigma _b^2{\mathbf{I}}} \right),{\mathbf{\Delta g}} \sim \mathcal{C}\mathcal{N}\left( {{\mathbf{0}},\sigma _e^2{\mathbf{I}}} \right).
\end{aligned}
\end{equation}

Considering the AN not constrained to be orthogonal to the information-bearing signal, the authors of \cite{MISO13} proposed a generalized AN scheme for MISOSE. To construct the AN-aided transmit signal, Alice generates an orthogonal basis $\mathbf{U}=[\mathbf{u}_{1},\mathbf{U}_{\mathrm{c}}]$,
where $\mathbf{u}_{1}$ denotes the direction of the main channel and $\mathbf{U}_{\mathrm{c}}$ represents the orthogonal complementary space of $\mathbf{u}_{1}$. The power of AN is distributed both in the space of $\mathbf{U}_{\mathrm{c}}$ and the direction of $\mathbf{u}_{1}$, yielding
\begin{equation}
\mathbf{x}=\sqrt{P\phi}\mathbf{u}_{1}\left(\sqrt{\theta}s+\sqrt{1-\theta}v\right)+\sqrt{P\left(1-\phi\right)}\mathbf{U}_{\mathrm{c}}\mathbf{r}.
\end{equation}
Furthermore, the SRM problem under the constraint of SOP was addressed with perfect CSI of the main channel and the distribution of a complex Gaussian random wiretap channel. However, numerical results showed that the optimal AN is always orthogonal to the information-bearing signal.

The authors of \cite{MISO3} explored two AN transmission schemes for MISOSE: (i) An on-off transmission scheme with a constant secrecy rate for all transmission periods. (ii) An adaptive transmission scheme with a varying secrecy rate during each transmission period. The transmitted signal is given by
\begin{equation}
\mathbf{x}=\mathbf{W}\left[t^T_\mathrm{IS} \; \mathbf{t}^T_\mathrm{AN}\right]^{T} =\mathbf{w}_\mathrm{IS}t_\mathrm{IS}+\mathbf{W}_\mathrm{AN}\mathbf{t}_\mathrm{AN},
\end{equation}
where $\mathbf{w}_\mathrm{IS}$ is used to transmit the information-bearing signal $t_\mathrm{IS}$ and $\mathbf{W}_\mathrm{AN}$ is used to transmit the AN $\mathbf{t}_\mathrm{AN}$. In order to degrade the quality of the received signals at Eve by transmitting AN in all directions except towards Bob, $\mathbf{w}_\mathrm{IS}$ is chosen as the principal eigenvector corresponding to the largest eigenvalue of ${\mathbf{h}}{{\mathbf{h}}^H}$, while $\mathbf{W}_\mathrm{AN}$ is comprised of the remaining ${N-1}$ eigenvectors of ${\mathbf{h}}{{\mathbf{h}}^H}$.

In the context of MISOME, the problem of optimal power allocation was tackled in \cite{MISO8} by minimizing the SOP under the secrecy rate constraint.
In this scenario, Alice knows the perfect CSI of Bob and the statistical CSI of Eve. The authors studied the problem of minimizing the secrecy outage probability given the secrecy rate and proved that the optimal power allocation can be conveniently obtained by using the method of bi-section search.

Meanwhile, the authors of \cite{MISO5} started with a generalized AN scheme for MISOME by solving an SOP-constrained SRM problem.  Unlike many existing works, which directly adopted the traditional null-space AN scheme, the authors started with a general assumption on the structure of the transmit signal ${\mathbf{x}}$. Specifically, Alice generates an orthogonal basis $\mathbf{U}=[\mathbf{u}_1,\mathbf{u}_2,\ldots,\mathbf{u}_N]$ and transmits the signal formulated as
\begin{equation}
\mathbf{x}=\sqrt{p_{1}}\mathbf{u}_{1}(\sqrt{\theta}s+\sqrt{1-\theta}r_{1})+\sum_{i=2}^{N}\sqrt{p_{i}}\mathbf{u}_{i}r_{i},
\end{equation}
where $s$ and $r_i$ denote the information-bearing signal and the AN signal in the direction of ${{\mathbf{u}}_i}$, respectively, ${p_i}$ denotes the transmit power in the direction of ${{\mathbf{u}}_i}$. Through generalizing several previously established findings related to scenarios with a single-antenna Eve, they demonstrated that the conventional null-space AN scheme remains optimal for an arbitrary number of Eve's antennas in terms of secrecy outage.

In the context of MISOME-CRNs, three AN-assisted beamforming schemes were designed in \cite{MISO6}, where Alice aims to transmit confidential information to two Bobs, namely legitimate PU and secondary user (SU) receivers, in the presence of SMA Eve in fast-fading environments.
By maximizing the achievable ESR in the large-antenna regime, the optimal power allocation was derived, and the performance in terms of the secrecy outage probability was also studied. In particular, it was shown that the interference threshold at the PU plays an important role in the beamforming design.

As an extension from MISOSE to MISOME, two AN transmission schemes were further introduced in \cite{MISO10} based on \cite{MISO3}, namely the on-off and adaptive transmission schemes. In the on-off transmission scheme, a channel-realization-independent secrecy rate was utilized for all transmission periods, leading to closed-form expressions for secure transmission probability, hybrid outage probability, and effective secrecy throughput. Conversely, in the adaptive transmission scheme, a channel-realization-dependent secrecy rate was employed for each transmission period, yielding closed-form expressions for secure transmission probability, the SOP, and effective secrecy throughput. Using these closed-form expressions, the authors optimized the power allocation and secrecy rate for both schemes to maximize the secure transmission probability and effective secrecy throughput.

In \cite{MISO14}, a randomized beamforming scheme named artificial fast fading (AFF) was proposed for MISOME. The AFF scheme employs a unique randomized weighting of information symbols across Alice's transmit antennas as
\begin{equation}
{\mathbf{x}} = {{\mathbf{w}}^H}s,
\end{equation}where ${{\mathbf{w}}^H} = \left[ {{w_1},{w_2}, \cdots ,{w_{{N_a}}}} \right]$ denotes the weighted coefficient vector. Note that the first ${{N_a}}-1$ coefficients ${w_n}\left( {n = 1,2, \cdots ,{N_a} - 1} \right)$ are randomly generated, while the last coefficient is calculated to compensate for the received signal at Bob to maintain the intended symbol $s$, yielding
\begin{equation}
x_{{N_a}}^ *  ={1 - \sum\limits_{n = 1}^{{N_a} - 1} {{h_n}w_n^ * } } .
\end{equation}
The AFF scheme facilitated the derivation of an exact secrecy rate expression for the single-antenna-Eve scenario, and a lower bound of the secrecy rate was derived for the multi-antenna-Eve case. Furthermore, a comparison between the AFF scheme and the AN scheme revealed that their relative advantages depended on the number of antennas held by Alice and Eve, i.e., when Eve has more antennas than Alice, the AFF scheme achieves a larger secrecy rate; otherwise, the AN scheme outperforms AFF.

A robust AN-aided SRM problem was considered in \cite{MISO1}, assuming perfect CSI of Bob's channel but imperfect CSI of non-colluding MSA Eves.
This scenario led to the formulation of an SRM problem with the objective of maximizing the worst-case secrecy rate by jointly designing the signal covariance ${\mathbf{W}}$ and the AN covariance ${\mathbf{\Sigma}}$ as
\begin{subequations}
\begin{align}
 &\mathop {\max }\limits_{{\mathbf{W}},{\mathbf{\Sigma }}}  \; {\mathop {\min }\limits_k  {{R_b} - \mathop {\max }\limits_{\left\| {{\mathbf{\Delta }}{{\mathbf{g}}_k}} \right\|\leq{\epsilon_e}} {R_{e,k}}} }  \\
&\text {s.t. }  {\text{Tr}}\left( {{\mathbf{W}} + {\mathbf{\Sigma }}} \right) \leqslant P, \\
&\;\;\;\;\; {\mathbf{W}}\underset{\raise0.3em\hbox{$\smash{\scriptscriptstyle-}$}}{\succ } {\mathbf{0}},{\mathbf{\Sigma }}\underset{\raise0.3em\hbox{$\smash{\scriptscriptstyle-}$}}{\succ } {\mathbf{0}},
\end{align}
\end{subequations}
where ${{\mathbf{\Delta }}{{\mathbf{g}}_k}}$ represents the deterministic uncertainty model for $k$-th Eve similar in (\ref{Uncertain:1}). To overcome the challenge brought by the CSI uncertainty, a reformulation was employed and the worst-case SRM problem was tackled by implementing a one-dimensional line search.

An AN-aided SRM problem was addressed in \cite{MISO12}, assuming perfect CSI of Bob but imperfect CSI of non-colluding MSA Eves. The authors considered the design of the transmit covariance ${\mathbf{W}}$ and AN covariance ${\mathbf{\Sigma}}$ under an achievable secrecy rate maximization formulation. Unlike the approach in \cite{MISO1}, the stochastic CSI uncertainties associated with Eves in \cite{MISO12} were handled using an outage-based formulation.

In the study presented in \cite{MISO7}, the SOP was employed as a metric to characterize the secrecy performance.
This study assumed the perfect CSI of Bob but lacked the CSI of non-colluding MSA Eves. Based on the SOP, a security region was defined, offering a spatial perspective on security.

In \cite{MISO9}, a semi-adaptive generalized AN scheme was proposed for MISOSE with the instantaneous CSI of Bob and the statistical CSI of non-colluding MSA Eves. The semi-adapative scheme fixes the achievable rate for Bob while it adaptively adjusts the secrecy rate and the power-allocation ratio based on the legitimate channel's CSI in order to maximize the secrecy rate.

The authors of \cite{MISO15} addressed the scenario of MISOME, where perfect CSI of Bob is available, along with either perfect or imperfect CSI of non-colluding MMA Eves.
They focused on the SRM problem by jointly optimizing the transmit and AN covariance matrices. With the perfect CSI of Eves, the SRM problem was formulated as
\begin{subequations}
\begin{align}
&\mathop {\max }\limits_{_{{\mathbf{W}} \succcurlyeq {\mathbf{0}},{\mathbf{\Sigma }} \succcurlyeq {\mathbf{0}},\beta  \geqslant 1}}  {C_b}({\mathbf{W}},{\mathbf{\Sigma }}) - \log \beta  \\
&{\text{s}}.{\text{t}}.  {C_{e,k}}({\mathbf{W}},{\mathbf{\Sigma }}) \leqslant \log \beta ,\quad \forall k \in \mathcal{K},\\
&\;\;\;\;\;\; {\text{Tr}}({\mathbf{W}} + {\mathbf{\Sigma }}) \leqslant P,\\
&\;\;\;\;\;\;  {\text{Tr}}({{\mathbf{\Phi }}_l}({\mathbf{W}} + {\mathbf{\Sigma }})) \leqslant {\rho _l},\forall l \in \mathcal{L},
\end{align}
\end{subequations}
where $\beta$ is the introduced slack variable.

In summary, in MISO systems, the existence of null space provides convenience for the generation of orthogonal AN, but the design of non-orthogonal AN based on the knowledge of wiretap channel is still worthy of attention and research.

\subsection{AN-aided mMIMO}

Due to the fact that the growth in the number of antennas at Alice enhances the DoF for AN, the AN-aided secure downlink transmission in mMIMO systems has sparked great research enthusiasm \cite{MIMO1,MIMO2,MIMO4,MIMO5,MIMO6}.

To address the SRM problem for mMIMO systems, an AN-assisted scheme was proposed in \cite{MIMO1} through optimizing the power allocation strategies for two distinct cases, which is, the case both Alice and Eve know the CSI of Bobs and the opposite case. The $k$-th Bob allocates a portion of $\alpha_{k}$ of its power to transmit the information-bearing signal, while the rest is used for AN transmission. Additionally, the authors proposed the optimization problem of power allocation between AN and data symbols with the objective of maximizing the secrecy rate.
Within each case, the authors also considered whether the accurate position of Eve is known to Alice. Furthermore, the study delved into the impacts of the non-ideal factors, including the channel estimation error and the uncertainty of Eve's position, on the power allocation strategies.

In a multicell mMIMO system, the study in \cite{MIMO2} focused on secure downlink transmission scenarios where the numbers of Alice's, Bobs', and Eve's antennas tend toward infinity. It is assumed that Eve's CSI is not available at Alice, so linear data precoding and AN are used to enhance confidentiality. Furthermore, to balance complexity and performance, they proposed linear precoder ${\mathbf{F}}_n$ for $n$-th BS based on matrix polynomials as
\begin{equation}
    \mathbf{F}_n=\frac1{\sqrt{N_T}}\hat{\overline{\mathbf{H}}}_{nn}^H\sum_{i=0}^\mathrm{J}\mu_i\left(\hat{\overline{\mathbf{H}}}_{nn}\hat{\overline{\mathbf{H}}}_{nn}^H\right)^i,
\end{equation}
where $\hat{\overline{\mathbf{H}}}_{nn}=\hat{\mathbf{H}}_{nn}/{\sqrt{N_T}},$ and $\boldsymbol{\mu}=[\mu_{0},\ldots,\mu_{\mathcal{J}}]^{T}$ are the real-valued coefficients of the precoder matrix polynomial which were optimized to minimize the sum of mean-squared errors and AN leakage to Bobs in the cell by using tools from free probability and random matrix theory.
Finally, the analytical and simulation results provided interesting insights for the design of secure multicell massive MIMO systems and revealed that the proposed polynomial data and AN precoders closely approach the performance of selfish RCI data and null-space-based AN precoders, respectively.


In \cite{MIMO4}, two AN-aided schemes were proposed to secure a mMIMO network with the imperfect CSI of Bob and without the CSI of Eve. In the first scheme, AN was injected into the downlink training signals to prevent Eve from accurately estimating its channel. In the second scheme, AN was utilized in both the downlink training phase and the payload data transmission phase to further deteriorate Eve's channel quality. It was found that deploying AN in the downlink training phase of massive MIMO networks has no impact on the channel estimation process at Bobs while suppressing the channel estimation process at Eve. Besides, implementing AN in both phases provided a flexible solution to improve its secrecy performance, albeit at the price of higher complexity.

For the sake of minimizing system complexity and reducing channel estimation overhead, a low-complexity beamforming and AN scheme was presented in \cite{MIMO5} for mMIMO systems over the Ricean fading channel. The AN scheme using only the specular component is described as
\begin{equation}
\mathbf{x}=\sqrt{p}\mathbf{W}\mathbf{s}+\sqrt{q}\mathbf{V}\mathbf{z}=\sum_{i=1}^{K}\sqrt{p}\mathbf{w}_{i}s_{i}+\sum_{i=1}^{M-K}\sqrt{q}\mathbf{v}_{i}z_{i},
\end{equation}
where $p$ and $q$ are the transmit data power and AN signal power. In addition, a tractable closed-form lower bound for the achievable ESR was derived, while the optimal power allocation was determined through asymptotic analysis with the aim of maximizing the achievable ESR. The analytical results revealed that the ESR improves as the increase of Ricean factor increases and converges to a specific constant as the number of antennas at Alice increases.

In the context of mMIMO-NOMA networks, the authors of \cite{MIMO6} examined an AN scheme utilizing minimum mean-squared-error estimated CSI. In the training phase, the BS uses the minimum mean-squared-error estimated technique to estimate the CSI. After the training phase, the precoder vector and the null-space AN vector are obtained through the estimated CSI. Following this, the authors proceeded to derive the ESR and proposed a joint power allocation strategy to maximize the ESR. The results indicated that the combination of the mMIMO technique and AN significantly enhances the secrecy performance of NOMA networks.

In summary, in mMIMO systems, AN-assisted downlink transmission techniques are widely used because the substantial increase in the number of transmit antennas provides a large number of spatial degrees of freedom, among which the design of precoding matrices and the optimization of power allocation need to be widely studied.

\subsection{AN-aided NOMA}

\begin{figure}[t]
\centering
\includegraphics[width=3.5in,height=2.6in]{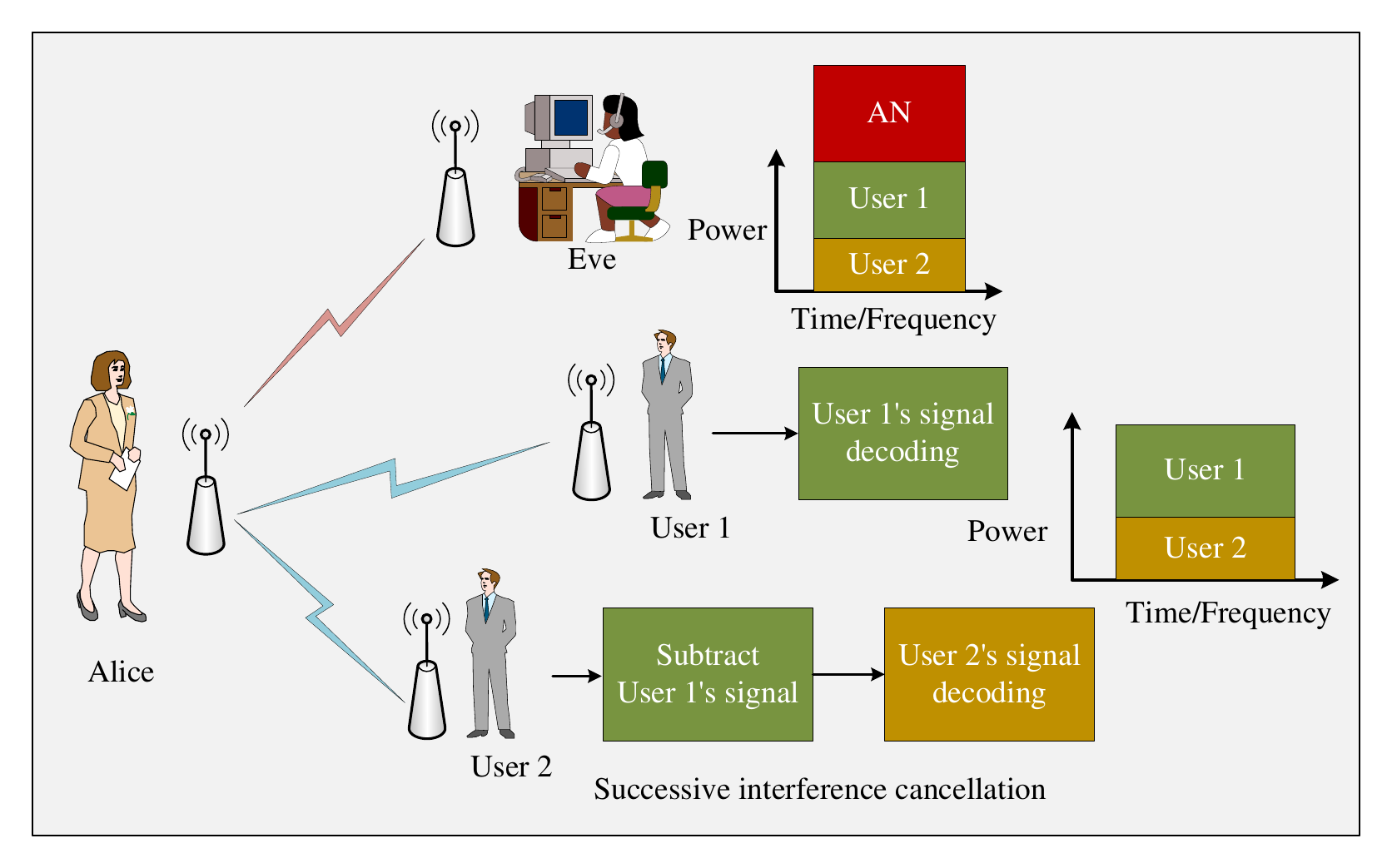}
\vspace{+3mm}
\caption{The principle of AN-aided NOMA.}
\label{fig_NOMA}
\vspace{-0em}
\end{figure}

NOMA has been envisioned as a promising multiple-access technique to improve spectral efficiency and support extensive connectivity in forthcoming wireless networks. To tackle the security concerns inherent in NOMA, the integration of AN with NOMA has garnered growing interest, as highlighted in various studies such as \cite{NOMA1,NOMA3,NOMA2,NOMA4}. As shown in Fig. \ref{fig_NOMA}, a specific user treats the remaining users' signals as interference and performs the successive interference cancellation to demodulate its own signal. At the same time, the AN is injected into the null space of the overall channel to avoid overlapping effects on the demodulation of multiple users.

Concretely, Ref. \cite{NOMA1} proposed a new two-phase FD-based AN scheme for NOMA in single-input multiple-output single-antenna-eavesdropper (SIMOSE) systems, with various relay selection methods, enabling simultaneous communicate with both a multi-antenna near-user and a far-user, where multiple FD DF relays are employed. In the first phase, using the NOMA technique, Alice conveys the information-bearing signal $s = \sqrt {{\alpha _1}} {s_1} + \sqrt {1 - {\alpha _1}} {s_2}$ to the near-user, where ${s_2}$ is the signal for the far-user that cannot be directly delivered. Meanwhile, the chosen relay ${{\text{R}}_i}$ emits the AN $v$ to confuse Eve with power ${P_1}$
\begin{equation}
\begin{gathered}
  {{\mathbf{y}}_1} = \sqrt {{P_s}} {{\mathbf{h}}_{a{b_1}}}s + \sqrt {{P_1}} {{\mathbf{h}}_{i{b_1}}}v + {{\mathbf{n}}_{{b_1}}}, \hfill \\
  {z_1} = \sqrt {{P_s}} {g_{ae}}s + \sqrt {{P_1}} {g_{ie}}v + {n_e}, \hfill \\ 
\end{gathered} 
\end{equation}
{where ${{\mathbf{h}}_{a{b_1}}}$ denotes the CSI between Alice and the near-user, ${{\mathbf{h}}_{i{b_1}}}$ represents that between the chosen relay and the near-user, ${g_{ae}}$ denotes that between Alice and Eve, and ${g_{ie}}$ represents that between the chosen relay and Eve.}
In the second phase, the selected relay performs exclusive OR (XOR) operation on the message intended for the far-user and the AN before broadcasting the combined signal
\begin{equation}
\begin{gathered}
  {y_2} = \sqrt {{P_2}} {{\mathbf{h}}_{i{b_2}}}\left( {{s_2} \oplus v} \right) + {n_{{b_2}}}, \hfill \\
  {z_2} = \sqrt {{P_2}} {g_{ie}}\left( {{s_2} \oplus v} \right) + {n_e}, \hfill \\ 
\end{gathered}  
\end{equation}
{where ${{\mathbf{h}}_{i{b_2}}}$ denotes the CSI between the chosen relay and the far-user.}
By utilizing the null-space beamforming, self-interference cancellation techniques, and the DF-XOR cooperative protocol, the scheme efficiently eliminated the AN at the near-user, far-user, and the selected relay, while simultaneously impairing the decoding capability of Eve.

Moreover, an examination of the SOP was conducted in \cite{NOMA3} for a large-scale downlink system featuring FD NOMA transmission supported by AN, where MSA Bobs are randomly distributed according to a homogeneous Poisson point processes in the presence of a certainly located single-antenna Eve. A secure cooperative communication scheme was proposed, in which nearby NOMA users operating in FD mode served as jammers, generating pre-shared pseudo random AN to enhance the PLS. Closed-form expressions in terms of the SOP for a user pair were acquired.

Additionally, the authors of \cite{NOMA2}  explored the integration of reconfigurable intelligent surface (RIS) and NOMA. Since Eve may enjoy similar performance gains brought by RIS as Bobs, an AN strategy was proposed to maximize the secrecy rate. The study revealed that the proposed strategy offers superior secrecy performance while requiring less AN power in comparison to the benchmark schemes. Meanwhile, the authors observed that increasing the number of RIS elements could reduce AN power, although this effect diminished as the number of RIS elements grew sufficiently large. Furthermore, they noted that an increase in the number of transmit antennas reduce AN power when Eve is in close proximity to Alice, but AN power increases when Eve is at a greater distance.

Furthermore, the study presented in \cite{NOMA4} delved into an AN-aided beamforming scheme for MISO-NOMA systems as
\begin{equation}
\begin{gathered}
\begin{gathered}
  {y_i} = {{\mathbf{h}}_i}{{\mathbf{w}}_i}x + {n_i}, \; i = 1,\; 2,\hfill \\
  {{\mathbf{y}}_e} = {\mathbf{G}}{{\mathbf{w}}_1}x + {\mathbf{GVr}} + {n_e}, \hfill \\ 
\end{gathered} 
\end{gathered}
\end{equation}
where $x = {\sqrt {{\alpha _1}\phi } {s_1} + \sqrt {\left( {1 - {\alpha _1}} \right)\phi } {s_2}}$, ${{\mathbf{w}}_1}$ denotes the beamforming for transmit signal, $0 < \phi  < 1$ represents the power allocation for information-bearing signal, $1 - \phi$ is the power of AN ${\mathbf{Vr}}$, while ${\alpha _1}$ and $1-{\alpha _1}$ stand for the power coefficients for users 1 and 2, respectively.
Emphasizing the practical scenario of the imperfect worst-case successive interference cancellation which is a distinctive characteristic in NOMA transmission, the authors derived a closed-form expression for the SOP to quantitatively represent the secrecy performance.

Generally, the deployment of AN in NOMA not only requires balancing the trade-off between the information-bearing signal and AN, but also demands allocating the power of different symbols for multiple users to satisfy the well-known successive interference cancellation technology.

\subsection{AN-aided OFDM}

Despite its utility, many AN techniques cannot be applied to MIMO systems when Alice has fewer antennas than Bob, which is due to null space constraints. However, in scenarios where the spatial DoF is non-existent, AN can be implemented in the time domain, a concept extensively explored in OFDM systems. Specifically, a time-domain AN generation technique was considered in \cite{OFDM8} for SISO-OFDM systems, which assumes OFDM as the transmission scheme, and generates AN in a different time slot by exploiting the redundancy derived from CP. This approach was extended to MIMO-OFDM in \cite{OFDM9}, where the time-domain AN also resides in a null space of the time-domain channel, which can be expressed as a Toeplitz matrix. Since this method uses the DoF derived from the CP, the constraint becomes ${N_a}\left( {1 + {N_{CP}}/N} \right) > {N_b}$, where ${{N_{CP}}}$ is the length of CP, and $N$ is the number of inverse fast Fourier transform points. For the design of time-domain AN, researchers in \cite{OFDM7,OFDM6,OFDM1} explored cases involving discrete inputs, multi-user OFDM channels, and AF relay systems, respectively. Moreover, note that time-domain and spatial-domain AN techniques can be combined in MIMO-OFDM systems \cite{OFDM5} to introduce additional optimization dimensions \cite{OFDM4}. Furthermore, the authors of \cite{OFDM3} proposed a frequency-domain AN-aided key generation transmission strategy for SWIPT in OFDM access systems, while the authors of \cite{OFDM2} assumed a shared AN signal scheme for hybrid parallel power-line/wireless OFDM communication systems.

In \cite{OFDM8}, AN-aided PLS was studied for SISO-OFDM systems, where temporal AN is injected in the time domain by exploiting the cyclic prefix CP. The impact of channel delay spread, CP length, power allocation, and precoder design on secrecy performance was analyzed. Notably, even without Eve’s instantaneous CSI, the proposed scheme achieved secrecy rates close to the full-CSI scenario.

Ref. \cite{OFDM7} studied an AN design for OFDM wiretap channels with discrete inputs. The authors formulated an SRM problem under a power constraint and showed how subcarrier power could be optimized accordingly. To further improve secrecy, a novel time-domain AN scheme was proposed, leveraging the CP to inject AN within the null space of the legitimate channel, effectively exploiting temporal degrees of freedom without affecting Bob's reception.

The authors of \cite{OFDM6} introduced a time-domain AN to an OFDM wiretap channel with multiple Bobs and a single-antenna Eve. The authors formulated a SSR maximization problem involving subcarrier allocation, power allocation, and AN design. They first optimized subcarrier allocation, then used a low-complexity Lagrange dual method for joint power and AN optimization. Simulation results validated the effectiveness of the proposed approach.

In \cite{OFDM1}, a time-domain AN scheme was proposed for an OFDM relay system, where Alice transmits AN-bearing signals to the relay, while the destination simultaneously sends jamming signals to disrupt eavesdropping. In the amplify-and-forward phase, the relay forwards the received signal to the destination. To maximize secrecy rate, joint power allocation for Alice and the destination was optimized. An iterative inner convex approximation algorithm was developed to efficiently solve the resulting non-convex problem.

For MIMO-OFDM systems in the presence of SMA Eve, hybrid AN injection along the temporal and spatial dimensions was investigated in \cite{OFDM5}. In the novel hybrid spatial and temporal AN design, the data is precoded in the frequency domain, and then data-spatial AN is injected into the direction orthogonal to the data vector. After performing IFFT variations and adding CP to the data-spatial AN, the signal sent by Alice was formulated by
\begin{equation}
{\mathbf{s}}_A={\mathbf{T}}^{cp}{\mathbf{F}}^{H}{\mathbf{P}}_{N_A}({\mathbf{A}}{\mathbf{x}}+{\mathbf{B}}{\mathbf{d}}^s)+{\mathbf{Q}}{\mathbf{d}}^t,
\end{equation}
where ${\mathbf{P}}_{N_A}$ is a permutation matrix that rear-ranges the precoded subcarriers, ${\mathbf{A}}$, ${\mathbf{B}}$, and ${\mathbf{Q}}$ are the precoders of information-bearing signal, spatial AN, and temporal AN. ${\mathbf{d}}^s$ and ${\mathbf{d}}^t$ are the spatial AN vector and the temporal AN vector modeled as complex Gaussian random vectors.
Assuming that Alice knows the perfect CSI of Bob but does not know the instantaneous CSI of the passive Eve, the closed-form expressions for the secrecy rate and average secrecy rate are derived, focusing on the asymptotic case with a large number of transmit antennas.

In the presence of MMA passive Eves, an existing AN technique was extended to MIMO-OFDM systems in \cite{OFDM4}. A multi-carrier optimization problem was formulated to jointly design the transmit beamforming and AN covariance matrices, ensuring reliable transmission to the legitimate user while limiting information leakage to Eves.

Ref. \cite{OFDM3} proposed a frequency-domain AN strategy for OFDM-based SWIPT systems to enhance PLS while ensuring energy harvesting at energy receivers. The authors formulated a weighted sum secrecy rate maximization problem involving power allocation, subcarrier assignment, and power splitting. Due to the non-convexity of the problem, they developed a Lagrange duality-based solution along with a low-complexity suboptimal algorithm.

The authors of \cite{OFDM2} introduced a novel AN scheme for indoor hybrid power-line/wireless OFDM systems, where Bob shares AN with Alice to enhance security against SSA Eve. Bob first transmits low-SNR AN to Alice, which is then amplified and combined with data for transmission back to Bob over high-SNR subchannels. The study also examined how transmit power and power allocation affect secure throughput under both one-link and two-link eavesdropping scenarios.

In summary, for OFDM systems, the introduction of time-domain AN promotes the diversity of AN generation methods. Furthermore, the combination of spatial-domain AN and time-domain AN deserves the attention of scholars in the future.

\subsection{AN-aided PLA}

Authentication technology plays a crucial role in maintaining secure communication, particularly in wireless environments, by allowing trusted users to verify the source of received transmissions. However, conventional authentication methods are typically implemented through cryptographic protocols in the MAC layer or above, which suffer from the disadvantage of the need for secret keys, the absence of information-theoretic guarantees, and the inability to provide covertness in contrast to physical layer authentication (PLA). The authors of \cite{Au1}  highlighted the potential of using AN to further obscure the message authentication code tag, thereby achieving information-theoretic security in PLA. They observed that AN significantly enhances security, with diminishing returns when the quality of CSI knowledge is poor. Additionally, the authors of \cite{Au2} demonstrated that AN-aided message authentication code can effectively resist key recovery attack, and they also proposed an AN-aided physical-layer phase challenge-response authentication scheme for practical OFDM transmission in \cite{Au3}. Furthermore, the authors of \cite{Au4} explored the use of AN added to received signals in message frames to enhance authentication performance in the presence of time-variant channels.

In the fingerprint embedding authentication framework, the authors of \cite{Au1} explored whether AN still improves security with only imperfect CSI available at Alice and Bob. Their findings indicated that diminishing returns in terms of security improvements when the quality of CSI knowledge is poor.

To construct new message authentication codes, the authors of \cite{Au2} introduced AN-aided message authentication codes. By formulating key recovery as a channel coding problem, the tag generation process is viewed as encoding a shared key using a message-specified code. The authors showed that AN can effectively resist key recovery attacks, even in scenarios where Eve has unlimited computational resources.

Ref. \cite{Au3} proposed an AN-aided physical layer phase challenge-response authentication scheme for OFDM systems. Using Tikhonov-distributed AN to protect phase-modulated keys, this scheme involves Alice sending a plot signal for Bob to estimate subcarrier phases, followed by Bob responding with a tagged signal embedding both the key and AN.

The authors of \cite{Au4} presented a CSI-based PLA method to counter spoofers with distinct spatial features. Under time-invariant channels, binary hypothesis testing was performed via power spectral density comparisons, while the use of AN was explored to enhance robustness in time-varying scenarios.

In a nutshell, in PLA systems, the role of AN is to increase the number of times that a single shared key can be used before being compromised. However, in practice, imperfect CSI caused by channel estimation errors, channel-constrained feedback, and delayed feedback may make AN generation more complicated.

\subsection{AN-aided RIS}

\begin{figure}[t]
\centering
\includegraphics[width=3.5in,height=3.5in]{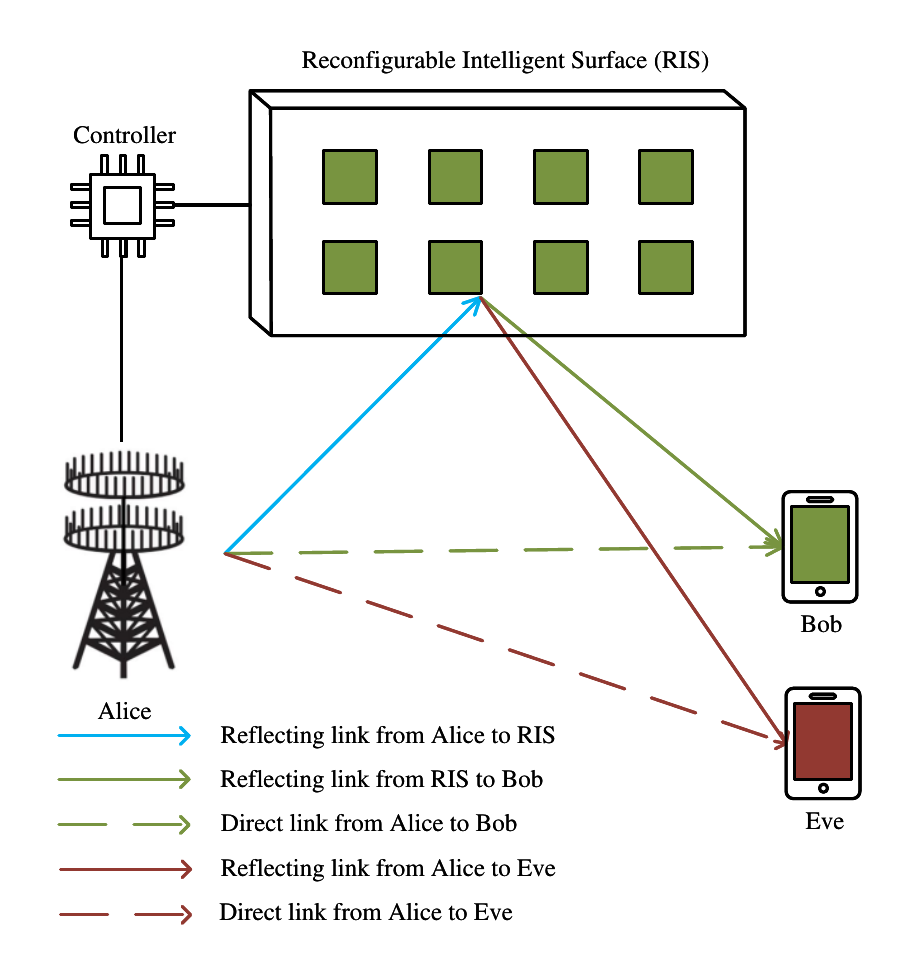}
\vspace{+3mm}
\caption{Framework of the AN-aided RIS wireless secure communications.}
\label{fig_RIS}
\vspace{-0em}
\end{figure}

The RIS, also known as the intelligent reflecting surface, has arisen as a potential candidate for the upcoming 6G wireless communications. Since the reflecting elements on RIS can manipulate the phase shifts of reflected signals, RIS benefits from considerable multi-path diversity gains without the expense of expensive hardware requirements. With the integration of RIS, AN can improve secrecy performance more effectively. As can be inferred from Fig. \ref{fig_RIS}, due to the introduction of RIS reflecting link, the joint design of transmit beamforming with AN and RIS phase shifts becomes a novel question for the AN-RIS systems \cite{RIS1,RIS2,RIS3,RIS4,RIS6,RIS5}.

First of all, the authors of \cite{RIS4} examined the effectiveness of AN in improving secrecy rates in RIS-assisted systems, where a RIS near single-antenna Bob helps SMA Alice communicate securely in the presence of multiple-antenna Eves. A joint optimization problem of transmit beamforming, AN, and RIS phase shifts was formulated and solved via an AO algorithm. Results showed that AN significantly enhances secrecy, especially when many Eves are located near the RIS.

In \cite{RIS1}, a RIS-assisted AN-aided secure MIMO system was studied, where Alice, Bob, and Eve have multiple antennas and full CSI is known at Alice. The secrecy rate maximization problem was solved by jointly optimizing the transmit precoding, AN covariance, and RIS phase shifts using a BCD algorithm, with Lagrangian and majorization-minimization (MM) methods for subproblems. The approach was also extended to a multi-Bob scenario.

To enhance the security of intelligent omni surfaces (IOS), which support both reflection and refraction, AN-aided beamforming was proposed in \cite{RIS2}. The SRM problem involving beamforming, AN, and IOS phase shifts was formulated under power and unit-modulus constraints. A BCD algorithm was used to solve it, with the Lagrangian dual method for beamforming and AN design, and QCQP for IOS phase shifts.

Additionally, the authors of \cite{RIS3} enhanced security in RIS- and AN-assisted mobile-edge computing (MEC) for IoT. By optimizing RIS phase shifts, receive beamforming, AN covariance, offloading time, transmit power, and local computing, they minimized Bobs' secure energy consumption. The nonconvex problem was solved via alternating optimization combining SDR and Dinkelbach’s method.

In \cite{RIS6}, the authors studied RIS-assisted MISO systems without Eve’s CSI. To enhance security under a total power constraint, they proposed a joint beamforming and jamming scheme. The optimization involved minimizing Alice’s transmit power while ensuring Bob’s QoS, formulated by
\begin{subequations}
\begin{align}
  &\mathop {\min }\limits_{{\mathbf{w}},{\mathbf{\Phi }}} {P_T} \hfill \\
  &{\text{s}}.{\text{t}}.{{|\left( {{\mathbf{h}}_{IB}^H{\mathbf{\Phi }}{{\mathbf{H}}_{AI}} + {\mathbf{h}}_{AB}^H} \right){\mathbf{w}}{|^2}}}/{{{\sigma _b}^2}} \geqslant \gamma , \hfill 
\end{align} 
\end{subequations}
where $\mathbf{w}$ is the beamforming vector, $\mathbf{\Phi}$ is the phase shift matrix of RIS. $\mathbf{h}_{IB}$, $\mathbf{h}_{AB}$, $\mathbf{H}_{AI}$ represent the channel from RIS to Bob, from Alice to Bob, and from Alice to RIS, respectively. $\sigma_b^2$ denotes the noise power at Bob side. This non-convex problem was addressed by using maximum ratio transmission (MRT) beamforming to maximize Bob’s channel gain, solved via oblique manifold (OM) or MM algorithms. The leftover power at Alice was then used to generate AN in Bob’s null space. Simulation results showed both algorithms have complexity $\mathcal{O}\left( {{N^2}} \right)$ per iteration, with OM requiring more iterations but achieving better performance.

In the context of RIS-MISO systems where Eve's CSI is unavailable, an AN-aided beamforming scheme was considered in \cite{RIS5}. Similar to \cite{RIS6}, the MRT beamforming was adopted in \cite{RIS5}, and the channel gain maximization problem was cast to optimize the phase shift matrix of RIS as
\begin{subequations}\label{RISfun:1}
\begin{align}\max_{\phi_{n}}&\left\|\mathbf{h}^{H}\mathbf{\Phi}\mathbf{H}+\mathbf{g}^{H}\right\|^{2},\\\mathrm{s.t.}&\left|e^{j\phi_{n}}\right|=1,\end{align}
\end{subequations}
where $\mathbf{h}$, $\mathbf{H}$, and $\mathbf{g}$ represent the reflect and direct channel. For the sake of alleviating the computational complexity, an alternating direction (AD) algorithm was invoked by determining a specific RIS unit through a closed-form solution while keeping the other configurations fixed.  By taking the first-order derivative of (\ref{RISfun:1}), the closed-form expressions for the optimal phase shift can be given by
\begin{equation}
\phi_n^1 =\arctan\left[\left({C-D}\right)/\left({E+F}\right)\right], 
\label{phi1RIS5}
\end{equation}
\begin{equation}
\phi_n^2 =\arctan\left[\left({C-D}\right)/\left({E+F}\right)\right]+\pi,
\label{phi2RIS5}
\end{equation}
where the specific expression of $C$ to $E$ can be found in \cite{RIS5}. The results from numerical simulations demonstrated that the AD algorithm exhibits faster and more stable convergence compared to the OM method, while maintaining the same level of computational complexity as OM for each iteration.

In a nutshell, the challenge of deploying AN in RIS lies in optimizing its use to enhance security without compromising system performance. Although AN improves communication privacy, it requires precise control to avoid degrading signal quality. Future trends are likely to focus on developing advanced algorithms and machine learning techniques to dynamically adapt real-time network conditions. This will ensure that RIS can effectively balance security and performance while adapting to varying communication needs.

\subsection{AN in SISO}

Since the SISO systems dissatisfy the requirement of equipping more antennas at the transmitter than at the receiver, the null space of the channel is non-existent, so that the space-domain AN cannot be directly used \cite{SISO2,SISO4,SISO3,SISO1,SISO5}. Therefore, common AN schemes in SISO may adopt novel protocols to relieve the impact of AN on Bob, including retransmission operations \cite{SISO2,SISO4,SISO3} or improving the signal processing capability of Bob \cite{SISO1,SISO5}.

In \cite{SISO2}, a novel time-domain AN strategy was proposed for single-antenna point-to-point systems over multi-path fading channels. This strategy was designed for not only confusing Eve, but also compensating inter-symbol interference arising from the multi-path propagation effects, i.e.,
\begin{equation}
\sum\limits_{l = 1}^L {{{\left| {{h_l}} \right|}^2}{r_n} + {\text{IS}}{{\text{I}}_n} = 0,} {\text{}}n = 1,2, \cdots ,N,
\end{equation}
{where $N$ denotes the total number of symbols in a frame, while ${h_l}$ represents the CSI of the $l$-th fading path.} The study evaluated the performance of this AN strategy by analyzing its impact on effective capacity and spectral efficiency. Additionally, it considered the tradeoff between these two metrics to formulate the achievable data rate of the proposed scheme, enabling the optimization of the design of the CP for enhanced system performance.

The authors of \cite{SISO4} proposed a joint physical/MAC layer security scheme combined automatic-repeat-request (ARQ) with adaptive AN that does not rely on null-space constraints. Upon retransmission requests, an AN-canceling signal is added to enable suppression via MRC. The authors derived secure throughput formulas and showed the scheme also reduces PAPR and out-of-band emissions in OFDM systems.

In \cite{SISO3}, a joint AN and repetition coding scheme with TDD protocol was designed for single-input single-output multiple-antenna-eavesdropper (SISOSE) systems. The transmission of the private information is divided into two time slots, where Alice sends a jamming-injected signal in the first slot and sends another well-designed jamming-injected signal in the second slot as
\begin{equation}
{r_2} =  - {{{h_1}{r_1}}}/{{{h_2}}}.
\end{equation}
By combining two signals at Bob, the influence of AN can be removed. Besides, the achievable secrecy rate was maximized by deriving the optimal power allocation with the proposed scheme.

For SISOSE over quasi-static fading channels, a novel AN injection scheme was proposed in \cite{SISO1}, in which an HD Bob broadcasts pseudo-random AN $r$ to Alice in phase 1 as
\begin{equation}
{y_{a,1}} = \sqrt {{P_b}} {h_b}r + n,
\end{equation}
while Alice forwards the received signal from phase 1 along with the information-bearing signal to Bob in phase 2 as
\begin{equation}
x = \sqrt \alpha  s + \sqrt {1 - \alpha } {y_{a,1}}/\left| {{y_{a,1}}} \right|,
\end{equation}
{where ${y_{a,1}}$ denotes the signal received by Alice from phase 1, $\alpha $ represents the power allocation parameter between the information-bearing signal and the AN, ${h_b}$ stands for the CSI in phase 1, while $s$ and $r$ are information-bearing signal and the AN, respectively.}
Since Bob knows the AN he generated in phase 1, he can cancel its effect in phase 2. The scheme’s performance was analyzed via SNR, SOP, outage probability, and throughput, with optimal power allocation derived to maximize throughput under SOP and outage constraints.

The authors of \cite{SISO5} proposed a reference-free AN waveform design and cancellation scheme for SISOSE. Specifically, the reference-free AN was generated at Alice by extracting the sign information of each information-bearing signal as 
\begin{equation}
r = \sqrt {{{{P_r}}}/{2}} \left\{{{\text{sign}}\left[ {\operatorname{Im} \left( s \right)} \right] + j{\text{sign}}\left[ {\operatorname{Re} \left( s \right)} \right]} \right\},
\end{equation}
{where ${\text{sign}}\left(  \cdot  \right)$ represents the sign function of a real number, while ${\text{Im}}\left(  \cdot  \right)$ and ${\text{Re}}\left(  \cdot  \right)$ denote the imaginary and real parts of a complex number, respectively.}
Bob can reconstruct and cancel this reference-free AN by subtracting it from the received signal. The authors derived the AN cancellation performance and secrecy capacity at Bob, along with their upper bounds.

In a word, since SISO cannot provide enough spatial DoFs, the application of AN in SISO is primarily categorized into two approaches, namely, adding AN in different time slots or pre-sharing the knowledge of AN to Bob. Therefore, researchers are expected to address the ensuing problems of reduced spectral efficiency and increased computational complexity.

\subsection{AN-aided SM}

The spatial modulation (SM) is heralded as an emerging MIMO approach, which activates only an antenna in each time slot and uses the index of the activated antenna to convey additional information, which owns the merit of reduced implementation complexity. As a PLS of the spatial domain, the AN has the potential to combine with the SM, which has also been investigated in several literature \cite{SM1,SM2,SM3,SM4}. 

Due to the incompatibility between AN and SM in antenna usage modes, the research of artificial-noise-aided spatial modulation (AN-SM) encounters many difficulties. Specifically, in order to improve the security of SM, a secure unitary coded spatial modulation (UC-SM) scheme was proposed in \cite{SM1}. However, the UC-SM scheme only supports a single receive antenna and transmits the same information in two time slots to eliminate the interference of AN, which lowers the spectral efficiency. Hence, an AN-SM scheme was proposed in \cite{SM2}. In the conventional SM scheme, only one antenna is activated, thus only one active RF chain is required at the transmitter. In order to improve the security of SM, AN was added to all the transmit antennas, which means that the AN-SM has to activate all the transmit antennas otherwise it cannot eliminate the interference of AN. Furthermore, the scheme was optimized in \cite{SM4} for MISO transmission by activating two antennas, where only an antenna is additionally activated. To address the incompatibility between AN and conventional SM, the combination of AN and GSM has emerged \cite{SM6,SM7}. As the extension of SM, the GSM activates more than one transmit antenna, which naturally solves the original incompatibility problem.

In order to prevent eavesdropping in MISO systems, a secure UC-SM was proposed in \cite{SM1} to achieve a second-order transmit diversity by using a single RF chain at two time slots. Specifically, SM combined with unitary codes was designed to provide a second-order diversity at the desired receiver, where the AN in two time slots ${\alpha _1}$ and ${{\alpha}_2}$ satisfies the cancellation criterion, i.e., 
\begin{equation}
{h_{{t_1}}}{\alpha _1} + {h_{{t_2}}}{\alpha _2} = 0.
\end{equation}
It was shown that the proposed scheme outperforms the existing secure SM scheme at the cost of spectral efficiency as transmitting the same signal at two time slots.

In \cite{SM2}, the secrecy performance of AN-SM over Rayleigh channels with imperfect CSI was analyzed. Closed-form lower bounds and approximations for ESR were derived. Results showed imperfect CSI reduces ESR due to estimation errors. However, unlike conventional SM, AN-SM activates all transmit antennas, increasing RF chain requirements.

The authors of \cite{SM3} studied a secure transmission using differential quadrature spatial modulation (DQSM) with a multi-antenna CJ for MISO wiretap channels. Alice uses a single RF chain, shifting antenna costs to the CJ, requiring perfect synchronization. The authors derived closed-form expressions for SOP and secrecy throughput under passive and active Eve scenarios. Simulations showed that increasing transmit power at Alice or CJ improves secrecy up to a point, beyond which performance degrades.

\begin{figure}[t]
\centering
\includegraphics[width=3.5in,height=2.5in]{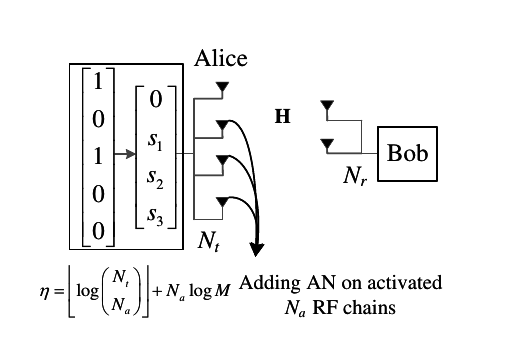}
\caption{Framework of AN-aided GSM wireless secure communications.}
\label{fig_GSM}
\vspace{-0em}
\end{figure}

Ref. \cite{SM4} proposed a combination of AN and antenna selection for the SM system. Specifically, Alice first chooses antennas for SM, and then, Alice transmits an AN through the active antenna. At the same time, another one of the remaining antennas was activated to transmit another AN. The ANs were designed by exploiting the CSI of Bob, and can only be canceled at Bob as
\begin{equation}
{h_j}{\beta _1}v + {h_i}{\beta _2}v = 0,
\end{equation}
{where $v$ is zero-mean unit-variance complex Gaussian AN, ${h_j}$ and ${h_i}$ denotes the CSI for the $j$-th and $i$-th antennas, respectively, while the coefficient ${\beta _2} =  - {h_j}/{h_i}$ is designed to cancel the AN at Bob.}
Furthermore, the secrecy rate of the proposed scheme was analyzed. Note that the proposed scheme just requires two activated transmit antennas, which means the incompatibility between AN and conventional SM still exists because the single RF chain required by SM is unattainable.

For the sake of alleviating the incompatibility between AN and conventional SM, the AN was first introduced to GSM systems \cite{SM6}. As the extension of SM, the GSM activates more than one transmit antenna, which naturally solves the original incompatibility problem by introducing a sparse null-space matrix corresponding to the index of activated antennas
\begin{equation}
{\mathbf{V}}\left( {{{\mathbf{x}}_{{l_x}}},:} \right) = {{\mathbf{V}}_0},
\end{equation}
{where ${{\mathbf{x}}_{{l_x}}}$ denotes the indices of activated antennas and ${{\mathbf{V}}_0}$ is the null space of the active antenna channel.}

As shown in Fig. \ref{fig_GSM}, GSM systems activate ${{N_a}}$ antennas and are similar to SM using the index to convey additional information bits. Thus, the AN is only added on the activated antennas, which avoids the incompatibility problem in SM.

Besides, a novel power minimized artificial noise (PMAN) was proposed in \cite{SM6} to improve the power efficiency while maintaining the jamming effect of AN as 
\begin{equation}
{\mathbf{r}} =  - {\mathbf{V}}_0^H{{\mathbf{s}}^i}.
\end{equation}
{where ${\mathbf{s}}^i$ denotes the GSM-modulated information-beraing signal.}
The power optimization ratio of PMAN was analyzed, while the theoretical bounds of the BER for both Bob and Eve were derived.

Based on a BD algorithm, a secure downlink multi-user GSM system with AN was investigated in \cite{SM7}. Specifically, the AN matrix was expressed as 
\begin{equation}
{\mathbf{V}} = [{{\mathbf{I}}_{{N_t}}} - {{\mathbf{H}}^H}{\left( {{\mathbf{H}}{{\mathbf{H}}^H}} \right)^{- 1}}{\mathbf{H}}].
\end{equation}
Particularly, the proposed MU-GSM system can achieve satisfying security performance even without AN, which is different from the single-user scenario. Finally, the simulation results showed that with AN protection and BD, the PLS can be significantly enhanced when compared with the random noise scheme and the multi-user precoding-aided SM.

In summary, the introduction of AN in the SM system may require an increased RF chains at the Alice, which needs attention from relevant researchers.

\section{Challenges and Road Ahead}

\subsection{Limited Training and Feedback}

\subsubsection{Challenge}
As indicated in \cite{Goel2}, the AN technology desires for the perfect CSI between Alice and Bob, which requires Bob to achieve perfect channel estimation based on pilot training and provide perfect channel feedback based on infinite feedback resources. However, both perfect training and perfect channel feedback are unattainable in practical applications and bring vitality to the research of limited training and limited channel feedback.
In practice, CSI is usually obtained through training-based receiver channel estimation and fed back to Alice, which is not perfect due to the estimation error. 

\subsubsection{Current Status}
In such scenarios, it is important to investigate the optimal resource tradeoff between the training phase and data transmission phase to maximize the secrecy rate \cite{T1}. By maximizing the achievable secrecy rate, the power allocation between signal and AN in both training and data transmission phases can be proposed for AN-assisted training-based schemes \cite{T2}.

Similarly, the channel estimation error \cite{CE1} and quantized channel feedback \cite{LF8} always bring imperfect CSI. For instance, with the statistics CSI of Eve, the authors of \cite{LF6} proposed a lower bound on ergodic secrecy capacity of the MIMOME wiretap channel. Their analysis showed the lower bound asymptotically matches the secrecy capacity as feedback bits and AN power approach infinity.

Furthermore, it has been observed in \cite{LF2} that enforcing stringent secrecy outage constraints puts higher requirements in terms of the number of feedback bits and the strength of the intended channel. For the sake of revealing the impact of the quantized channel feedback \cite{LF9}, several studies focused on the AN schemes with limited feedback \cite{LF5,LF7,LF4,LF3,LF1}. Specifically, the authors of \cite{LF9} revealed that only when quantized channel direction is available, the AN that was originally intended to jam Eve may not leak into Bob's channel. In addition, the study in \cite{LF5} emphasized that the channel estimation error can drastically reduce the secrecy sum-rate, and more power needs to be allocated for AN when the channel estimation error grows larger.

The impact of quantized channel feedback on the secrecy capacity achievable with AN was studied \cite{LF8}, revealing that the number of feedback bits must increase logarithmically with the transmission power to maintain consistent performance levels. Encouragingly, the authors of \cite{LF7} demonstrated that, for sufficiently high transmit power, a positive secrecy capacity can still be attained, thereby addressing the challenge of unfavorable CSI at Bob compared to Eve.

In order to enhance the secrecy performance with limited feedback, researchers in \cite{LF4} proposed a novel on-off transmission scheme to perform secure transmission and derive a closed-form expression for the secrecy throughput. Moreover, the authors of \cite{LF3} presented an adaptive transmission strategy that judiciously selects the wiretap coding parameters, as well as the power allocation between the AN and the information signal. Their simulation results indicated that allocating approximately 20$\%$ of the feedback bits to quantize channel gain information, with the remainder for channel direction quantization, yields robust performance irrespective of secrecy outage constraints. Additionally, they found that 8 feedback bits per transmit antenna achieve approximately 90$\%$ of the throughput achievable with perfect feedback. In the context of multi-cell multi-antenna networks, coordinated beamforming was employed in \cite{LF1} for multiple BSs with the assistance of limited feedback AN beamforming.

\subsubsection{Future Direction}
However, current methods primarily emphasize resource allocation challenges in the context of conventional AN scheme, whereas neglect the exploration of near-optimal channel estimation approaches and the  associated optimization problems for AN. Consequently, there is a pressing need to address these underexplored areas, particularly focusing on the development of low-complexity yet near-optimal channel estimation methods within the framework of AN. Furthermore, the AN optimization in the presence of channel estimation errors should be paid more attention, along with the theoretical analysis on the influence of channel estimation errors on AN. These issues constitute vital dimensions in the domain of AN and deserve significant attention from researchers and practitioners alike.

\subsection{Channel Correlation}

\subsubsection{Challenge}
The presence of correlation between the legitimate channel and the wiretap channel may introduce a reduction in the effectiveness of AN. This correlation is particularly pronounced in scenarios where the locations of Bob and Eve are in close proximity. When such correlation exists, it becomes imperative to develop strategies to compensate for the performance degradation that arises from this issue.

\subsubsection{Current Status}
One approach to mitigating the impact of correlation-based performance loss is to implement a correlation-based power allocation scheme, as proposed in \cite{CC1}, specifically tailored to situations where transmitter-side correlation is prevalent. Additionally, in cases characterized by high reception correlation, an AN-aided beamforming scheme was introduced in \cite{CC2}. The findings of these studies confirm that channel correlation does indeed diminish the efficacy of AN. However, this adverse effect can be partially mitigated by allocating higher power to AN.

\subsubsection{Future Direction}
It is worth mentioning that the degree of channel correlation is highly contingent on the relative positions of Bob and Eve. This dependency is especially pronounced when LoS paths dominate the overall channel as opposed to non-light-of-sight (nLoS) paths. Consequently, the modeling of channel correlation can be substantially enhanced by incorporating positioning algorithms designed to accurately determine the locations of Eve \cite{localization1,localization2,localization3}. Such positioning algorithms play a crucial role in optimizing AN strategies by providing more precise information about the spatial relationship between Bob and Eve, thus enabling more effective countermeasures against the negative impact of channel correlation.

\subsection{Adversarial Scheme}

\subsubsection{Challenge}
As mentioned before, the evaluation of secrecy performance in most research papers has primarily revolved around metrics like secrecy capacity or secrecy rate, which effectively gauge the security of information transmission by measuring the gap in channel qualities between Bob and Eve. In this conventional framework, the AN is typically treated as an interference source that degrades signal quality by reducing the equivalent SINR. However, it is important to recognize that Eve may employ various countermeasures to mitigate the detrimental effects of AN. When Eve's strategies allow for the neutralization of AN, it becomes necessary to adapt the evaluation metrics accordingly.

\begin{figure}[t]
\centering
\includegraphics[width=3.5in,height=2.5in]{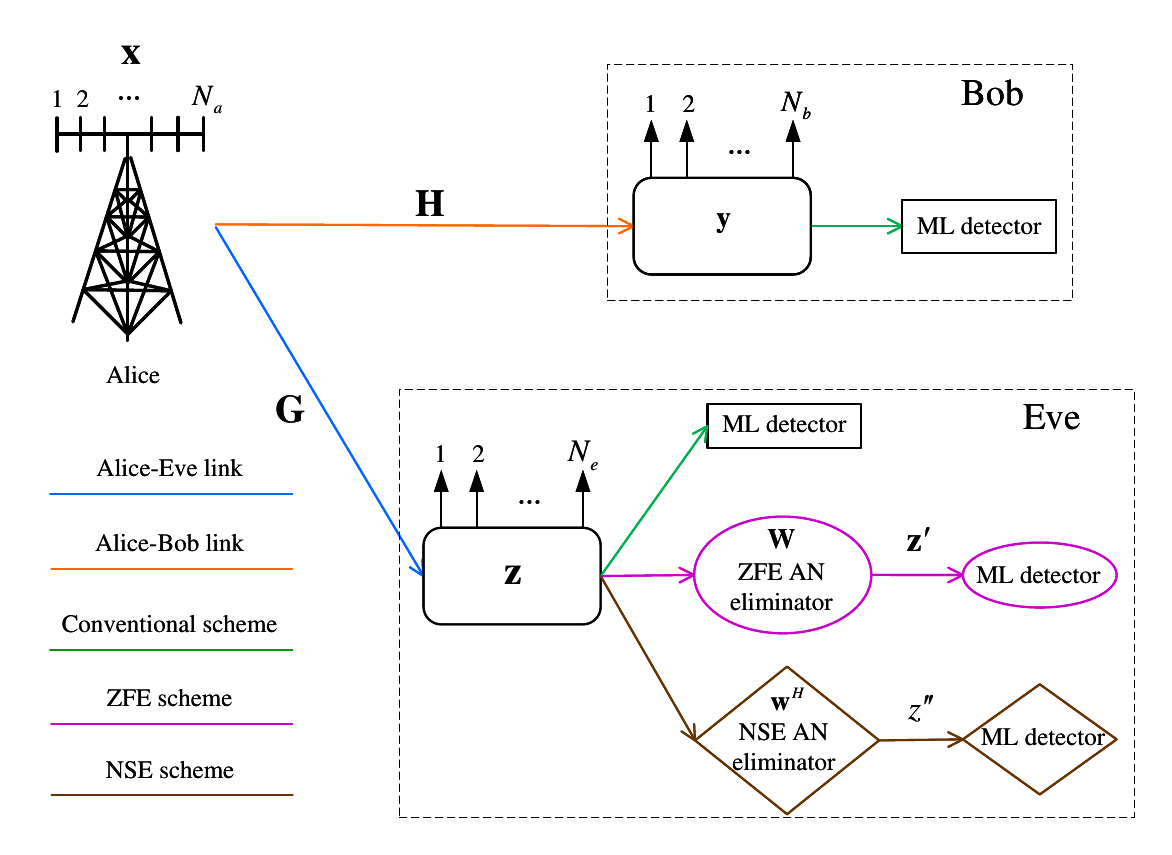}
\caption{Adversarial Model of AN.}
\label{fig_ad}
\vspace{-0em}
\end{figure}

\subsubsection{Current Status}
As shown in Fig. \ref {fig_ad}, the adversarial scheme adopted by Eve is to project the received signal to minimize the jamming effect caused by AN.

Specifically, the ZF-aided artificial noise elimination (ANE) method, as presented in \cite{ANE3}, offers a solution for Eve to mitigate the impact of AN. With the CSI of Alice-Bob link, Eve is capable of mitigating the influence of AN by multiplying the project matrix
\begin{equation}
{\mathbf{W}}{\text{= }}{\mathbf{H}}{\left( {{\mathbf{GG}}} \right)^{- 1}}{{\mathbf{G}}^H}
\end{equation}
at the left side of received signal, which proves particularly effective when Eve has more antennas than Alice, i.e., ${N_e} \geqslant {N_a}$. In \cite{ANE2}, theoretical analysis is derived to characterize the secrecy rate of the AN scheme even with the adversarial scheme. Simulation results from this work revealed that increasing AN power has little effect on increasing secrecy rate when the number of antennas at Eve exceeds that of Alice.

Moreover, the null-space elimination, outlined in \cite{ANE1}, also use the CSI of Alice-Bob link to assist Eve to achieve ANE. Compared to \cite{ANE3}, the work in \cite{ANE1} reduces the hardware requirements of Eve to ${N_e} \geqslant {N_a} - {N_b} + 1$ and provides better performance for ANE by solving
\begin{subequations}
\begin{align}
&\mathop {\max }\limits_{\mathbf{w}} \frac{{\left\| {{{\mathbf{w}}^H}{\mathbf{Gp}}} \right\|}}{{{{\mathbf{w}}^H}{\mathbf{w}}}} \\
&{\text{s}}{\text{.t}}{\text{. }} {{\mathbf{w}}^H}{\mathbf{GV}} = {\mathbf{0}}.
\end{align}
\end{subequations}

With the development of the adversarial model, it can be found that in theory Eve can mitigate the interference of AN at the cost of higher hardware requirements. However, in practice, it is difficult for Eve to cooperate with Alice and Bob to obtain their CSI.

Against this background, the authors of \cite{ANE4} proposed the hyperplane clustering algorithm to achieve ANE without the legitimate CSI. The basic principle is to use numerous AN-mixed received signals to extract signal features, which provides superior anti-AN performance in comparison to conventional techniques such as multiple signal classification.

Additionally, also with multiple AN-mixed received signals, Ref. \cite{ANE5} demonstrated that machine-learning-based ANE algorithms are effective for Eve under the constraint of 
\begin{equation}
M > {N_a} - {N_b},
\end{equation}
where $M$ denotes the number of received signals. Furthermore, the authors of \cite{ANE6} also defined the artificial noise to signal ratio (ANSR) as a metric to quantify the effectiveness of ANE without the legitimate CSI, i.e.,
\begin{equation}
{\text{ANSR}} = {{{{\left\| {{{\mathbf{w}}^H}{\mathbf{GV}}} \right\|}^2}}}/{{{{\left\| {{{\mathbf{w}}^H}{\mathbf{Gp}}} \right\|}^2}}}.
\end{equation}

\subsubsection{Future Direction}
It is worth mentioning that the countermeasures discussed above typically require Eve to commit more resources and incur higher computational complexity than Bob. These additional resources might encompass a greater number of RF chains and higher computational complexity. This observation implies that the deployment of AN as a security mechanism effectively coerces Eve into increasing hardware and computational requirements in her attempts to undermine the security of the communication.

Given these considerations, there is a compelling need to develop more robust evaluation standards that factor in the interplay between AN and specific countermeasures employed by Eve. Such standards should facilitate a comprehensive and accurate assessment of the secrecy enhancement provided by AN in diverse scenarios, accounting for the evolving landscape of secure communication protocols.

\subsection{Practical Optimization Algorithms}

\subsubsection{Challenge}
A fair number of AN optimization strategies tended to maximize the secrecy capacity or secrecy rate so as to enhance the overall secrecy performance. However, these approaches often rely on the availability of Eve's CSI, which is not always feasible in practice. Note that the instantaneous CSI of Eve may be challenging to obtain, except two main scenarios: i) Eve is active, allowing the BS to monitor her behavior and obtain its CSI, or ii) the CSI of a passive Eve can be obtained by exploiting the power leakage from her local oscillator through the received RF front end of RF receiver \cite{IE1,IE2}. The same assumption aligns with similar considerations in the field of PLS, as seen in \cite{OFDM3,MISO15}.

\subsubsection{Current Status}
An example of utilizing Eve's instantaneous CSI is presented in \cite{IE3}, where a multi-antenna Alice aims to transmit a confidential message to a single-antenna IR while transferring wireless energy to multiple multi-antenna ERs. Assuming imperfect instantaneous CSI of all channels at the transmitter, the covariances of confidential information and AN were jointly optimized to maximize the secrecy rate at the IR with the constraint that each ER receives a prescribed amount of wireless energy. Similarly, the authors of \cite{IE4} considered an AN-aided wireless powered backscatter communication system in which an FD Alice transmits multi-sinewave signals to power backscatter devices.

However, compared to the instantaneous CSI scenarios, a more operational assumption is that Alice only obtains the statistical CSI of Eve, which may occur when the location of Eve can be confirmed \cite{StaE2,StaE6}. In such cases, AN can still provide better performance than the case without any knowledge CSI of Eve, although it may not achieve the same level of performance as when instantaneous CSI is available. This difference in performance is observed in terms of secrecy rate \cite{StaE4} and energy efficiency \cite{StaE3}. In addition, some other technologies, such as antenna grouping \cite{StaE1} and cooperative jamming \cite{StaE5}, can be compatible with this scenario. For instance, researchers in \cite{StaE7} studied the AN-aided distributed antenna systems with statistical CSI for all channels by maximizing the ESR under a per-antenna power constraint.

\subsubsection{Future Direction}
While the aforementioned approaches are challenging to implement without knowledge of Eve's CSI, they often come with high computational complexity and deviate from the original intention of AN (to provide secure communication with low complexity). Therefore, practical AN optimization algorithms with low computational complexity and high jamming intensity are desired. Achieving this may involve developing novel quantization standards for situations where Eve's CSI is unavailable and pursuing closed-form solutions for AN design. These efforts could bridge the gap between theoretical developments and practical implementation of AN.

\subsection{Wider Usage Scenarios and Technological Combination}

\subsubsection{Future Direction}
The commercialization of the 5G mobile communications has brought forth new application scenarios that demand both communication and sensing capabilities, leading to the consensus on the ISAC in future wireless networks. In the long term, ISAC not only requires the provision of sensing services such as localization and imaging based on the acquired sensing information but also demands a more reliable transmission by utilizing the information in turn to enhance robust security and communication performance for the integration of architectures and waveforms. In this context, AN, with its unique ability to artificially generate interference, holds significant potential for ISAC applications.

Simultaneously, the 6G era is expected to introduce space-air-ground integrated network to provide global coverage, necessitating support for a diverse array of emerging applications in high-mobility and hostile environments. Conventional AN schemes, often used in slow fading channels \cite{SFC1}, may encounter performance degradation due to significant Doppler shifts. This issue can be addressed by integrating AN with a novel 2D modulation techniques, such as orthogonal time frequency space \cite{OTFS1}, tailored for high-mobility use cases in future communication systems.

Furthermore, due to its flexible strategies and excellent performance, the AN finds application across a wide range of scenarios and can be combined with various technologies, including but not limited to interference channel \cite{IC1}, \cite{IC2}, broadcast channel \cite{BC1}, multi-casting \cite{MC1}, \cite{MC2}, vehicle communications \cite{VC1}, cognitive wireless sensor networks \cite{CWSN1}, large-scale spectrum sharing  networks\cite{LSSS1}, layered PLS \cite{LPLS1}, IA-based networks \cite{IAN1}, time-reversal-based transmission \cite{TRBT1}, adaptive scheme \cite{AANS1}, secure state estimation \cite{SSE1}, antenna selection \cite{AS1}, space-time line code \cite{STLC1,STLC2}, stacked intelligent metasurfaces \cite{SIM1,SIM2}, fluid antenna systems \cite{FAS1,FAS2,FAS3}, cell-free MIMO \cite{CF1,CF2,CF3}, and reconfiguring wireless environments \cite{RWE1,RWE2,RWE3,RWE4}, etc.

\section{Conclusions}

In this paper, we introduced the emerging concept of AN and its evolution, along with generic system models and technical backgrounds. Furthermore, a comprehensive survey of the current state of research on various AN-enabled scenarios and AN-combined technologies was provided. Finally, we discussed the most significant research issues and challenges to tackle.


\ifCLASSOPTIONcaptionsoff
  \newpage
\fi

\newpage

\begin{IEEEbiography}[{\includegraphics[width=1in,height=1.25in,clip,keepaspectratio]{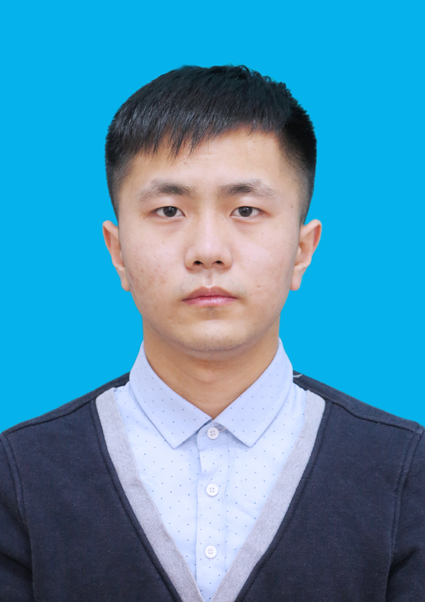}}]{Hong Niu}
received the B.E. and Ph.D. degrees in Information and Communication Engineering from the University of Electronic Science and Technology of China, Chengdu, China, in 2018 and 2024, respectively. He has been a Research Fellow with the School of Electrical and Electronic Engineering, Nanyang Technological University, Singapore. He has published over 10 journals and has participated in several projects. His current research interests include physical-layer security, reconfigurable intelligent surfaces, stacked intelligent metasurfaces, localization, and quantum computing.
\end{IEEEbiography}

\begin{IEEEbiography}[{\includegraphics[width=1in,height=1.25in,clip,keepaspectratio]{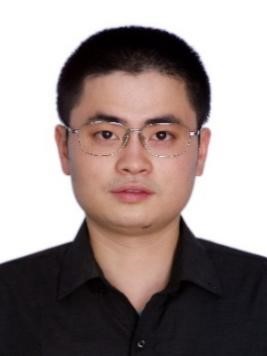}}]{Yue Xiao}
received the Ph.D. degree in communication and information systems from the University of Electronic Science and Technology of China (UESTC) in 2007. He is currently a Professor with National Key Laboratory of Science and Technology on Communications, UESTC. He has published more than 100 international journals and has been in charge of more than 20 projects in the area of Chinese 3G/4G/5G wireless communication systems. He is an inventor of more than 50 Chinese and PCT patents on wireless systems. His research interests include system design and signal processing toward future wireless communication systems. He serves as a Senior Associate Editor for IEEE COMMUNICATIONS LETTERS and an Associate Editor of IEEE OPEN JOURNAL OF THE COMMUNICATIONS SOCIETY.
\end{IEEEbiography}

\begin{IEEEbiography}[{\includegraphics[width=1in,height=1.25in,clip,keepaspectratio]{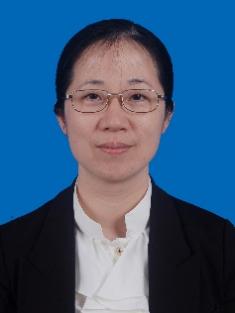}}]{Xia Lei}
received the Ph.D. degree in communication and information systems from the University of Electronic Science and Technology of China, Chengdu, China, in 2005. She is currently a Professor and supervisor of PhD candidates with University of Electronic Science and Technology of China. She has published more than 40 international journals and been involved in several projects. Her research interests are in the area of wireless broadband communication system.
\end{IEEEbiography}

\begin{IEEEbiography}[{\includegraphics[width=1in,height=1.25in,clip,keepaspectratio]{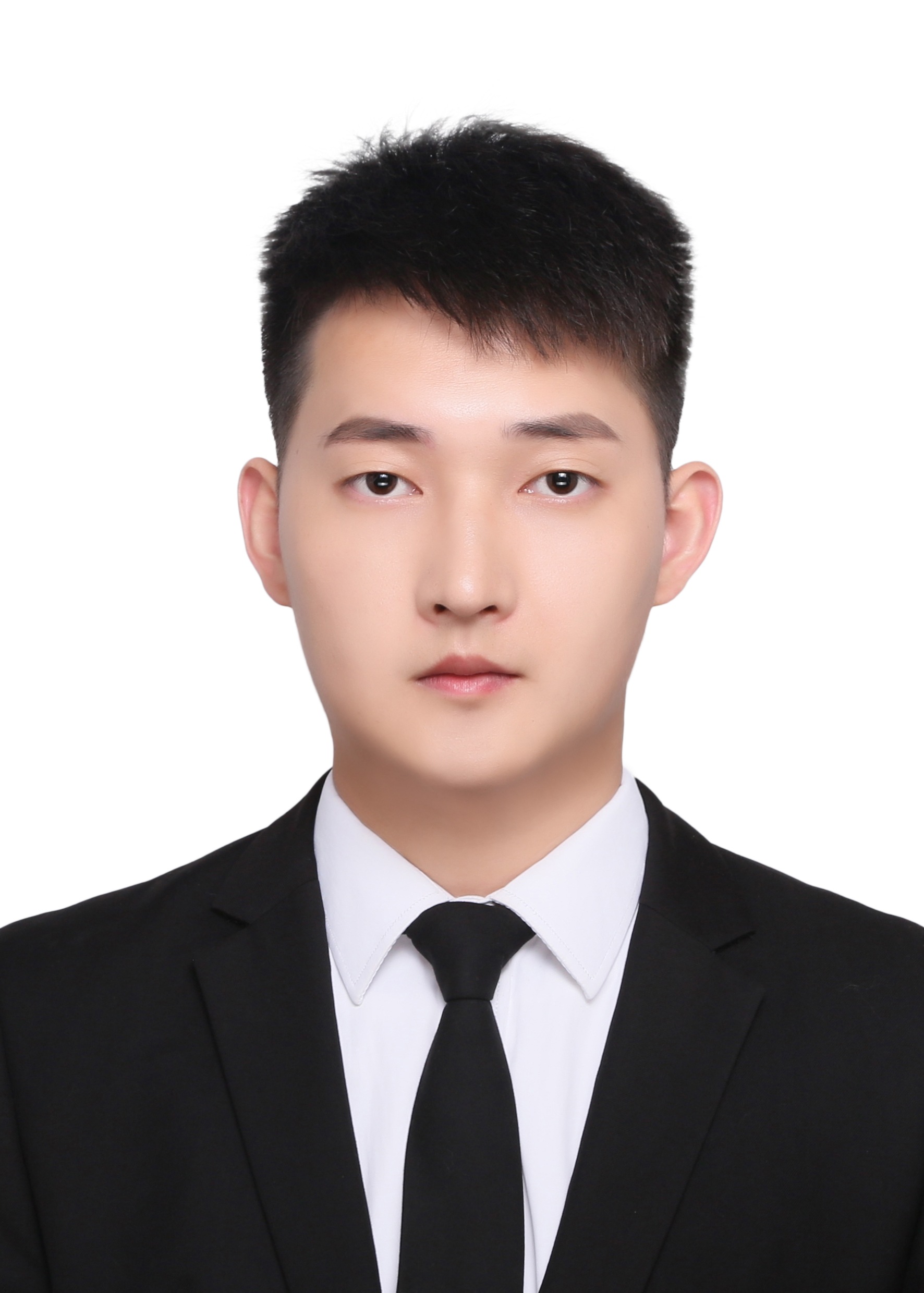}}]{Jiangong Chen}
received the M.S. degree from the University of Electronic Science and Technology of China (UESTC) in 2022, where he is currently pursuing the Ph.D. degree with the National Key Laboratory of Wireless Communications, UESTC. From 2024 to 2025, he was a Visiting Scholar with the Department of Electronic and Electrical Engineering, University College London, London, UK. His current research interests include MIMO systems, reconfigurable intelligent surface, integrated sensing and communication, and physical layer security.
\end{IEEEbiography}

\vspace{10pt}
\begin{IEEEbiography}[{\includegraphics[width=1in,height=1.25in,clip,keepaspectratio]{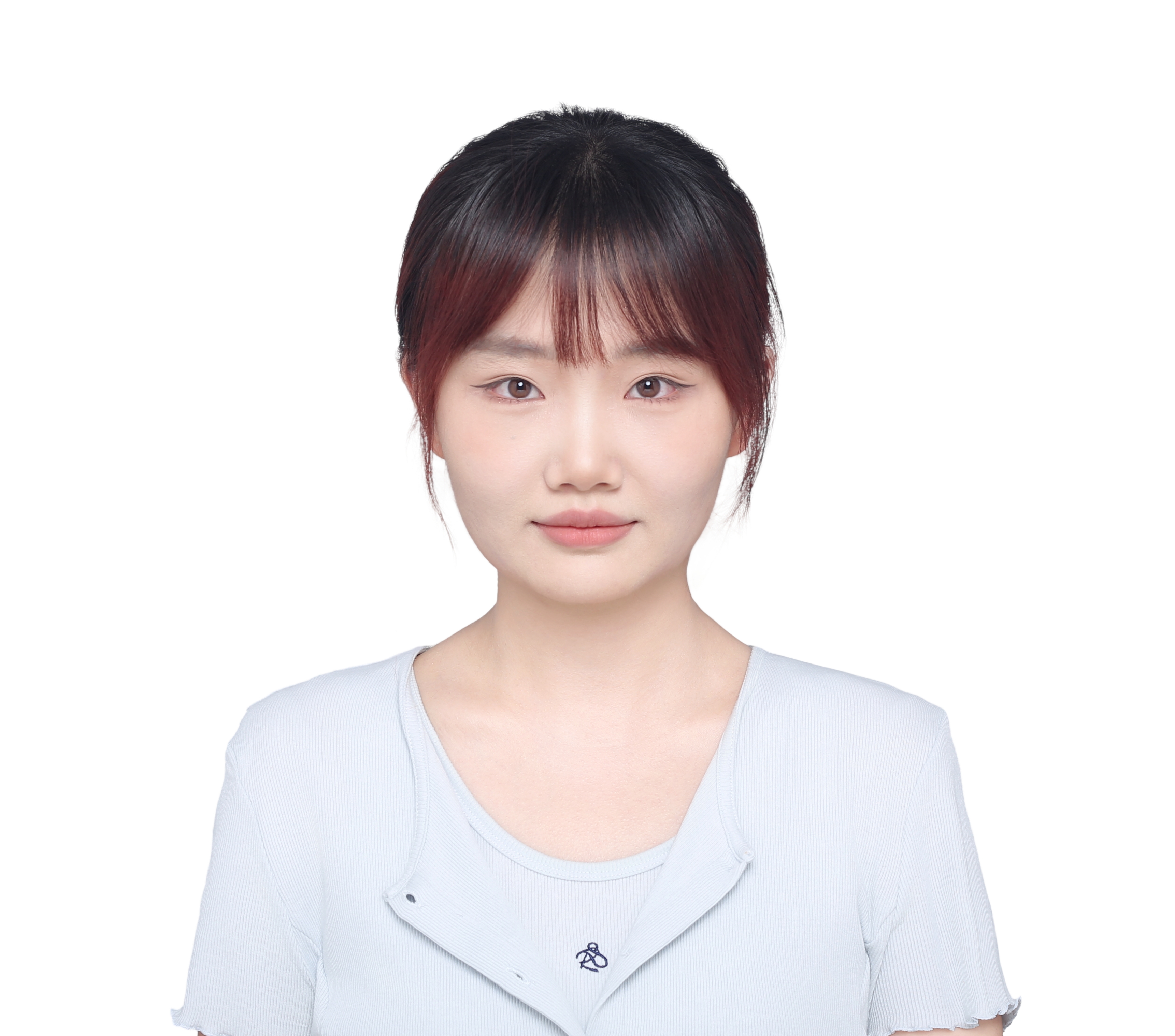}}]{Zhihan Xiao}
received her master degree in Information and Communication Engineering at the National Key Laboratory of Wireless Communications, University of Electronic Science and Technology of China, Chengdu, China, in 2025. Her research interests include wireless communications and physical layer security.
\end{IEEEbiography}

\begin{IEEEbiography}[{\includegraphics[width=1in,height=1.25in,clip,keepaspectratio]{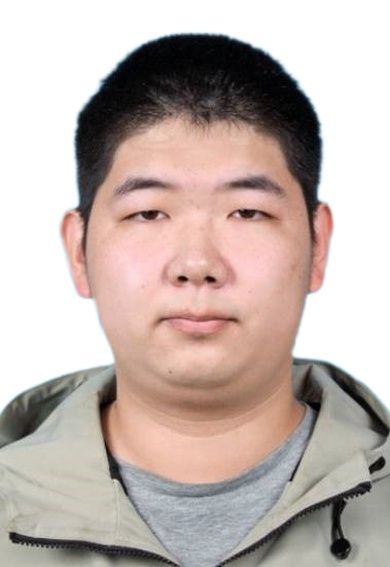}}]{Mao Li}
is currently pursuing the Ph.D. degree in Information and Communication Engineering at the National Key Laboratory of Wireless Communications, University of Electronic Science and Technology of China, Chengdu, China. His research interests include wireless communications, physical layer security, and integrated sensing and communication.
\end{IEEEbiography}

\begin{IEEEbiography}[{\includegraphics[width=1in,height=1.25in,clip,keepaspectratio]{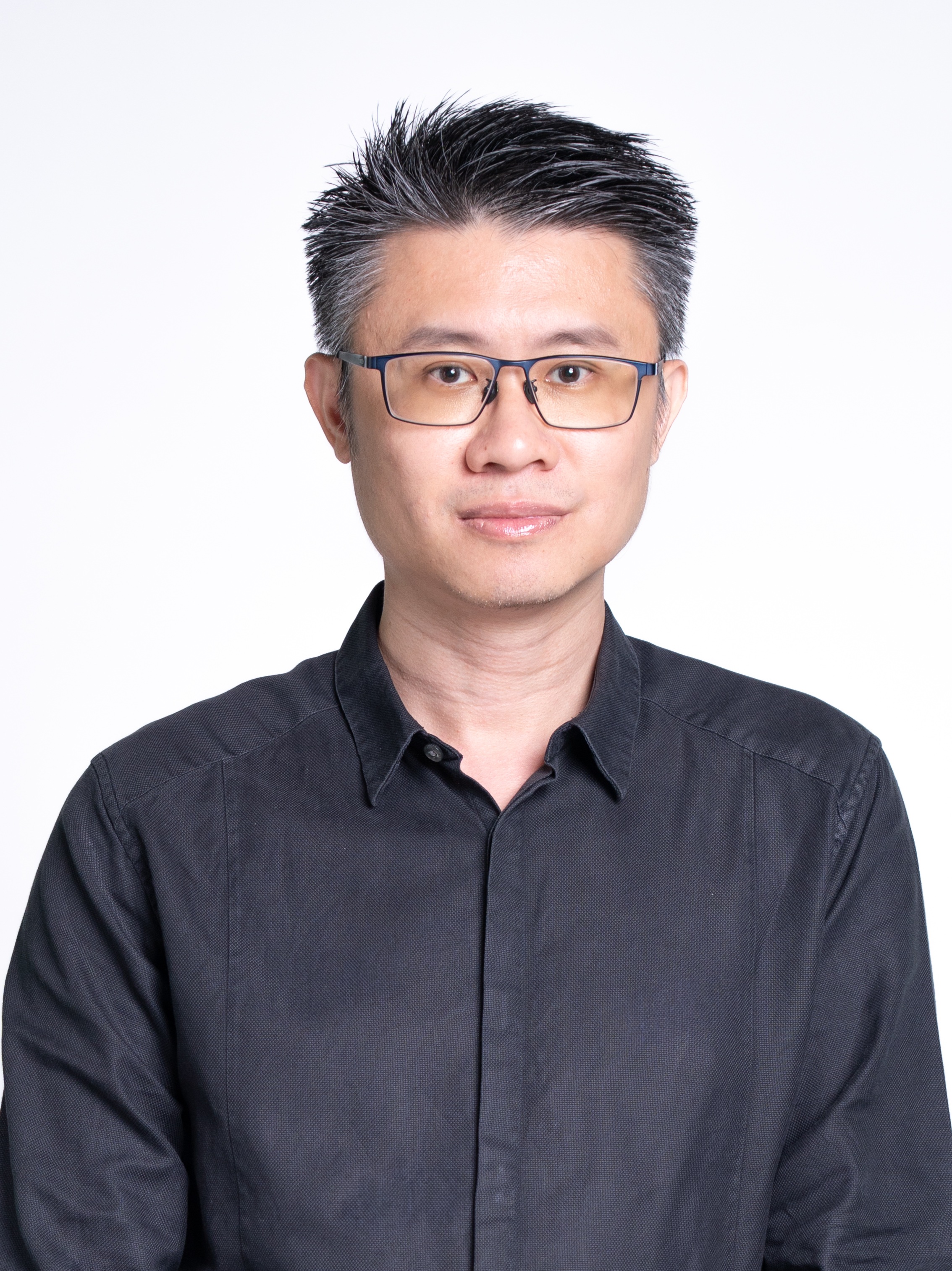}}]{Chau Yuen}
(S02-M06-SM12-F21) received the B.Eng. and Ph.D. degrees from Nanyang Technological University, Singapore, in 2000 and 2004, respectively. He was a Post-Doctoral Fellow with Lucent Technologies Bell Labs, Murray Hill, in 2005. From 2006 to 2010, he was with the Institute for Infocomm Research, Singapore. From 2010 to 2023, he was with the Engineering Product Development Pillar, Singapore University of Technology and Design. Since 2023, he has been with the School of Electrical and Electronic Engineering, Nanyang Technological University, currently he is Provost's Chair in Wireless Communications, Assistant Dean in Graduate College, and Cluster Director for Sustainable Built Environment at ER@IN.

Dr. Yuen received IEEE Communications Society Leonard G. Abraham Prize (2024), IEEE Communications Society Best Tutorial Paper Award (2024), IEEE Communications Society Fred W. Ellersick Prize (2023), IEEE Marconi Prize Paper Award in Wireless Communications (2021), IEEE APB Outstanding Paper Award (2023), and EURASIP Best Paper Award for JOURNAL ON WIRELESS COMMUNICATIONS AND NETWORKING (2021).

Dr. Yuen current serves as an Editor-in-Chief for Springer Nature Computer Science, Editor for IEEE TRANSACTIONS ON VEHICULAR TECHNOLOGY, IEEE TRANSACTIONS ON NEURAL NETWORKS AND LEARNING SYSTEMS, and IEEE TRANSACTIONS ON NETWORK SCIENCE AND ENGINEERING, where he was awarded as IEEE TNSE Excellent Editor Award 2024 and 2022, and Top Associate Editor for TVT from 2009 to 2015. He also served as the guest editor for several special issues, including IEEE JOURNAL ON SELECTED AREAS IN COMMUNICATIONS, IEEE WIRELESS COMMUNICATIONS MAGAZINE, IEEE COMMUNICATIONS MAGAZINE, IEEE VEHICULAR TECHNOLOGY MAGAZINE, IEEE TRANSACTIONS ON COGNITIVE COMMUNICATIONS AND NETWORKING, and ELSEVIER APPLIED ENERGY.

He is listed as Top 2\% Scientists by Stanford University, and also a Highly Cited Researcher by Clarivate Web of Science from 2022. He has 4 US patents and published over 500 research papers at international journals.
\end{IEEEbiography}

\end{document}